\documentclass[prd,twocolumn,superscriptaddress,showpacs,floatfix,longbibliography,nofootinbib]{revtex4-1}
\usepackage[utf8]{inputenc}
\usepackage{graphicx,amsfonts,amssymb,amsmath,xspace,physics}
\usepackage{comment}
\usepackage[pdftex,breaklinks=true,colorlinks=true]{hyperref}
\usepackage{epstopdf}
\usepackage{tabularx}
\usepackage{array}
\usepackage{multirow}
\usepackage{diagbox}
\usepackage{enumitem}
\usepackage[dvipsnames]{xcolor,colortbl}
\usepackage{stmaryrd}
\usepackage{tikz}
\usepackage{ulem}
\usepackage{soul}
\usetikzlibrary{quantikz}
\definecolor{g}{rgb}{.1,0.4,.1} 
\definecolor{b}{rgb}{0,0.2,1}
\definecolor{rouge}{rgb}{0.82,0.,0.}
\definecolor{vert}{rgb}{0.,0.82,0.}
\definecolor{darkgreen}{rgb}{0.,0.75,0.}

\definecolor{orange}{rgb}{1,0.5,0.}
\definecolor{bleu}{rgb}{0.,0.,0.82}
\definecolor{m}{rgb}{0.82,0.,0.82}
\definecolor{vert2}{rgb}{0.,0.5,0.}
\definecolor{rougeclair}{rgb}{1.0,0.7,0.7}
\definecolor{gris}{rgb}{0.8,0.8,0.8}
\newcolumntype{g}{>{\columncolor{gris}}m}

\newcommand{\beq}{\begin{equation}}
\newcommand{\be}{\begin{equation}}
\newcommand{\beqn}{\begin{eqnarray}}
\newcommand{\eeq}{\end{equation}}
\newcommand{\ee}{\end{equation}}
\newcommand{\eeqn}{\end{eqnarray}}

\newcommand{\bem}{\begin{pmatrix}}
\newcommand{\eem}{\end{pmatrix}}

\newlength{\ldag}
\newcommand{\new}[1]{{\leavevmode\color{black} #1}}
\newcommand{\older}[1]{{\leavevmode\color{blue}}}

\newcommand{\newest}[1]{{\leavevmode\color{black} #1}}
\newcommand{\old}[1]{{\leavevmode\color{blue}}}

\begin{document}
\title{
Mitigating crosstalk errors by randomized compiling: \\ Simulation of the BCS model on a superconducting quantum computer
}

\author{Hugo Perrin\normalfont\textsuperscript{a}}
\email{hugo.perrin@kit.edu}
\affiliation{Karlsruhe Institute of Technology, Institut für Theorie der Kondensierten Materie, TKM, 76049, Karlsruhe, Germany}

\author{Thibault Scoquart\normalfont\textsuperscript{a}}
\email{thibault.scoquart@kit.edu}
\affiliation{Karlsruhe Institute of Technology, Institut für Theorie der Kondensierten Materie, TKM, 76049, Karlsruhe, Germany}

\author{Alexander Shnirman}
\affiliation{Karlsruhe Institute of Technology, Institut für Theorie der Kondensierten Materie, TKM, 76049, Karlsruhe, Germany}
\affiliation{Karlsruher Institut f\"ur Technologie, Institut für Quantenmaterialien und Technologien, IQMT, 76021, Karlsruhe, Germany}

\author{J\"org Schmalian}
\affiliation{Karlsruhe Institute of Technology, Institut für Theorie der Kondensierten Materie, TKM, 76049, Karlsruhe, Germany}
\affiliation{Karlsruher Institut f\"ur Technologie, Institut für Quantenmaterialien und Technologien, IQMT, 76021, Karlsruhe, Germany}

\author{Kyrylo Snizhko}
\affiliation{Univ. Grenoble Alpes, CEA, Grenoble INP, IRIG, PHELIQS, 38000 Grenoble, France}

\date{\today}

\begin{abstract}
We develop and apply an extension of the randomized compiling  (RC) protocol that includes \new{a special treatment of} neighboring qubits and dramatically reduces crosstalk \new{effects caused by the application of faulty gates on}\old{ between} superconducting qubits in IBMQ quantum computers (\texttt{ibm\_lagos} and \texttt{ibmq\_ehningen}). Crosstalk errors, stemming from CNOT two-qubit gates, are a crucial source of errors on numerous quantum computing platforms. For the IBMQ machines, their effect on the performance of a given quantum computation is \newest{often overlooked}\older{usually underestimated by the benchmark protocols provided by the manufacturer}. Our RC protocol turns coherent noise due to crosstalk into a depolarising noise channel that can then be treated using established error mitigation schemes,  such as noise estimation circuits.  We apply our approach to the quantum simulation of the non-equilibrium dynamics of the Bardeen-Cooper-Schrieffer (BCS) hamiltonian for superconductivity, a particularly challenging model to simulate on quantum hardware because of the long-range interaction of Cooper pairs. With 135 CNOT gates, we work in a regime where  crosstalk, as opposed to either trotterization or qubit decoherence, dominates the error.  Our twirling of neighboring qubits is shown to dramatically improve the noise estimation protocol without the need to add new qubits or circuits and allows for a quantitative simulation of the BCS model. 
\end{abstract}

\maketitle
\def\thefootnote{\normalfont a}\footnotetext{These two authors contributed equally to this work}\def\thefootnote{\arabic{footnote}}

\date{\today}
\section{Introduction}
\label{sec:intro}

\par Despite the long lasting quest for a noiseless and scalable quantum computer (QC) that could achieve quantum advantage \cite{Preskill2012}, currently available platforms  require  the development and improvement of the so-called Noisy Intermediate-Scale Quantum Computer (NISQ) \cite{Preskill2018}. In these devices, qubit operations are subject to noise, which produces errors at a level that remains beyond the scope of fault-tolerant quantum error correction codes \cite{Knill1998,Lidar2013,Nielsen2012}. Nevertheless, several implementations of digital quantum computers are now accessible, like the IBMQ digital quantum computers \cite{IBMQ2021}, which can implement a universal set of quantum operations on arrays of superconducting transmon qubits (available devices containing up to 128 qubits). In order to make the most of these already existing machines, numerous error mitigation solutions have been proposed over the past decade \cite{Cai2022,Endo2021,Czarnik2021, McArdle2019, Temme2017,Li2017}, with the ultimate goal of pushing the efficiency and precision limit of NISQ until they surpass the performances of classical computers on a set of well-known quantum algorithms: quantum simulation, quantum optimization and quantum Fourier transform-based algorithms (for detailed reviews on quantum algorithms see \cite{Bharti2022,Leymann2020}). While the existing quantum hardware is quickly approaching the number of qubits theoretically required to do so, noise-induced errors remain very significant on NISQ \cite{Zhou2020,Ayral2022}, and practical solutions on how to reduce these errors to a minimum are lacking.

\par On IBMQ devices, quantum gate implementations produce errors that can mainly be attributed to the 2-qubit (CNOT) gate~\cite{Kandala2021}, with an announced error rate of 1\%, obtained by Clifford randomized benchmarking \cite{Magesan2011,Magesan2012}. However, little is known about the structure of the quantum noise channels acting on the qubits \cite{Georgopoulos2021,Johnstun2021,Georgopoulos2021b}. The noise is known to be a mixture of random, incoherent noise channels, experimentally linked to imperfect gate application and qubit decoherence, as well as coherent noise channels \cite{Iverson2020}, induced by the weak residual couplings between a qubit and its environment (neighboring qubits, readout resonator and residual electromagnetic fields for instance). Coherent noise may be correlated, both in time and in space, and lead to systematic deviations from the perfect output. One of the identified origins of coherent noise on a given qubit is linked to unwanted resonances while applying radio frequency (RF) pulses to implement a quantum gate on neighboring qubits. This effect, dubbed crosstalk \cite{Note1}, is known to be important in superconducting qubit architectures~\cite{Zhao2022,Ketterer2023}\old{, in which a weak residual coupling between neighboring qubits usually remains}. Crosstalk severely limits the performance of these NISQ implementations, in spite of considerable experimental efforts~\cite{Kandala2021,Rudinger2021,Tripathi2022} and software mitigation approaches~\cite{Ding2020,Murali2020, Xie2021}. Finally, we emphasize that previous experiments have shown that the effective noise acting on the transmon qubits of IBMQ QCs is very unstable in time, and may change drastically over the scale of a few hours/days \cite{Woitzik2023}, which limits the possibilities for benchmarking the noise channels and reusing this information.

\begin{figure*}
    \centering
    \includegraphics[width=\linewidth]{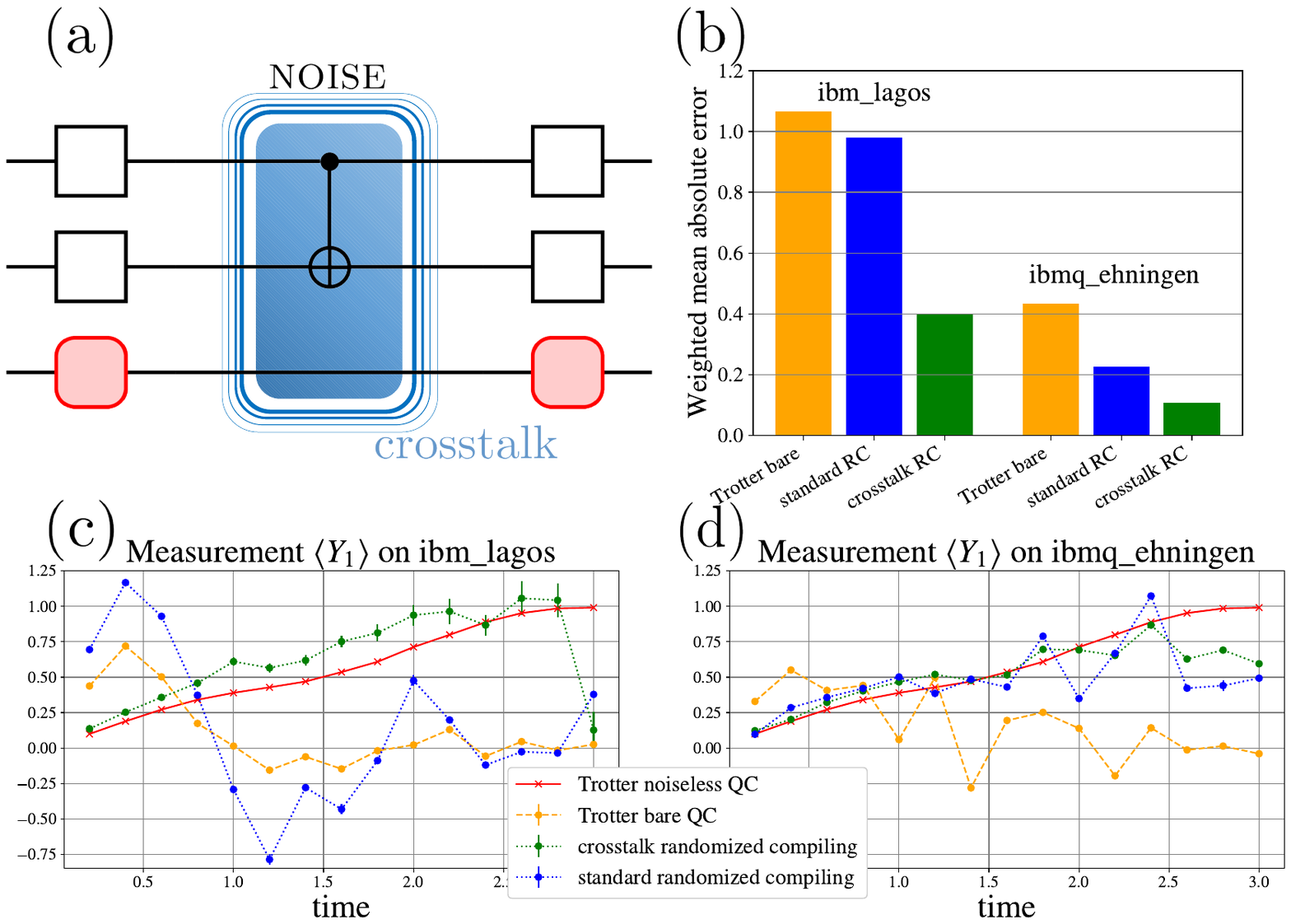}
    \caption{\textbf{Illustration of the importance of crosstalk mitigation}. (a): Pictorial representation of noise occuring during the application of a CNOT gate (blue cloud). \old{The noise can affect neighboring qubits known as crosstalk effects}\new{Crosstalk may extend the effect of this noise to the neighboring qubits}. Gates applied before and after the CNOT gates depict the RC protocol (white gates are used for the standard RC procedure, adding the red gates constitutes the crosstalk RC). A more detailed description is given in Sec.~\ref{sec:twirl}. \old{and Fig.~\ref{fig:twirling}.} (b):
    \older{Average over all quantum runs on two different IBMQ devices of the absolute error, weighted by the theoretical expectation value $\epsilon=\left|(n-p)\times \frac{p}{S}\right|$, where $p$ is the noiseless expectation value, $n$ is the value measured on the noisy QC, and the $S=\sum_{p}|p|$ the normalization for the weights\normalfont\textsuperscript{c}. $\epsilon$ is plotted without error mitigation protocols (orange bars), and with error mitigation using standard RC (blue bars) and crosstalk RC (green bars). This quantity, averaged over all the time steps and observables that we simulated (see Appendix~\ref{ap:evolution}) captures the general trend of our results and shows that crosstalk RC brings a substantial improvement on the effectiveness of our error mitigation protocol.} \newest{Figure of merit for the simulation accuracy when using different error mitigation techniques. Clearly, the effect of mitigating crosstalk is not negligible. The figure of merit is defined as $\epsilon = S^{-1} \sum_j |i_j| |(n_j-i_j)|$, where $j$ enumerates all simulated evolution durations and observables measured, $i_j$ is the ideal noiseless expectation value for the respective observable, $n_j$ is the value measured on the noisy QC; $S = \sum_j |i_j|$ represents the normalization factor, as $|i_j|$ weights prioritize the contributions of observables with expectation value away from 0.}\old{Mean relative errors without error mitigation protocols (orange bars) and with error mitigation protocols using standard (resp. crosstalk)  RC protocol (blue bars) (resp. green bars). The relative error is defined as the absolute difference between the noiseless and the corresponding noisy expectation values divided by the latter quantity. They are averaged over all the time steps and observables computed throughout the article (see Appendix~\ref{ap:evolution}) and shown separately for both IBMQ devices used.} (c)-(d): Evolution of the expectation value $Y$ of the second qubit out of three under the 3-body BCS hamiltonian (see Eq.~\eqref{eq:BCS_pauli}) with energy levels $\{-1,0,+1\}$ and coupling constant $g=0.5$ respectively on \new{both IBMQ devices}. \old{the IBQM device: \texttt{ibm\_lagos} and  \texttt{ibmq\_ehningen}.} The red curve is the noiseless trotterized dynamics, the dashed orange curve represents the simulation on a real, noisy QC. The dashed blue (resp. green) curve shows the simulated evolution when using a number of error mitigation protocols: readout-error correction, standard randomized compiling (resp. crosstalk randomized compiling), and NEC. While not allowing for perfect agreement with the ideal red curve, crosstalk randomized compiling dramatically improves the accuracy of the quantum simulation.}
    \label{fig:intro}
    \end{figure*}

    \par Error mitigation schemes are, therefore, essential in order to make NISQ a reliable platform for quantum computing. Over the past decade, several of those techniques have been proposed with the aim of turning noisy outputs into practical and reliable results. They can be classified into two categories: noise parameter-dependent and noise parameter-free error mitigation protocols. The former, which include e.g. Probabilistic Error Cancellation (PEC)~\cite{Temme2017,Li2017}, rely on a precise characterization of the noise parameters, which requires a substantial initial overhead~\cite{BlumeKohout2013,BlumeKohout2017,Endo2018,Nielsen2021}. Noise parameter-free error mitigation schemes, on the other hand, usually assume a model of noise but do not require any characterization of their parameters. Therefore, they are free from this initial overhead but fail when noise model assumptions are not met. The most prominent example is known as Zero Noise Extrapolation (ZNE), which consists of performing several experiments with increasing noise strength (which can be achieved both with experimental or algorithmic approaches), and extrapolating the results to the zero noise limit~\cite{Temme2017,Li2017,Dumitrescu2018,He2020,GiurgicaTiron2020}. Another example is the so-called Noise Estimation Circuits (NEC) technique, introduced in~\cite{Urbanek2021}. \newest{A common feature of both parameter-dependent and parameter-free methods is an exponential scaling of the required number of circuit runs with the number of noisy gates~\cite{Quek2022}}.

\par The NEC method was originally designed for noisy qubits affected by a very simple one-parameter noise channel called depolarising noise. This assumption is unrealistic for NISQ, and particularly for IBMQ quantum computers, as discussed above. Fortunately, the noise structure can be simplified using a procedure called Randomized Compiling (RC) ~\cite{Kern2005,Wallman2016, Hashim2021}. RC consists in averaging a given circuit output over different randomized versions of the original circuit in a way which effectively maps any noise channel to a simpler Pauli noise channel~\cite{Cai2019}, bringing it closer to the depolarising channel required to apply the NEC method.

\par RC is  typically performed only on qubits directly affected by CNOT gates~\cite{Rosenberg2022,Kurita2022,Vazquez2023,Kim2023}, therefore ignoring \old{Such methods, however, ignore} crosstalk errors. These are typically assumed to be insignificant, yet turn out to be harmful \new{to our simulation results} in practice. 
\new{Methods for mitigating crosstalk via RC have been previously proposed~\cite{Wallman2016} and demonstrated to be effective~\cite{Hashim2021}. In this article, we propose an extension of the twirling gate set  (see Fig.~\ref{fig:intro}-a or Fig.~\ref{fig:twirling}) which further maps crosstalk errors onto a depolarizing noise channel.}
\old{we show that an extension of the standard RC protocol to neighboring qubits together with an extension of the twirling set (see Fig.~\ref{fig:intro}-a or Fig.~\ref{fig:twirling} for a more detailed version) can turn crosstalk errors stemming from CNOT application into an effective depolarising noise channel on neighboring qubits.}We then demonstrate that using this technique drastically improves the performances of the NEC mitigation scheme implementation on IBMQ QC. We expect that similar improvements can be achieved for other error mitigation schemes that assume a depolarising noise model. This is relevant for all QC architectures affected by crosstalk. In addition, our results confirm that crosstalk errors on IBMQ devices cannot be neglected (we find that they lead to \newest{an} effective error rate \newest{per CNOT gate} of up to 5\%) and need to be mitigated.

\par To benchmark the RC and NEC methods used in this article, we perform a quantum simulation of the BCS (Bardeen-Schrieffer-Cooper) hamiltonian, the historical model describing superconductivity through the emergence of electronic Cooper pairs \cite{Bardeen1957}. Starting from an initial quenched state, we simulate the out-of-equilibrium dynamics of the system using a simple Trotter algorithm. The evolution is then monitored by measuring time-dependent expectation values of Pauli strings observables. \older{(see Fig.~\ref{fig:intro}-c,d). While not allowing for perfect agreement with the ideal red curve, crosstalk randomized compiling  dramatically improves the accuracy of the quantum simulation for generic observables (see Fig.~\ref{fig:intro}-b) without the need of adding new qubits or circuits.} Specific features make this model very suitable to benchmark the performance of QCs. First, due to the hardcore boson statistics of the Cooper pairs, the model has a natural description in terms of Pauli matrices. Second, the BCS model is integrable, so that the few-particle dynamics can be completely solved efficiently on a classical computer~\cite{Richardson1963,Richardson1964,Dukelsky2004, Rombouts2004, Faribault2009}. Finally, the BCS interaction term contains an all-to-all coupling between spins, which makes it challenging to implement on a low connectivity platform like IBMQ QC (which has a quasilinear layout of qubits). Previously, the BCS model has been implemented on IBMQ with the purpose of finding its ground state energies using variatonal quantum algorithms~\cite{Lacroix2020,Khamoshi2020,Guzman2022}. Also, an implementation of the BCS evolution operator for a NV-center-based QC with a star-shaped connectivity has been proposed recently~\cite{Ruh2023}.
\par \newest{Our key results are summarized in Fig.~\ref{fig:intro}. Specifically, Fig.~\ref{fig:intro}-b shows that crosstalk randomized compiling dramatically improves the accuracy of the quantum simulation without the need of adding new qubits or circuits (compared to the standard randomized compiling). While Fig.~\ref{fig:intro}-b presents the results averaged over multiple observables and simulation times, Fig.~\ref{fig:intro}-c,d demonstrates how the results of applying different techniques compare for one specific observable as a function of the simulated evolution time (i.e. number of Trotter steps applied). It is clear that the contribution of crosstalk is not negligible.}

\par This article is organized as follows: In Section~\ref{sec:BCS}, we Trotterize the BCS time evolution, and show how to map the corresponding unitary operator to a quantum circuit. We run naive simulations of the model on IBMQ QCs, illustrating the dramatic effect of the noise on the results. In Section~\ref{sec:twirl}, we present our improved RC procedure, and perform a comparative analysis between standard RC and crosstalk RC, by applying them to our BCS simulations. Section~\ref{sec:mitig} is devoted to our implementation of the NEC mitigation scheme, \newest{and readout error correction}. Finally, we provide in Section~\ref{sec:combining} an in-depth comparison of various levels of error mitigation --- from no mitigation to NEC assisted with crosstalk RC. We discuss the observed discrepancies between classical simulations assuming a simple depolarising noise channel and the results from real QC.

\section{Quantum simulation of the BCS model on a superconducting QC}
\label{sec:BCS}
\begin{figure}
    \centering
\includegraphics[width=\linewidth,trim={0 3cm 0 3cm},clip]{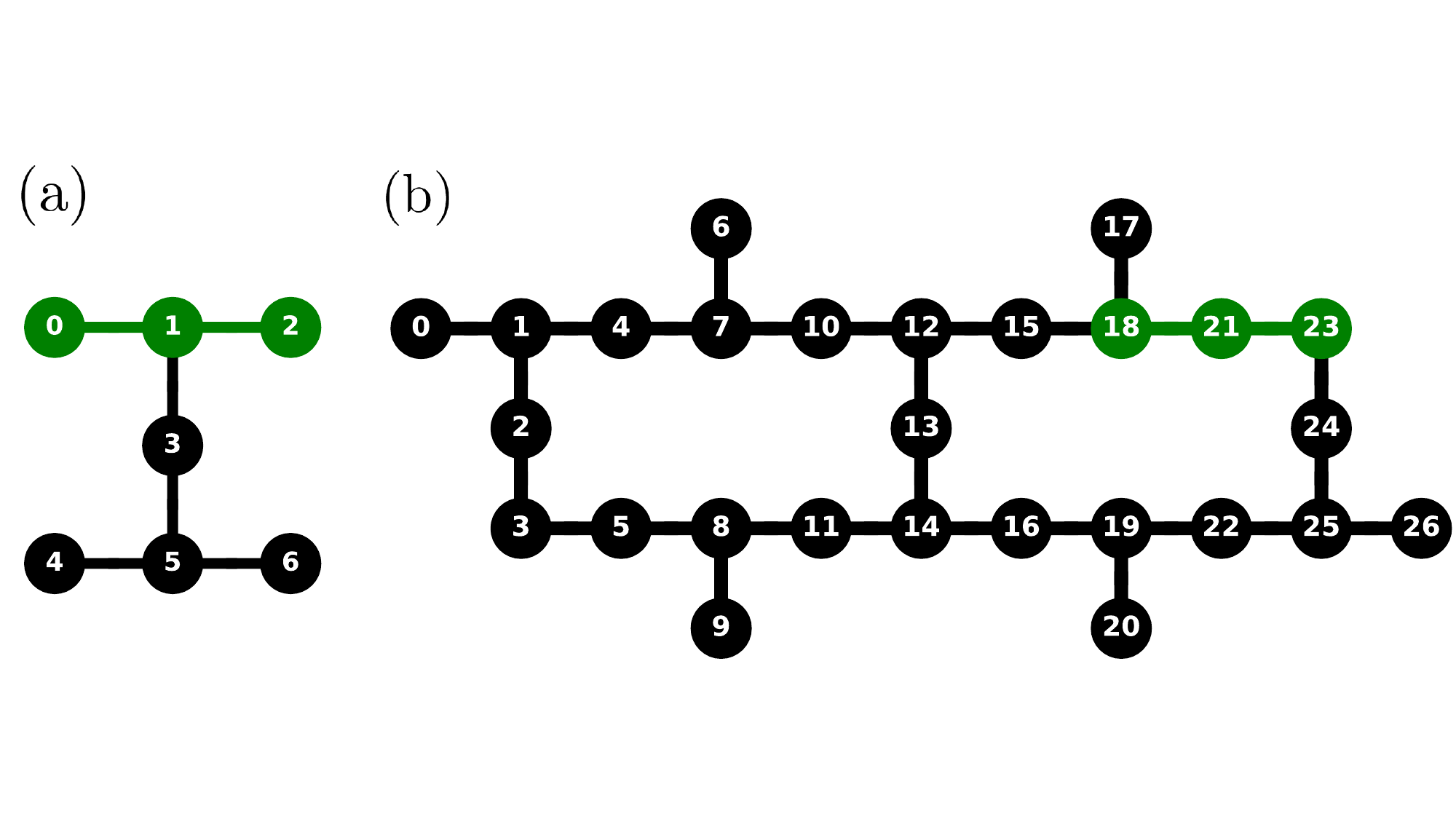}
    \caption{Coupling map of (a): \texttt{ibm\_lagos} and (b): \texttt{ibmq\_ehningen} machines. Links represent the physical junctions on which 2-qubit gates can be realised. The green qubits are used for the simulation and black qubits are the idle qubits.}
    \label{fig:coupling}
\end{figure}

\new{Throughout this work, we model a quench in the BCS model with 3 Cooper pairs in order to gauge the efficiency of both RC techniques and their combination with NEC. The dynamical evolution of various local observables is the quantitative outcome of these simulations. In this section, we show how the BCS quench can be simulated on a quantum computer and present the results of a basic simulation - without any error mitigation - setting a reference for the performance of more advanced methods.

We execute the quantum code on 3 linearly connected qubits of \texttt{ibmq\_lagos} and of \texttt{ibmq\_ehningen}, cf. Fig.~\ref{fig:coupling}. Throughout the paper, unless stated otherwise, results for \texttt{ibmq\_lagos} are presented. Results for \texttt{ibmq\_ehningen} are gathered in Appendix.~\ref{ap:evolution}.}

\subsection{Problem definition}
\label{subsec:prob_definition}

\new{The BCS hamiltonian can be represented in the spin language via 

\begin{align}
\mathcal{H}_\text{BCS}&=-\sum_{j=0}^{L-1}(\epsilon_j -\frac{g}{2})\sigma_j^z
-\frac{g}{2}\sum_{ \substack{i,j = 0\\i<j}}^{l-1}\left(\sigma_i^x \sigma_j^x + \sigma_i^y \sigma_j^y\right),
\label{eq:BCS_pauli}
\end{align}
(see Appendix~\ref{ap:BCS_to_Pauli}) where $L = 3$ is the number of energy levels and $\sigma_j^a$, $a\in \{x,y,z\}$, refers to the operator such that the Pauli matrix $\sigma^a$ is applied on the energy level $j$ and the identity operator on the other energy levels. The  eigenstates $\ket{0},\ket{1}$ of $\sigma_j^z$ denote respectively the absence or presence of a Cooper pair on the energy level $j$.

The system is initialized in state $\ket{\psi_0}$, corresponding to the mean field solution of the above hamiltonian, cf. Appendix~\ref{ap:BCS_to_Pauli}. The aim is to simulate the evolution of the system up to time $t$, described by the evolution operator $\smash{U(t)=e^{-i\mathcal{H}_\text{BCS} t}}$, and measure expectation values of observables $\bra{\psi_0}\mathcal{O}\ket{\psi_0}$, where
$\ket{\psi(t)} = U(t) \ket{\psi_0}$,
and the observables $\mathcal{O}$ are products of Pauli operators on the qubits.}

\subsection{Trotter algorithm for the unitary evolution}
\label{sec:Trotter step}

\par First, we need to implement the unitary evolution operator $\smash{U(t)=e^{-i\mathcal{H}_\text{BCS} t}}$ on the quantum computer, i.e. to break down $U(t)$ into a quantum circuit comprised of 1- and 2-qubit unitary gates that can be directly applied on the QC (native gates). This operation, dubbed transpilation, is a difficult problem that usually cannot be tackled exactly, even for simple integrable systems. To circumvent this issue, we approximate the exact unitary evolution operator using the Trotter algorithm. We slice the full time evolution into $r$ small time steps of length $\smash{\Delta t=t/r}$:
\begin{equation}
e^{-i\mathcal{H}t}=\left(e^{-i\mathcal{H}\Delta t}\right)^r.
\end{equation}
 For each time step, we use the Suzuki-Trotter formula \cite{Suzuki1976} to expand the complex exponential  \new{ at first order in $\Delta t$.}\old{, and neglect all the non-zero commutators between the  two terms of the hamiltonian such that we can write each of them in a separate exponential.} For the BCS hamiltonian this yields:
 
\begin{equation}
e^{-i\mathcal{H}_\text{BCS}\Delta t}\simeq\prod_{j=0}^{L-1}e^{i(\epsilon_j-\frac{g}{2})\sigma_j^{z}\Delta t} \hspace{-0.5cm} \prod_{0\leq i<j\leq L-1}\hspace{-0.5cm} e^{i\frac{g}{2}(\sigma_i^x\sigma_j^x+\sigma_i^y\sigma_j^y)\Delta t}.
\label{eq:Trotter}
\end{equation}

\par For each Trotter step, the error accumulated by this approximation has an upper bound of order $\mathcal{O}(\Delta t^2)$. This error scales  as $ \mathcal{O}(t \Delta t)$ for the whole dynamics (see e.g. ~\cite{Childs2018,Seetharam2021}). In our particular case however, this upper bound is not reached, and the first order Trotter algorithm remains a good approximation of the exact dynamics over the time range we consider. \new{Thus, in what follows we compare the results obtained on a quantum computer to the result of perfect Trotter evolution, and not to the exact result of continuous evolution.}

\par The exponential terms in Eq.~(\ref{eq:Trotter}) can now be straightforwardly transpiled into a quantum circuit comprised of the native 1-qubit and 2-qubit gates of the IBMQ machines. More details about this procedure can be found in Appendix~\ref{ap:transpilation}, and the resulting quantum circuit is displayed in Fig.~\ref{fig:Trotter_circuit}, where the two first exponential factors in Eq.~(\ref{eq:Trotter}) are highlighted with red boxes as an example. The rightmost one, in particular, implements the interactions between qubits $q_0$ and $q_1$. It is then simply repeated for the interaction $q_1-q_2$. For the last interaction however, a SWAP gate (composed of a series 3 alternating CNOT gates) is required, to account for the linear layout of the quantum computers. This gate exchanges the logical information contained in physical qubits $q_1$ and $q_2$, thus allowing to implement the interaction $q_0-q_2$ using physical qubit $q_1$ as an intermediary. Finally, we obtain the equivalent quantum circuit representation of an elementary Trotter step from Fig.~\ref{fig:Trotter_circuit}, which requires 9 CNOT gates in total. To simulate the dynamics up to an arbitrary time $t$, we simply concatenate $r = t/\Delta t$ copies of this circuit \cite{Note2}.

\subsection{Parameters}
\label{sec:Params}

\par We choose the following model parameters: on-site energies $\epsilon_j\in\{-1,0,+1\}$ and a coupling constant $g=0.5$\old{, which yields the value $\Delta\approx 0.46$ for the superconducting energy gap}. We then perform the Trotter evolution up to time $T=3.0$ with a time step $\Delta t=0.2$. Unless stated otherwise, quantum simulations shown throughout the main article are performed on qubits $q_0$, $q_1$ and $q_2$ of the \texttt{ibm\_lagos} machine. We choose $q_0$ as the first qubit with energy $\epsilon_0=-1$, $q_1$ as the second qubit with energy $\epsilon_1=0$ and $q_2$ as the third qubit with energy $\epsilon_2=+1$.

\subsection{Observables}
\label{subsec:Observables}

\par \new{After performing the evolution, we monitor the dynamics of the system by evaluating the expectation values of different observables, which are tensor products of Pauli matrices. These observables may bear little physical interest, but their expectation values are straightforwardly estimated from simple measurements, and their exact evolution is easily obtained classically.} We made two different experiments: in the first one we select observables from set $\mathcal{S}_1 \equiv \{Z_0, Z_1, Z_2, Z_0 Z_1, Z_1 Z_2, Z_0 Z_2, Z_0 Z_1 Z_2\}$, and in the second one - from set $\mathcal{S}_2 \equiv  \{X_0, Y_1, Z_2, X_0 Y_1, Y_1 Z_2, X_0 Z_2, X_0 Y_1 Z_2\}$ (capital letters correspond to the direction of the Pauli observable, and subscripts refer to the qubit they are measured on). We therefore have access to a total of 7+7 different observables. All the measurement outputs are sampled over 32000 repetitions.

\begin{figure*}
    \centering
    \includegraphics[width=
    0.95\textwidth,trim={0 3cm 0 3cm},clip]{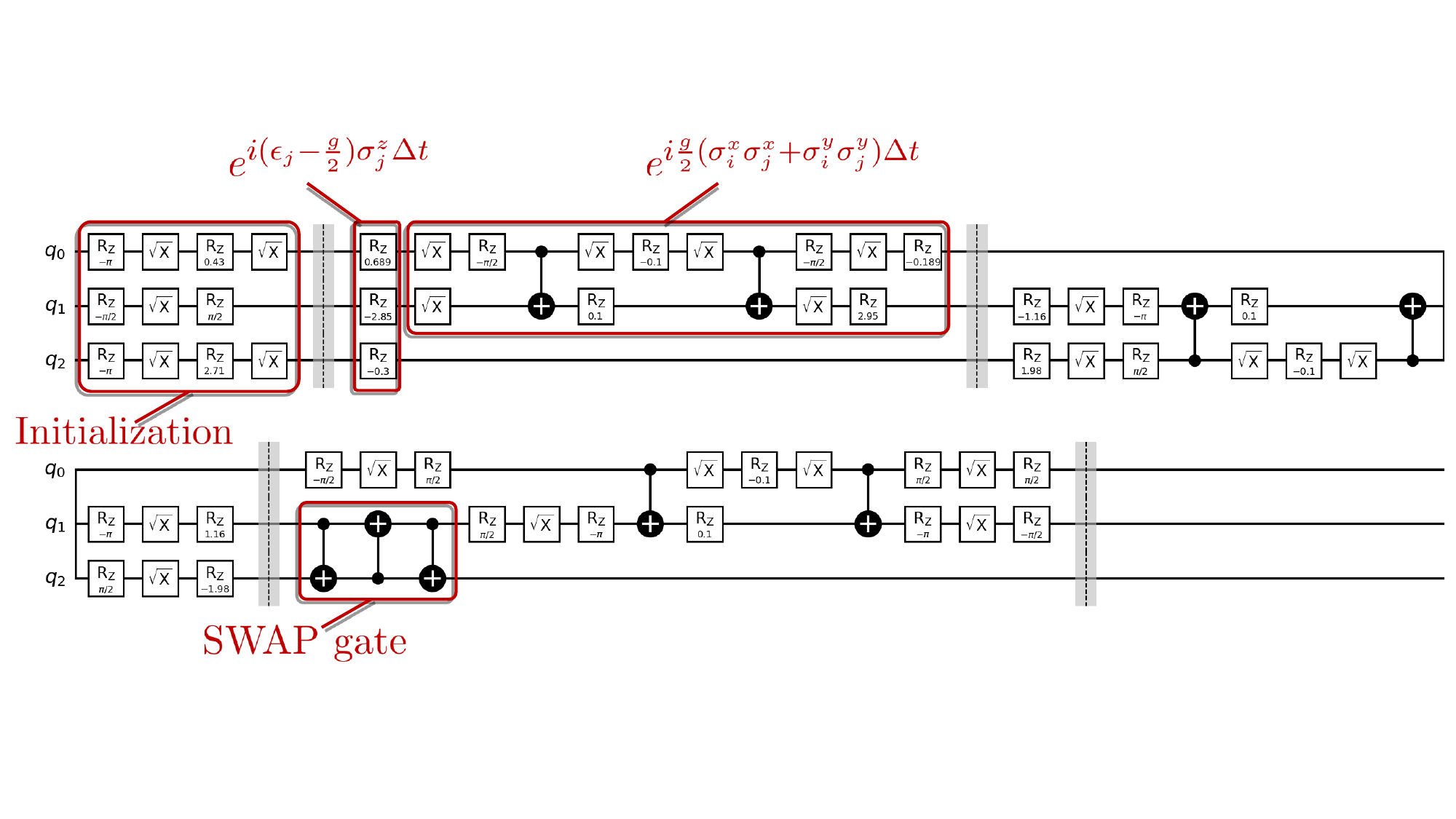}
    \caption{Transpiled quantum circuit for a single Trotter step \eqref{eq:Trotter}, preceded by the initial state preparation. The on-site energy and the interaction between the first two qubits are highlighted in red. The highlighted SWAP gate is required to ensure the interaction between the first and last qubit on a linear layout (see main text). Quantum circuits for the evolution up to arbitrary time $t = r \Delta t$ are built by concatenation of this elementary, circuit $r$ times (without the initialization), and thus contain $9r$ CNOT gates.}
    \label{fig:Trotter_circuit}
\end{figure*}

\subsection{Raw Results}
\label{subsec:raw_results}

\par Despite the small number of qubits, the 3-qubit BCS circuits are challenging to simulate on a noisy QC due to their depth. Their execution is subject to significant experimental noise, mainly originating from imperfect (CNOT) gate applications (IBM estimates the CNOT gate's error rate at $\sim 10^{-2}$ while the single-qubit gate's error is of the order of $10^{-4}$). 

\par
\begin{figure}
\centering
\begin{tabular}{l}
\includegraphics[width=0.8\linewidth]{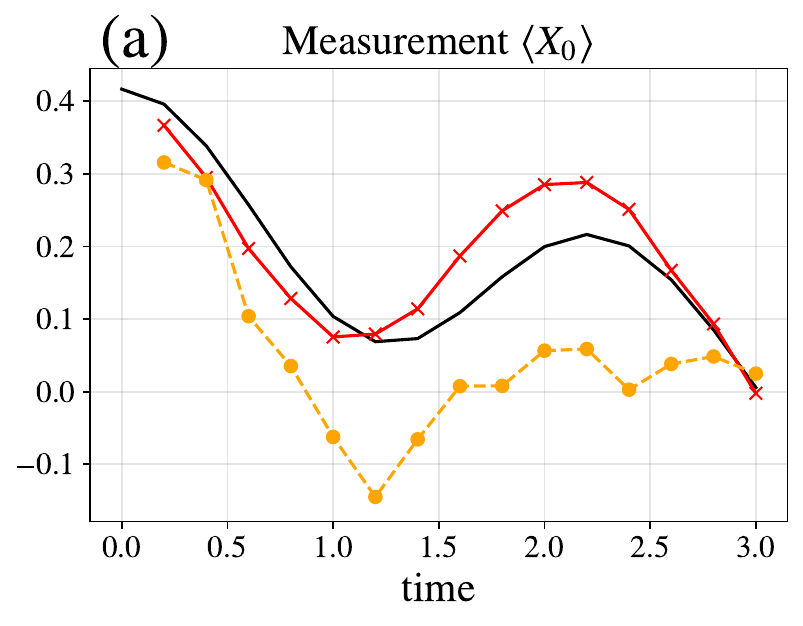}   \\     \includegraphics[width=0.8\linewidth]{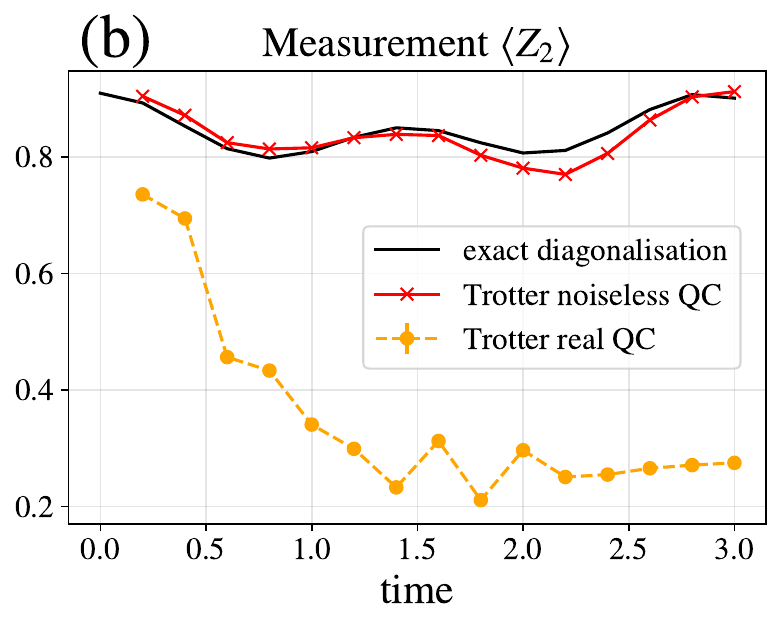} \\
\includegraphics[width=0.8\linewidth]{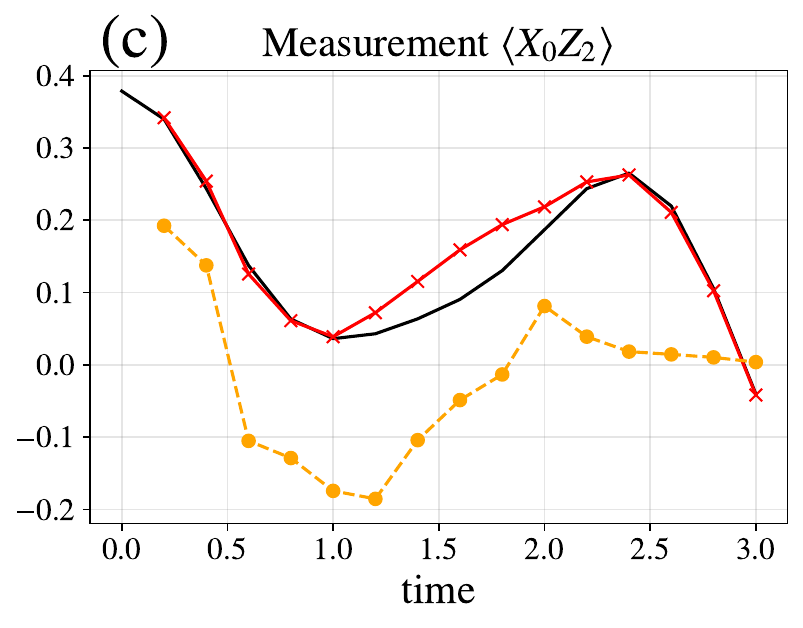} 
\end{tabular}
    
    \caption{Time evolution of the first qubit's $x$-component $\langle X_0\rangle$ (top panel), the third qubit's $z$-component $\langle Z_2\rangle$  (middle panel) and their joint evolution $\langle X_0 Z_2\rangle$ (bottom panel) under the exact BCS hamiltonian evolution  (black), the Trotter noiseless circuits (red) and the Trotter circuit on real noisy QC (orange). Error bars due to shot noise are smaller than the plot marker size $\sigma_s \sim 1/\sqrt{N_s}=5.10^{-3}$ where $N_s=32000$ is the number of shots. }
    \label{fig:Trotter_evolution}
\end{figure}

 We have simulated the dynamics with circuits containing up to 15 Trotter steps with 9 CNOT gates per step, i.e with a maximum of 135 CNOT gates. We plot in Fig.~\ref{fig:Trotter_evolution} the time evolution for three of the fourteen available observables (the other observables are plotted in Appendix~\ref{ap:evolution}): the first qubit's $x$-component $\langle X_0\rangle$ (see Fig.~\ref{fig:Trotter_evolution}-a), the third qubit's $z$-component $\langle Z_2\rangle$ (see Fig.~\ref{fig:Trotter_evolution}-b) and their joint evolution $\langle X_0 Z_2 \rangle$ (see Fig.~\ref{fig:Trotter_evolution}-c). We compare their evolution with the exact BCS dynamics (black solid line), the trotterized dynamics obtained on a noiseless classical computer (red solid line) and on a real, noisy, IBM QC (\texttt{ibm\_lagos}) (orange dotted line). We first emphasize that the first-order trotterized dynamics is indeed a good approximation of the exact evolution here, so we can expect the algorithmic Trotter error to be negligible compared to errors stemming from the noise acting on the QC. 

\par When the Trotter circuits are run on a real, noisy QC, we observe that the measured expectation values of all observables decay with a priori unpredictable fluctuations. Despite the fact that the CNOT error rate is around $1\%$, the effect of noise is already dramatic after a few Trotter steps, and no interesting features of the dynamics can be extracted from these results. This illustrates the need for error mitigation strategies to fight the effects of the noise, which we can expect to be even more dramatic as the circuit depth increases. For the sake of comparison, simulation of this system as well as the following error mitigation protocols have also been performed on qubits $q_{18}$, $q_{21}$ and $q_{23}$ of the \texttt{ibmq\_ehningen} device measured respectively in the direction x, y and z. The results are qualitatively similar and are shown in Fig.~\ref{fig:evolution appendix XYZ ehningen} in Appendix~\ref{ap:evolution}.

\par Small systems like the one we consider, whose full dynamics is easy to simulate on a classical computer, constitute a good platform to benchmark different noise-tailoring and quantum error mitigation protocols. In what follows, we implement a few of these protocols and assess their efficiency to improve the results of our quantum simulation.

\section{Simplifying the noise structure : Crosstalk randomized compiling}
\label{sec:twirl}
\par Errors in QC have multiple experimental origins. They can originate for instance from imperfect measurement during a qubit state readout. They can also stem from the experimentally imperfect application of certain quantum gates, in particular CNOT gates. In addition, qubit decoherence constitutes another source of errors, which sets an upper bound on the size of the circuits than can be run. On superconducting chips of IBM, the decoherence times $T_1$ and $T_2$ are typically of the order of $\sim100\mu$s while the application of the CNOT gates is $\sim300$ns~\cite{Kandala2021,Jurcevic2021} (data taken from \text{IBMQ backend properties}). Because we apply no more than $135$ CNOT gates, we consider decoherence of qubits negligible compared to the noise introduced by CNOT gates.
\par \newest{To address errors arising from noisy CNOT gates, understanding and simplifying their noise structure is crucial. Randomized compiling (RC) was introduced in~\cite{Wallman2016} as a method for converting any coherent noise channel into a Pauli (incoherent) noise channel. The original proposal recommended applying the twirling protocol to all qubits, but in practice, this approach is often performed only on the active qubits of the CNOT gates~\cite{Rosenberg2022,Kurita2022,Vazquez2023,Kim2023}, therefore overlooking spillover crosstalk effects. 

\par In the present Section, we discuss the importance of applying RC to neighboring qubits as well, and improve the procedure to simplify crosstalk noise even further. We start, in Sec.~\ref{sec:standardtwirl}, by presenting in details the standard RC protocol on active qubits. Then in Sec.~\ref{sec:twirlcrosstalk}, we propose an extension of the quantum gate twirling set on idle qubits, which turns any coherent noise channel into a depolarising (incoherent and unbiased) noise channel. By implementing the crosstalk RC protocol in our BCS simulation, we demonstrate its efficiency on IBMQ devices. Unlike for the standard RC protocol, we show that final results can be faithfully reproduced with classical simulations of noisy quantum circuits assuming a simple depolarising noise channel.}

\subsection{Standard randomized compiling protocol}
\label{sec:standardtwirl}
\par Before mitigating errors introduced by the application of CNOT gates on IBM QCs, we first need a way to model them. The effect of external markovian noise on the physical qubits can be described by a superoperator called a quantum error channel \cite{Nielsen2012}, which is usually applied after the noisy gate, and acts on the full density matrix of the system. A general error channel for two qubits \older{is characterized by 240\old{255} parameters and}can be written as follows:
\begin{equation}
    \mathcal{E}^{\text{gen.}}_2(\rho)=\sum_{i,j,k,l=0}^3 p_{ijkl}
    (\sigma^i\otimes \sigma^j).\rho.(\sigma^k\otimes\sigma^l)
    \label{eq:general_noise}
\end{equation} 
with $p^*_{ijkl}=p_{klij}$, and $\{\sigma^0,\sigma^1, \sigma^2, \sigma^3\} \equiv \{\mathbb{I}_2,\sigma^x, \sigma^y, \sigma^z\}$\new{. The trace-preserving condition yields 16 equations:
\begin{gather}
    \sum_{k,l}p_{klkl}=1,\\
   \forall i\neq0,\forall j\neq0,\sum_k p_{kjk0}=\sum_k p_{ik0k}=p_{ij00}=0,
\end{gather}}
\newest{reducing the naive number of $4^4 = 256$ independent parameters to $240$.}\old{and $\sum_{i,j,k,l} p_{ijkl}=1$} We do not know a priori the values for the $p_{ijkl}$, and determining them would require quantum state tomography protocols with unrealistic overhead. Moreover, this noise originates from many different experimental sources, and is not only different for different qubits, but can also change in time, making it impossible to fully characterize. 

\par Randomized compiling is a computational method which \new{allows for the reduction in
the complexity of the noise channel}\old{allows to reduce the complexity of the noise channel} (Eq.\eqref{eq:general_noise}) to only 15 free parameters, at the expense of running more circuit experiments. The resulting, equivalent noise channel reads:
\begin{equation}
\mathcal{E}^{\text{pauli}}_2(\rho)=\sum_{i,j=0}^3 p_{ij}(
    \sigma^i\otimes \sigma^j).\rho.(\sigma^i\otimes\sigma^j)
    \label{eq:pauli_noise}
\end{equation} where  $0\leq p_{ij}\leq 1$ and $\sum_{i,j} p_{i j}=1$. Such an error channel is called a Pauli noise \new{channel}.
\par RC consists in inserting a random layer of single-qubit gates (considered as noiseless) before and after each CNOT gates such that the overall action is still equivalent to a CNOT gate:
\begin{equation}
    (U_1\otimes U_2).\text{CNOT}. (U_3\otimes U_4)=\text{CNOT}
    \label{eq:CNOT_twirl}
\end{equation} 
\newest{where $U_i$ ($i\in\{1,2,3,4\}$) are random single-qubit gates, with combinations ${U_1,...,U_4}$ deliberately chosen to satisfy Eq.~\eqref{eq:CNOT_twirl}. The set of allowed combinations comprises $16$ options that can be constructed by choosing $U_1$ and $U_2$ arbitrarily from the set $\mathcal{T}_p=\{\mathbb{I}_2,\sigma^x,\sigma^y,\sigma^z\}$, and then choosing $U_3$ and $U_4$ from the same set such that Eq.~\eqref{eq:CNOT_twirl} is satisfied.} Table I of~\cite{Urbanek2021} lists the different possibilities for $U_i$'s. 

\par \newest{In the presence of noise, one can show~\cite{Cai2019} that averaging over all twirling configurations $\{U_i\}$ convert the general error channel, Eq.~\eqref{eq:general_noise}, to an effective Pauli noise, Eq.~\eqref{eq:pauli_noise})}. In other words, the average over the expectation values measured from all possible RC circuits coincides with the value that would be obtained from the original circuit with a Pauli noise channel on each CNOT gate.

\newest{The number of circuits with different twirl configurations scales exponentially, though: $16^{n_{\text{CNOT}}}$ where $n_{\text{CNOT}}$ is the number of CNOT gates in the circuit.}\old{However, the number of possibilities scales exponentially with the number of CNOT gates: $16^{n_{\text{CNOT}}}$ where $n_{\text{CNOT}}$ is the number of CNOT gates in the circuit.} Performing an average over this number of circuit experiments is unrealistic in practice. \old{Fortunately, we have found that, for our circuits, averaging over more than 300 twirling configurations does not change the outcome quantitatively (see and compare Fig.~\ref{fig:mitigated evolution} and Fig.~\ref{fig:evolution appendix XYZ} for $t>1.8$). In the following, each measurement result is thus averaged over 300 randomly compiled circuits.} \newest{However, in the assumption of noise Markovianity (i.e., the noise on different CNOT gates is not correlated), one can sample very few twirl configurations. For an observable $\in\left[-1,+1\right]$, the deviation from the exact average scales $\mathcal{O}(1/\sqrt{N_{RC}})$ where $N_{RC}$ is the number of RC circuits with different twirl configurations --- in the spirit of Monte Carlo sampling. For $N_{RC}=300$, error bars in Fig.~\ref{fig:randomized_evolution} are below the plot marker size, underscoring the method's accuracy.}

\newest{We note in passing that the above discussion concerns the cost of "converting" the error channel by means of RC. Further use of error mitigation techniques (such as NEC in Sec.~\ref{sec:combining}) on the measured results would require exponential accuracy of the measured results (cf.~Fig.~\ref{fig:mitigated evolution}) and thus impose exponential growth on the required $N_{RC}$ with the number of CNOT gates. Yet, in practice, this is not the limiting factor. For example, in Fig.~\ref{fig:evolution appendix XYZ}, the number of RC circuits has been increased to 600 for $t>1.8$, yet this adjustment does not bring about significant changes in comparison with Fig.~\ref{fig:mitigated evolution}.}

\subsection{Randomized compiling for crosstalk}
\label{sec:twirlcrosstalk}
\par So far, by restricting ourselves to 2-qubit error channel, we have supposed that the noise acting on CNOT gates only affects the qubits on which the gate is applied. In practice, the application of a CNOT is known to also produce errors on neighboring qubit, an effect known as crosstalk~\cite{Murali2020,Sarovar2020,Xie2021,Zhao2022}. Since the precise noise structure induced by crosstalk is also unknown, we aim to simplify it as well by extending RC to these neighboring qubits. 
\par
\begin{figure}[h]
\includegraphics[width=.9\linewidth]{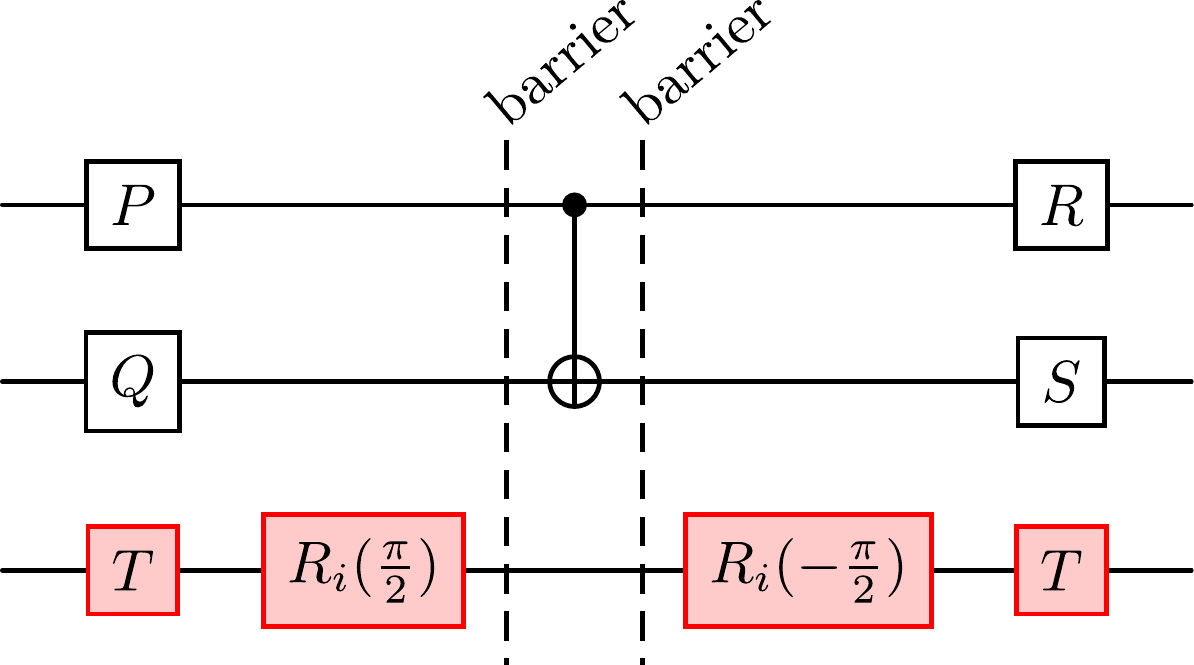}
\caption{Representation of the crosstalk RC protocol used for our 3-qubit simulations. The white gates are used in the standard RC protocol. In order to twirl crosstalk errors, we extend this scheme by adding the red gates. Gates $P$, $Q$, $R$, $S$, and $T$ are either the identity matrix or one of the three Pauli matrices. $P$, $Q$, $T$ are drawn randomly from $\mathcal{T}_p$, and $R$ and $S$ are chosen such that the circuit is stricly equivalent to an isolated CNOT. Gates $R_i(\pi/2)$ are $\pi/2$ rotations around the axis $i\in\{x,y,z\}$, drawn at random. Barriers indicate the order of application of gates and prevent annihilation between single-qubit gates.}
\label{fig:twirling}
\end{figure}

 \newest{The crosstalk RC (cRC) procedure for 3 qubits is represented in Fig.~\ref{fig:twirling}.} Unlike \newest{for} the \older{former case} \newest{two active qubits on which the CNOT is applied}, no logical operation is performed on the neighboring qubits. As a consequence, any single-qubit gates can be part of the twirling set (\newest{imagine replacing CNOT with identity in Eq.~\eqref{eq:CNOT_twirl}; then once could choose $U_3=U_1^{-1}$ and $U_4=U_2^{-1}$ for any $U_1$ and $U_2$m not only those belonging to $\mathcal{T}_p$}). We can thus simplify the structure of the noise channel \textit{on the neighboring qubits} even further than for the active qubits by using two \newest{distinct} twirling sets. The first one is the same twirling set $\mathcal{T}_p$ as in the previous section. The second is composed of the gates $\mathcal{T}_R=\{R_x(\pi/2), R_y(\pi/2), R_z(\pi/2)\}$. We perform two RC procedures in a row. First, we apply the gate drawn randomly from the set $\mathcal{T}_p$, then a second gate drawn from $\mathcal{T}_R$ and undo their action by choosing their corresponding inverse gate after application of the CNOT on the neighboring qubits. \older{The whole procedure is represented in Fig.~\ref{fig:twirling}.}

\par In Appendix ~\ref{ap:noise}, we derive the effective \newest{general $n$-qubit error channel ($n-2$ neighbors that can experience crosstalk), as obtained after performing cRC (see Eq.~\eqref{eq:crosstalkRC noise})}. The restriction of this channel to our 3-qubits BCS simulations \newest{(with only one neighbor)} has $\newest{16\times2^{n-2}-1=31}\older{2^{n+2}-1=31}$ free parameters:
\newest{
\begin{gather}
    \mathcal{E}_3^{\text{cRC}}(\rho)=\frac{1}{3}\sum_{i,j=0}^3\sum_{s=0,1}p_{i,j,s}\sum_{a=1,2,3}P^{i,j,s,a}.\rho.P^{i,j,s,a},\\    
    \text{with }P^{i,j,s,a}=\sigma_1^i\otimes\sigma_2^j\otimes\left(\sigma_3^{a}\right)^{s},\ \frac{1}{3}\sum_{i,j=0}^3\sum_{s=0,1}p_{i,j,s}=1, \nonumber
\end{gather}
where indices $1$ and $2$ stand for the control and target qubits of the CNOT, $s$ indicates if an error has occured on the neighboring qubit ($s=1$) or not ($s=0$), and $a$ encodes whether it was an X, Y or Z error ($a=1,2$ or $3$).}
Note, in particular, that taking the partial trace \new{of the density matrix} over the neighboring qubit \new{subspace} yields\older{ for} \newest{the same Pauli channel as in the standard RC protocol presented in the last section as the effective error channel for the active qubits} (see Eqs.~\eqref{eq:pauli_noise},~\eqref{eq:pauli_noise2}). Likewise, tracing over the active qubits yields a 1-qubit depolarizing noise channel for the neighboring qubit \newest{because $p_{i,j,s}$ does not depend on $a$}:
\begin{eqnarray}
    \mathcal{E}^{\text{dep.}}_1(\rho)&=&(1-p)\rho +\frac{p}{3}\sum_{i=1}^3    \sigma^i.\rho.\sigma^i
    \label{eq:dep_noise}\\\nonumber
    &=&(1-\lambda)\rho +\lambda\frac{\mathbb{I}}{2}
    \label{eq:dep_noise2}
\end{eqnarray}
where $0\leq\lambda=\frac {4 p}{3}\leq\frac{4}{3}$,\newest{$p=\sum_{i,j=0}^3p_{i,j,1}$} is the error rate and $\lambda$ the depolarizing parameter. The key point about cRC is that it simplifies the error channel of \new{neighboring} qubits because it symmetrizes the effect of noise on the Bloch sphere \old{of the individual qubits. When using RC over $\mathcal{T}_p$ the noise is symmetrized between the two eigenstates of $\sigma_i$. Indeed, the application of a $\sigma_j$ matrix before the CNOT gate rotates the state on the Bloch sphere around the j-axis with an angle $\pi$. Therefore, eigenstates  of the other Pauli matrices $\{\ket{0}_i,\ket{1}_i\}_{i\neq j}$ are swapped $\ket{0}_i\leftrightarrow\ket{1}_i$. The second RC over $\mathcal{T}_R$ symmetrizes the effect of the noise} with respect to the different directions. Application of $R_i(\pi/2)$ permutes the eigenstates of $\sigma_{j}$ and $\sigma_{k}$: $\ket{0}_j\rightarrow\ket{0}_k\rightarrow\ket{1}_j\rightarrow\ket{1}_k\rightarrow\ket{0}_j$. Indices are such that  $\epsilon_{ijk}=+1$, where $\epsilon$ is the the Levi-Civita tensor. As a result, the depolarizing noise \new{channel} is obtained by equally mixing the different coefficients of the Pauli noise \new{channel}.

\par \older{Note, adding Pauli gates on neighboring qubits to achieve a Pauli noise \new{structure} was discussed in Refs.~\cite{Wallman2016,Hashim2021}. 
Here we focus  on crosstalk errors. In particular, our proposal consists in adding a second set of gates with $\frac{\pi}{2}$ rotation around the $x,y,z$-direction, to transform the Pauli noise channel into a depolarising noise one.}
\subsection{Comparing standard and crosstalk RC protocols through BCS simulation}
\label{sec:twirlBCS}
\par  
\par
\begin{figure*}[!!h]
\centering
\begin{tabular}{ll}
\includegraphics[width=0.4\linewidth]{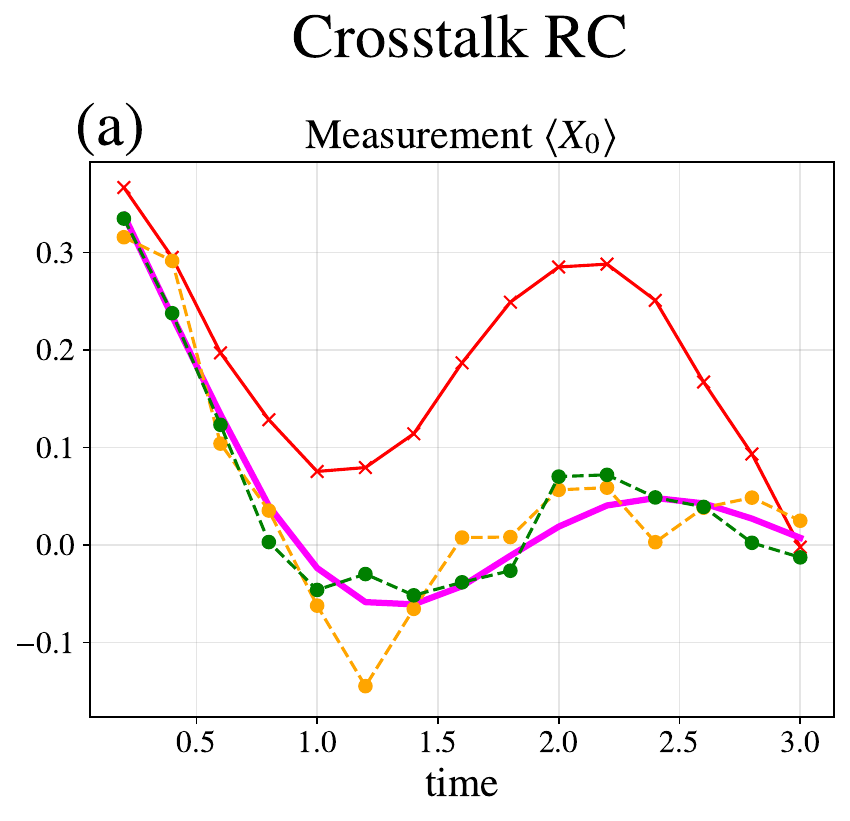} & \includegraphics[width=0.4\linewidth]{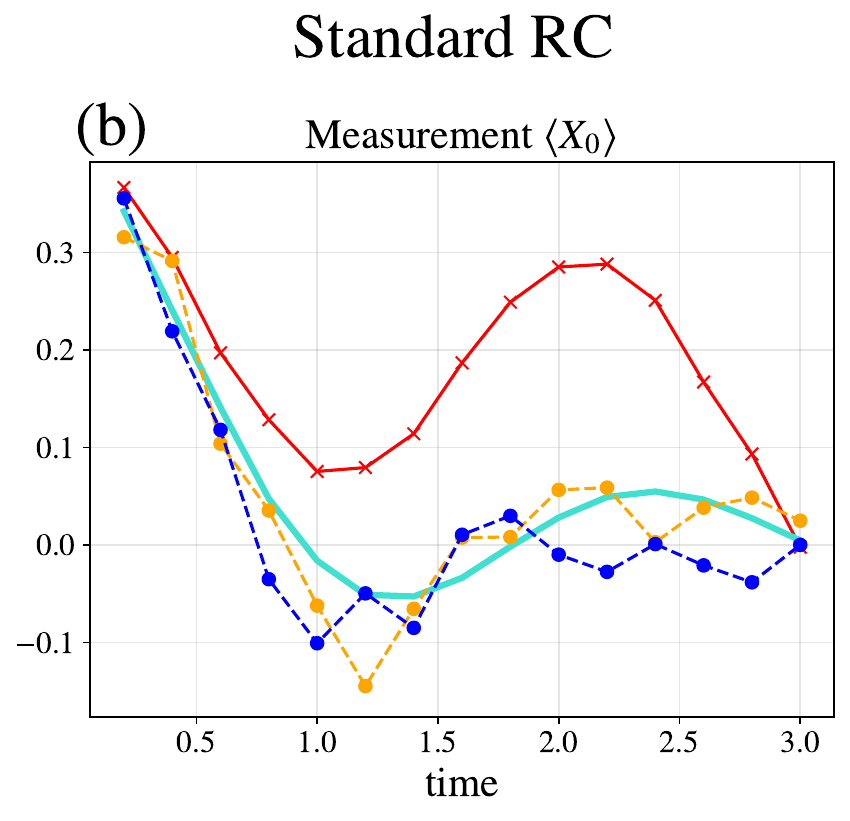}  \\    \includegraphics[width=0.4\linewidth]{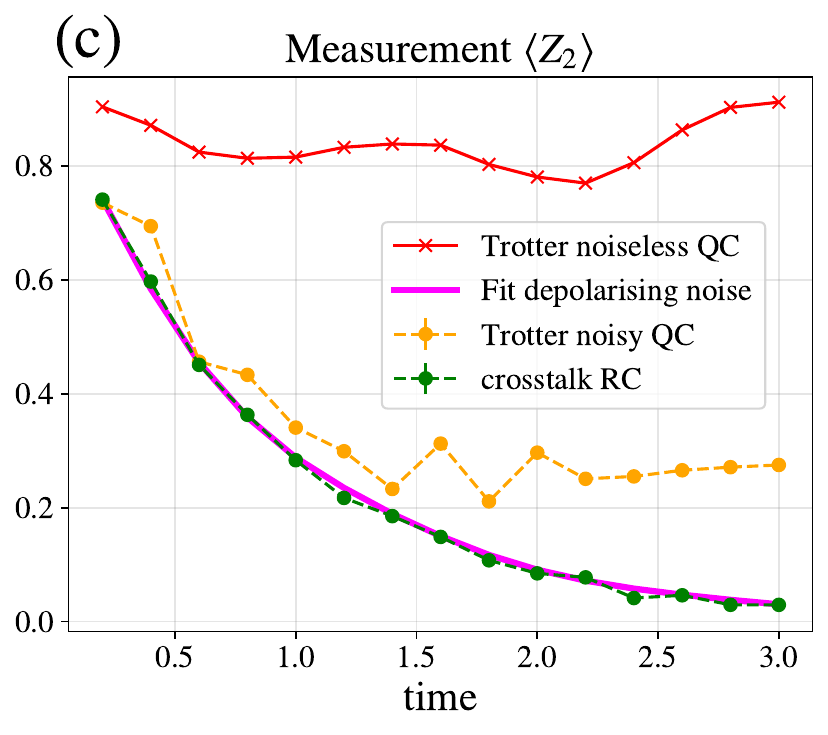} & \includegraphics[width=0.4\linewidth]{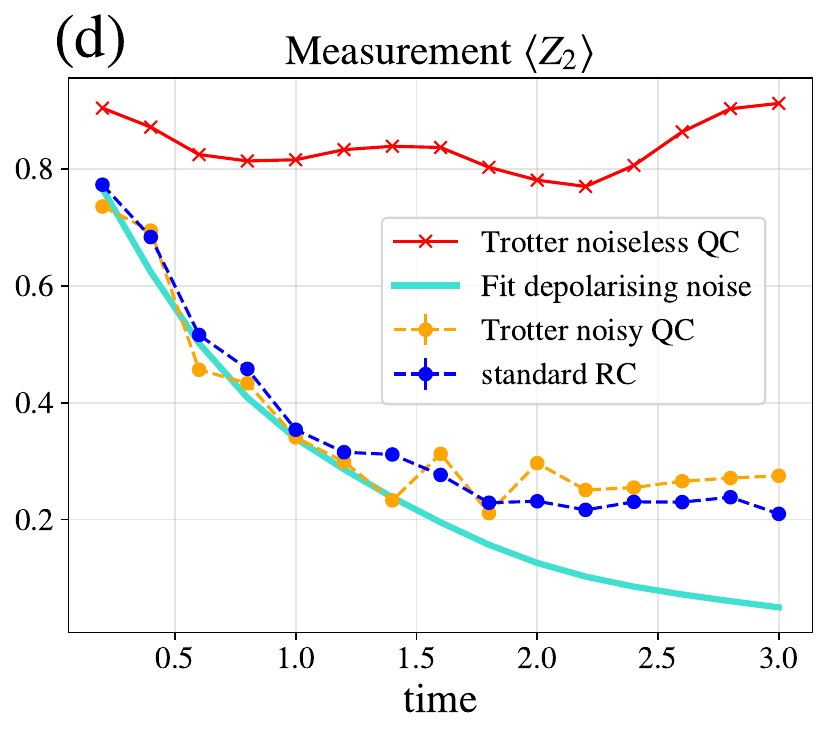}   \\
\includegraphics[width=0.4\linewidth]{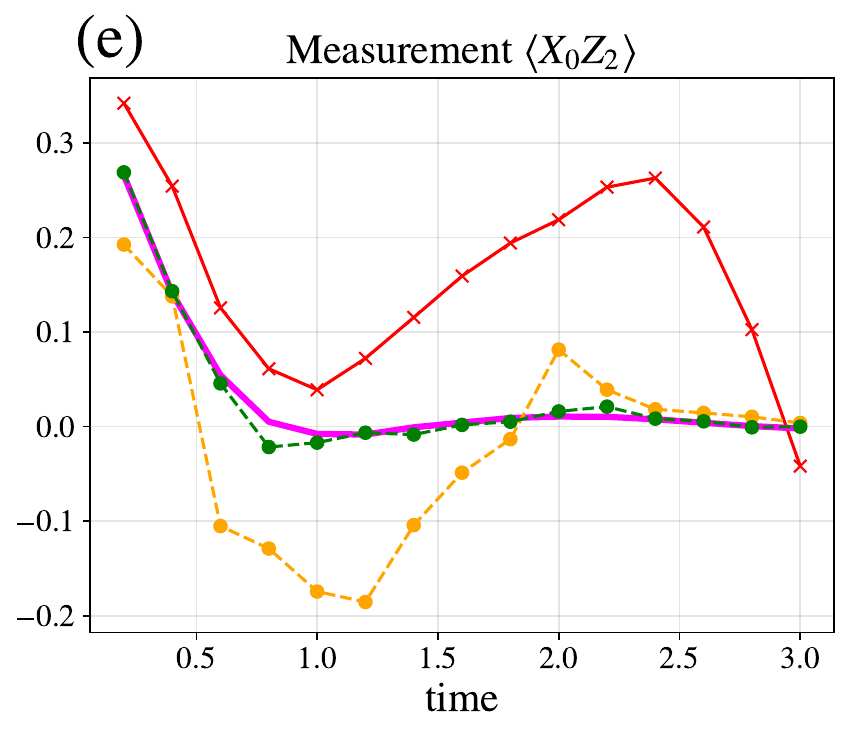} & \includegraphics[width=0.4\linewidth]{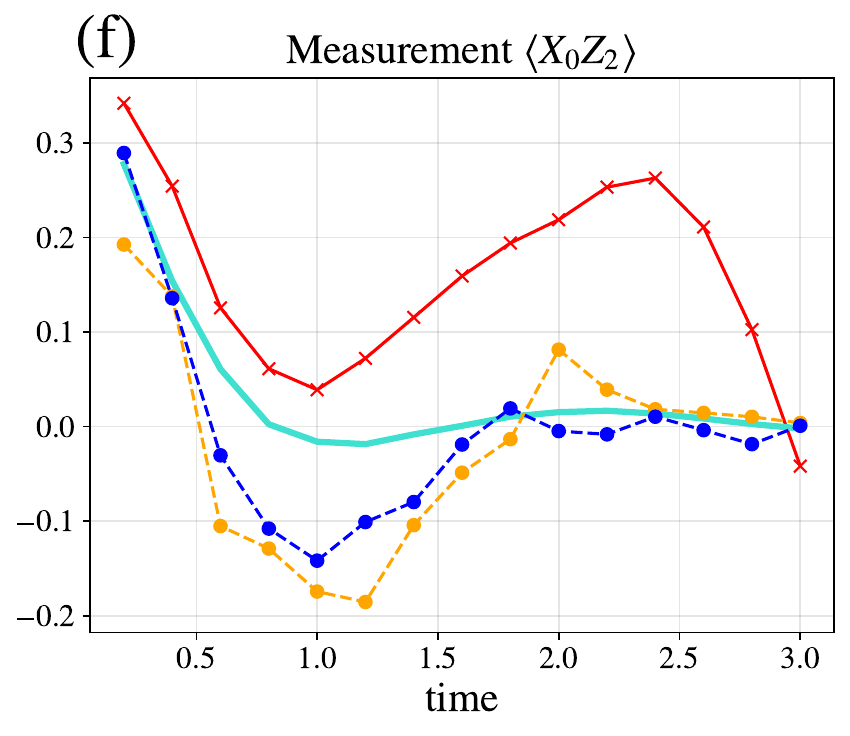}    
\end{tabular}
    \caption{Time evolution of $\langle X_0\rangle$ (top panels), $\langle Z_2\rangle$  (middle panels) and $\langle X_0 Z_2\rangle$ (bottom panels). The green (resp. blue) dashed curves on the left (resp.right) panels correspond to the crosstalk (resp. standard) RC protocol. The RC protocol has been averaged over $300$ different circuits. The magenta (resp. cyan) solid curves on the left (resp. right) panels  are the result of a fit of the green (resp. blue) curves made using classical simulations of the circuit with the noise model of Eq.~\eqref{eq:3qubitchannel} (see main text). Error bars due to shot noise and finite sampling of random twirl configurations are computed in Appendix~\ref{ap:uncertainty}. They are smaller than the plot marker size. The red and orange curves correspond respectively to the noiseless and noisy trotterized dynamics and are plotted here for comparison.}
    \label{fig:randomized_evolution}
\end{figure*}

 \older{We applied the 3-qubit  RC method} \newest{We apply both the standard and crosstalk RC methods} from the last sections to the BCS simulation algorithm described in Sec.~\ref{sec:BCS}. \newest{For the crosstalk RC protocol (resp. standard RC protocol),} results are shown in Fig.~\ref{fig:randomized_evolution}-a,c,e \newest{(resp. Fig.~\ref{fig:randomized_evolution}-b,d,f)}. As expected, both RC protocols do not improve the accuracy of the results on their own, as they simplify the CNOT noise structure, but don't \newest{eliminate the noise}\old{affect the noise strength a priori}. However, we observe that they tend to flatten the curve and attenuate the unpredictable fluctuations of the expectation values due to the complex structure of the original noise. This could be explained by the fact that the coherent part of the noise, typically responsible for \newest{such fluctuations}, has indeed been turned incoherent. 

\par \newest{To further interpret these results, we try to replicate the QC's output with classical simulations of our Trotter quantum circuits using a simple noise channel}. In the following, we assume that the noise acting on the quantum device is close to isotropic (in the sense of the qubit basis), so that the Pauli noise channel restricted to the active qubits can be approximated by a 2-qubit depolarising noise channel. This assumption is rather standard~\cite{Urbanek2021} for quantum error mitigation protocols because it allows to relate noisy and noiseless expectation values analytically. Under this assumption, the $(n=3)$-qubit error channel can be described by only $2^{n-1}-1=3$ free parameters $\{\lambda_{\text{CNOT}},\lambda_{\text{neigh.}},\lambda_{\text{glob.}}\})$:
\begin{eqnarray}
\mathcal{E}_3(\rho)&=&(1-\lambda_{\text{CNOT}}-\lambda_{\text{neigh.}}-\lambda_{\text{glob.}})\rho \nonumber\\
&+&\lambda_{\text{CNOT}}\frac{\mathbb{I}_{01}}{4}\otimes\text{Tr}_{01}(\rho)\nonumber\\
&+&\lambda_{\text{neigh.}}\text{Tr}_{2}(\rho)\otimes\frac{\mathbb{I}_{2}}{2}+\lambda_{\text{glob.}}\frac{\mathbb{I}_{012}}{8}
\label{eq:3qubitchannel}
\end{eqnarray}
where $0$ and $1$ denote the indices of the active qubits and $2$ the index of the neighboring qubit, $\text{Tr}_{\{k\}}$ is the partial trace over the set of qubits $\{k\}$. We call this noise channel a quasi-local depolarising noise.

\par As displayed in Fig.~\ref{fig:randomized_evolution}-a,c,e (magenta curves), the QC results can be well reproduced by classical simulations (using the \texttt{density matrix} method of the noise simulator Qiskit Aer) performed with a quasi-local depolarising noise (see Eq.~\eqref{eq:3qubitchannel}) applying on every CNOT of each junction (between qubits 0 and 1 and between qubits 1 and 2). This leaves us with 5 free parameters, $\lambda_{\text{CNOT}}^{01}$, $\lambda_{\text{neigh.}}^{01}$,$\lambda_{\text{CNOT}}^{12}$, $\lambda_{\text{neigh.}}^{12}$ and $\lambda_{\text{glob.}}$ that we assume are the same for each Trotter step. We find a set of parameters that minimizes the square distance between the classical simulation results and all the data points for the 7 different observables ($7\times 15=105$ data points). The resulting fit parameters are: $\lambda_{\text{CNOT}}^{01}\simeq 0$, $\lambda_{\text{CNOT}}^{12}\simeq 0.014$, $\lambda_{\text{neigh.}}^{01}\simeq 0.05$, $\lambda_{\text{neigh.}}^{12}\simeq 0.01$ and $\lambda_{\text{glob.}}\simeq0.002$, with the regression standard error $\chi^2 =2.10^{-4} $. From the fit parameters, it is clear that the noise associated to \textit{neighboring qubits} cannot be neglected. Rather, they seem to confirm that crosstalk plays an important role in IBM's quantum devices and that the scheme we are proposing is essential for the implementation of efficient error mitigation protocols. The noise parameters derived differ noticeably from the CNOT gate error rate (around $0.005-0.01$) announced by IBM. We believe that this discrepancy stems from the method used to estimate these parameters. IBM uses the randomized benchmarking protocol~\cite{Magesan2011,Magesan2012}, which does not take crosstalk into account because each junction is considered separately. The resulting error rate does not seem to relate easily to the Pauli noise \new{channel} parameters that we obtain through the cRC method.

\par For comparison, we also plot in Fig.~\ref{fig:randomized_evolution}-b,d,f, fit by classical simulations of the results we obtain using the standard RC protocol (no twirling of neighboring qubits). The resulting fit parameters are: $\lambda_{\text{CNOT}}^{01}\simeq 0.007$, $\lambda_{\text{CNOT}}^{12}\simeq 0$, $\lambda_{\text{neigh.}}^{01}\simeq 0.07$, $\lambda_{\text{neigh.}}^{12}\simeq 0.009$ and $\lambda_{\text{glob.}}\simeq0.0002$ and the regression standard error is $\chi^2\simeq 0.02$. The quality of the fit is not as good as in the former case, indicating that the noise model of Eq.~\eqref{eq:3qubitchannel} is a rough approximation of the real noise occurring on the QC.
\par \newest{The above shows that cRC is \textit{essential} to accurately describe the noise on the system with Eq.~\eqref{eq:3qubitchannel}, ensuring that the effect of crosstalk is reduced to a depolarising noise channel on the neighboring qubits. In turn, the simpler noise structure facilitates error mitigation, as we illustrate in the next Section.} \older{Therefore, the quality of the mitigation protocol (see Sec.~\ref{sec:estimation}) which relies on a simple noise model will be drastically affected (see Fig.~\ref{fig:mitigated evolution}-b,d,f).}

\section{Error mitigation techniques}
\label{sec:mitig}
\par In order to illustrate the practical implications of our cRC protocol, \newest{we implement two auxiliary error mitigation schemes. First, to tackle errors stemming from measurements, we apply a simple readout error correction (REC) procedure detailed in~\cite{Nachman2020}. Second, to mitigate the noise affecting CNOT gates, we use the noise estimation circuits (NEC) method proposed in \cite{Urbanek2021}.}

\newest{\subsection{Readout error correction (REC)}}
\par Measurement is expected to introduce an error rate of $2\%$ (data taken from \text{IBM backend properties}). \new{In what follows, we systematically} apply the \new{iterative Bayesian} unfolding method presented in \cite{Nachman2020} to correct readout errors \new{(REC). Note that we invert the whole $2^3 \times 2^3$ measurement matrix, taking into account measurement error correlations. This method scales poorly with the number of measured qubits. However, we have found that perfoming the inversion qubit-per-qubit, discarding the correlations, yields similar results.} \old{ A detailed description of the readout error correction (REC) protocol used is given in Appendix~\ref{ap:readout}.}

\subsection{Noise estimation circuits (NEC)}
\label{sec:estimation}

The NEC technique aims to restore the noiseless value of an observable by dividing the noisy expectation value of an observable by a certain factor which estimates the circuit fidelity. This factor is obtained by running additional circuits derived from the original one, \newest{while preserving} the same CNOT structure, and in practice allows one to partly get rid of the errors stemming from depolarizing noise channels. 

\begin{figure*}
    \centering
    \includegraphics[width=\linewidth,trim={0 0 0 -1cm},clip]{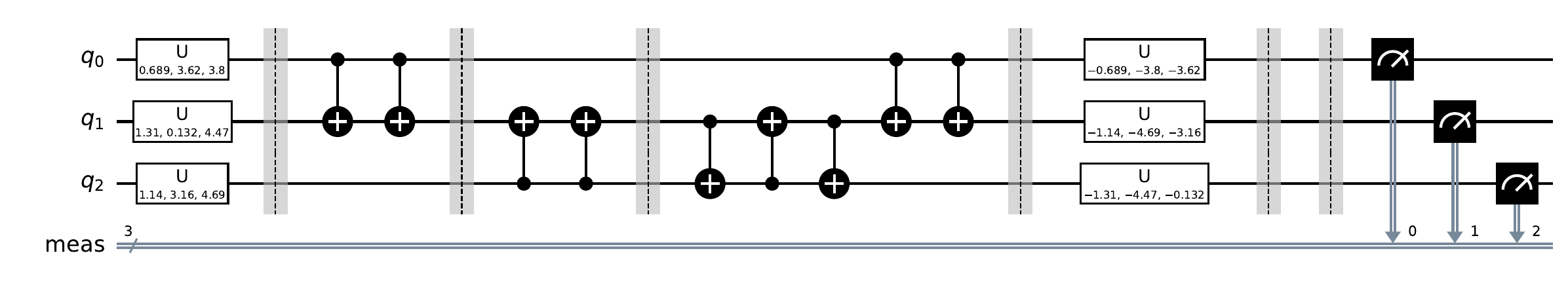}
    \caption{Example of an estimation circuit for the first Trotter step. \new{To build it, we remove the initial state initialization and single-qubit gates from the example circuit in Fig.~\ref{fig:Trotter_circuit}}. A random initial product state is added before the circuit and undone at the end (\new{white boxes}). We also perform RC on each CNOT gate (not shown here).}
    \label{fig:estimation_circuit}
\end{figure*}

\par To understand the above statement, first, one has to relate the noisy and the noiseless expectation values of an observable. In the same way as done in ~\cite{Urbanek2021} for local and global depolarising noise (see also Appendix~\ref{ap:mitigation}), we expand the noisy outcome up to the first order in the number of errors for the \newest{quasi-local depolarising} noise channel (see Eq.~\eqref{eq:3qubitchannel}), which yields:

 \begin{widetext}
 \begin{equation}
\langle O \rangle_{\text{noisy}}\simeq(1-\lambda_\text{glob.})^{n_\text{CNOT}}\langle O \rangle_{0}
\left[1+\lambda_\text{CNOT}\left(\frac{\langle O\rangle _1}{\langle O \rangle_{0}}-n_\text{CNOT}\right)+\lambda_\text{neigh.}\left(\frac{\langle O'\rangle _1}{\langle O \rangle_{0}}-n_\text{CNOT}\right) \right]
+\mathcal{O}(\lambda_\text{CNOT}^2,\lambda_\text{neigh.}^2),
\label{eq:noise obs real}
\end{equation}
 \end{widetext} 
where $n_\text{CNOT}$ is the number of CNOT gates in the circuit, and $\langle O \rangle_0$ is the noiseless outcome. $\langle O \rangle_1$ is the sum of expectation values obtained from all circuits where an error occurred on a single CNOT gate, putting the active qubits of the CNOT in the maximally mixed state $\smash{\mathbb{I}/4}$. $\langle O' \rangle_1$ is the sum of expectation values obtained from all circuits where an error occurred on a single CNOT gate, putting the \textit{neighboring} qubit of the CNOT in the maximally mixed state $\smash{\mathbb{I}/2}$. \newest{For exposition simplicity}, we consider here that noise parameters for different junctions are equal, but the result can be generalized to CNOT gates with different noise parameters for different junctions.
 
\par To mitigate these errors, it has been proposed in~\cite{Urbanek2021} to run extra circuits (called noise estimation circuits) whose outcome yields a part of the error terms appearing in Eq.~\eqref{eq:noise obs real}. To construct these NEC, one simply removes all single-qubit gates from the original Trotter circuits (see Fig.~\ref{fig:estimation_circuit}). 

\par In our Trotter circuits, we can distinguish two types of CNOT gates. The first type is the CNOT gates encoding the interaction terms. After removing all the single-qubit gates, they are now applied in a row with the same control and target qubits. As they always come by pairs, they are now logically equivalent to the identity operator. The second type is the CNOT gates implementing the SWAP gate. Their action does not change after removing single-qubit gates. The overall unitary operation performed by the estimation circuits, for any number of Trotter steps, is thus the identity operator up to some SWAP operations. 

Applying Eq.~\eqref{eq:noise obs real} to the NEC and dividing \newest{the noisy Trotter circuits outcome by the fidelity estimated via NEC} leads to the mitigated expectation value:
\begin{widetext}
 \begin{equation}
  \langle O \rangle_{\text{mitigated}} =\frac{  \langle O \rangle_{\text{noisy}}}{\langle E \rangle_\text{noisy}}\simeq\langle O \rangle_{0}
\left[1+\lambda_\text{CNOT}\left(\frac{\langle O\rangle _1}{\langle O \rangle_{0}}-\langle E\rangle_1\right)+\lambda_\text{neigh.}\left(\frac{\langle O'\rangle _1}{\langle O \rangle_{0}}-\langle E' \rangle_1\right) \right]
+\mathcal{O}(\lambda_\text{CNOT}^2,\lambda_\text{neigh.}^2)
\label{eq:noise mitigated real}
\end{equation}   
\end{widetext}
where  $\langle E \rangle_1$ and $\langle E' \rangle_1$ are the NEC measured expectation values correspond respectively to $\langle O \rangle_1$ and $\langle O' \rangle_1$ in the original circuits and $\langle E\rangle_0=1$. Eqs.~\eqref{eq:noise mitigated real} shows that the NEC mitigates exactly the noise associated with the global part of our quasi-local depolarising channel. For the active qubits and neighboring depolarising channels, it is expected that the coefficients in front of $\lambda_\text{CNOT}$  and $\lambda_\text{neigh.}$ are strongly suppressed.

\par To show this, we propose in Appendix~\ref{ap:Trotter noise} an expansion in terms of the Trotter time step $\Delta t$ (the computation is done in the case of a local depolarising noise). This computation takes into account any number of errors in the circuit but only relates the noisy result to the noiseless terms at zeroth an first order $\Delta t$. Applying the equation found (see Eq.~\eqref{eq:formula1}) to the case of NEC gives an exact result as the only non vanishing term is of order $0$ in $\Delta t$. Therefore, one finds that the main term of the noisy Trotter circuit is reproduced by \newest{the fidelity estimated using} NEC. The result holds for any number of errors. In particular, at small time (i.e. for a few Trotter steps), we have: \begin{equation}
    \frac{\langle O \rangle_1}{\langle O\rangle_0}\simeq\langle E \rangle _1
\end{equation}

As a consequence, NEC is an efficient method to remove noise at small \newest{evolution} times, or equivalently for small-depth circuits. The results can be generalized also for quasi-local depolarising noise. At longer \newest{evolution} times, we expect \newest{the noise estimated by NEC to differ from the noise on the Trotter circuits}. To remove this remaining noise, it has been proposed in~\cite{Urbanek2021} to use the zero noise extrapolation method~\cite{Li2017, Temme2017}, which consists in artificially increasing the noise of sets of CNOT gates~\cite{He2020} in order to extrapolate the result to the case of zero noise. \older{This technique also assumes a depolarising channel on the CNOT gates.} In this article, we do not apply this additional mitigation scheme because as we will see, at longer  \newest{evolution} times the QC result deviates from the classical simulations using depolarising noise, indicating that \newest{the cRC noise channel can no longer be described accurately as an effective quasi-local depolarising noise channel}. 

\par \newest{We remind the reader making the noise describable in terms of ~\eqref{eq:3qubitchannel} requires the application of RC techniques. Therefore, any circuits we run involve twirling applied in agreement with either RC or cRC protocols described in Sec.~\ref{sec:twirl}.}

\par \newest{Finally, in order to stabilize the fidelity-estimation output over different initial states, we generate random initial product states and undo them at the end after the application of the noise estimation circuit by inserting the corresponding inverse gates and taking into account the action of the SWAP gates. We produce \newest{$N_{RC}=300$} such noise estimation circuits, each of which has a different random initial state and a different twirling configuration for the RC, cf.~Fig.~\ref{fig:estimation_circuit}.}

\section{Combining crosstalk RC and error mitigation}
\label{sec:combining}

\newest{In order to obtain the error-mitigated results and thus assess the ultimate benefit of replacing RC with cRC, we combine all the techniques described above: REC, cRC/RC, and NEC. Namely, }
\old{To obtain our final results,} we divide the Trotter result obtained in Sec.~\ref{sec:twirl} --- after averaging over 300 RC circuits and performing REC --- by the result of the noise estimation circuits as described in Sec.~\ref{sec:mitig}, also averaged over 300 RC and with REC.

\begin{figure*}[!!h]
\centering
\begin{tabular}{ll}
\includegraphics[width=0.4\linewidth]{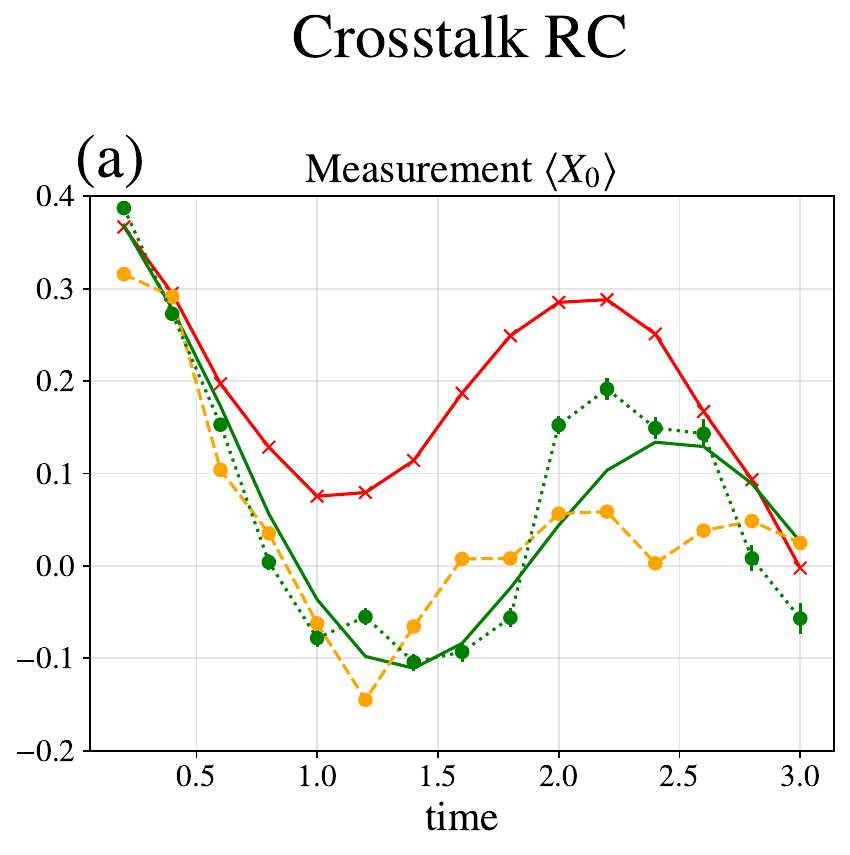}&  \includegraphics[width=0.4\linewidth]{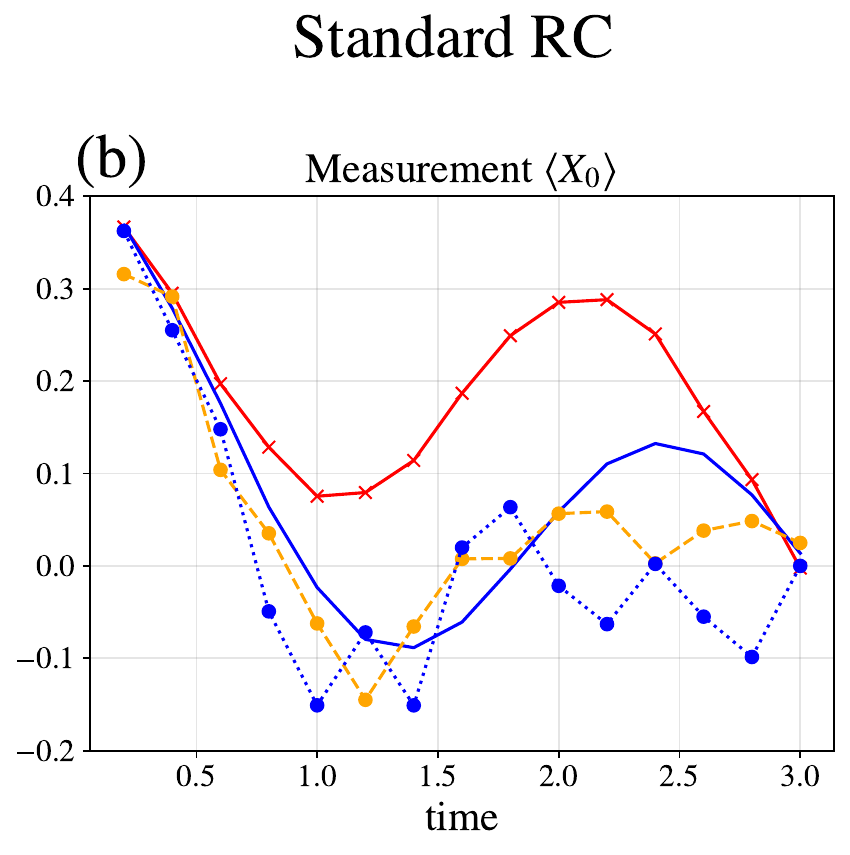}  \\    \includegraphics[width=0.4\linewidth]{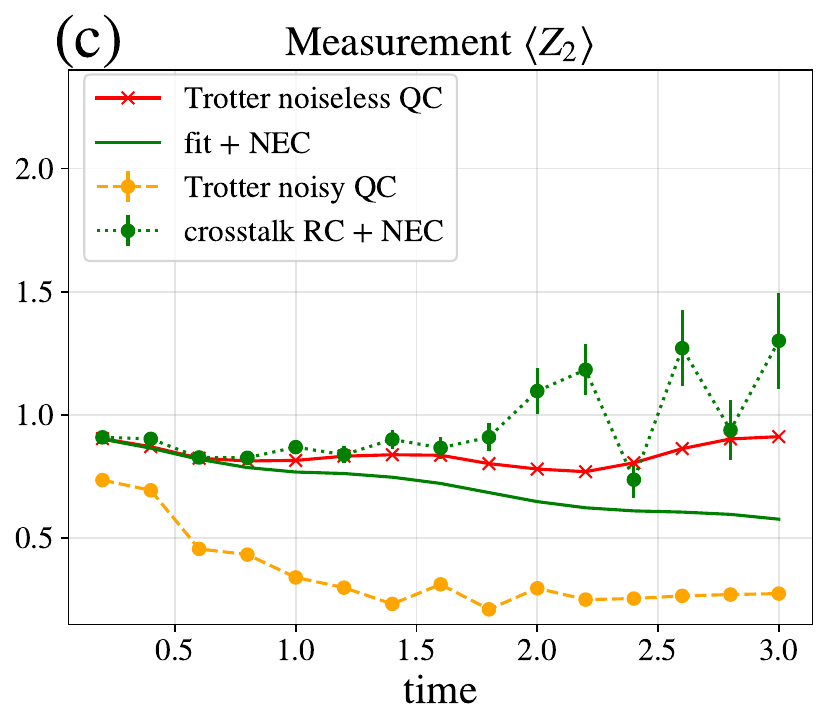}&\includegraphics[width=0.4\linewidth]{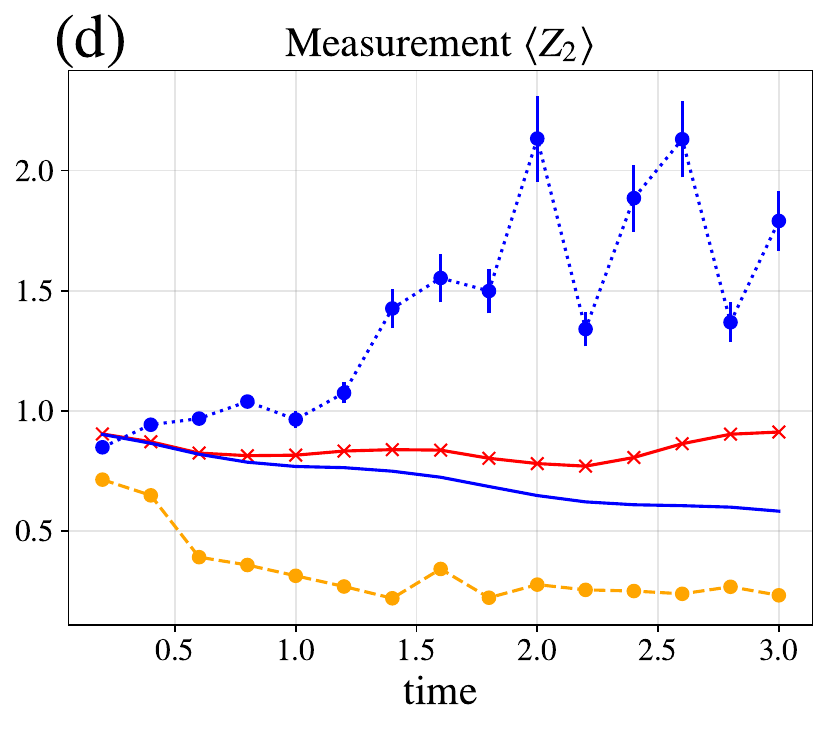} \\
\includegraphics[width=0.4\linewidth]{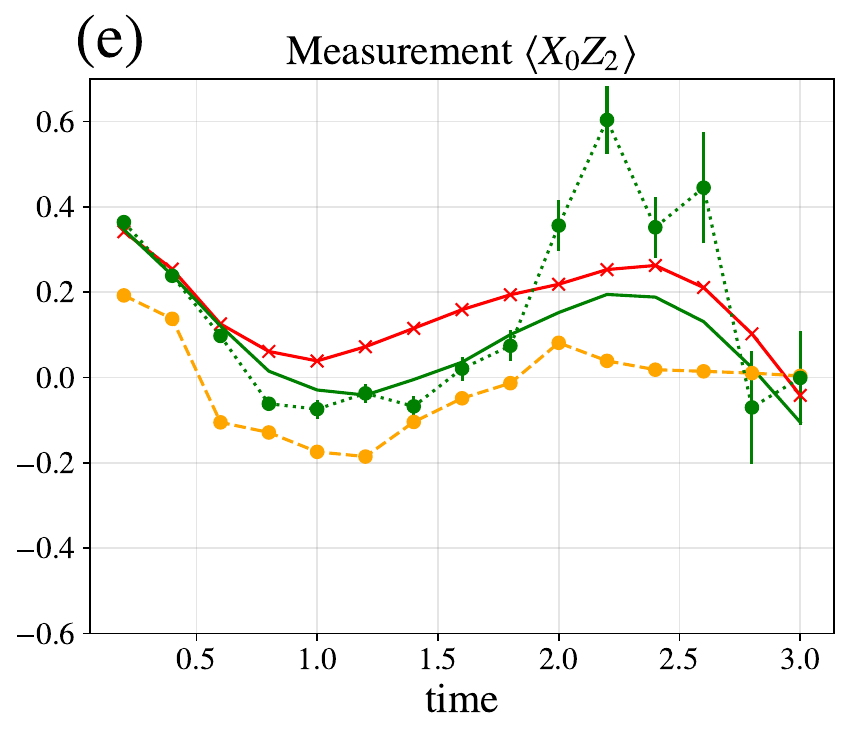}&\includegraphics[width=0.4\linewidth]{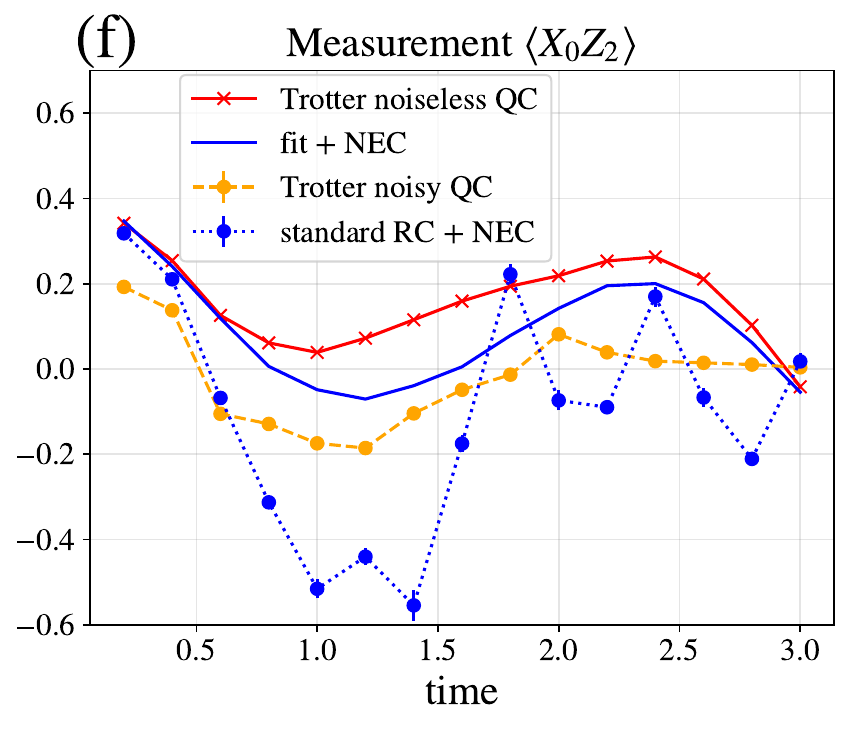}
\end{tabular}
   \caption{Evolution of $\langle X_0\rangle$ (top panels), $\langle Z_2\rangle$  (middle panels) and $\langle X_0 Z_2\rangle$ (bottom panels) after quantum error mitigation protocols (NEC + cRC/RC + REC). Green (resp. blue) dotted curves on the left (resp. right) panels are the results of the real IBMQ device using crosstalk (resp. standard) RC protocols. Green (resp. blue) solid curves are the fits obtained in Fig.~\ref{fig:randomized_evolution}-a,c,e (magenta curves) (resp. Fig.~\ref{fig:randomized_evolution}-b,d,f (cyan curves)) mitigated using classical simulation of NEC. Error bars correspond to the uncertainty due to the limited number of shots and the finite sampling of twirl configurations.}
    \label{fig:mitigated evolution}
\end{figure*}

The results for the observables $\langle X_0\rangle$, $\langle Z_2\rangle$ and $\langle X_0 Z_2\rangle$ using the crosstalk RC technique are plotted in Fig.~\ref{fig:mitigated evolution}-a,c,e. The results for the standard RC protocol are shown in Fig.~\ref{fig:mitigated evolution}-b,d,f. \newest{While not perfect, the results using cRC are clearly better than the results using the standard RC.}

\newest{Note the error bars that are sizeable in Fig.~\ref{fig:mitigated evolution}, as opposed to the negligible ones in Fig.~\ref{fig:randomized_evolution}. This is due to dividing by the fidelity estimate in the NEC protocol, which amplifies the expectation value, but also the errors. For the same reason, the expectation values of the observables may lie outside their range $\left[-1,1\right]$. In practice, such behavior indicates that we have reached the regime where the efficiency of NEC falls off, as unaccounted sources of noise kick in.}
\old{Note that in these figures and throughout this paper, some of the NEC-mitigated data points for the expectation values lay outside of the $\left[-1,1\right]$, which should be impossible. This stems from the fact that the raw results $\langle O\rangle_\text{noisy}$ are divided by the NEC output $\langle E\rangle_\text{noisy}$. Being an estimate of the total fidelity of the circuit, the latter can become arbitrarily small when the noise is too strong. This can yield nonphysical results like going outside of the observable range. In practice, this indicates that we reached the regime of noise where the efficiency of NEC falls off.}

Results for all the remaining observables $\langle X_0\rangle$, $\langle Y_1\rangle$, $\langle Z_2\rangle$, $\langle X_0Y_1\rangle$, $\langle X_0Z_2\rangle$, $\langle Y_1Z_2\rangle$ and $\langle X_0Y_1Z_2\rangle$ are gathered in Fig.~\ref{fig:evolution appendix XYZ} of Appendix~\ref{ap:evolution}. \newest{The conclusion of cRC producing better results than RC holds for these as well. This is quantitatively demonstrated in Fig.~\ref{fig:intro}-b by comparing the accuracy averaged over all the observables. Below we provide an in-detail analysis of the results.}

\subsection{Global accuracy improvement}

\par A comparison between the two columns of Fig.~\ref{fig:mitigated evolution} clearly shows that cRC, i.e. the twirling of neighboring qubits \new{using rotation gates}, qualitatively improves the accuracy of noise estimation protocol, while not requiring to run more circuits\new{, compared to the standard RC protocol}. \newest{These experiments demonstrate that, at least in the context of superconducting qubit architectures like IBMQ QC, this method should systematically be preferred.}

\par \newest{The error mitigation, however, remains imperfect, as simulations still deviate from the noiseless result. These deviations can be separated into two categories.}\old{However, the result remains imperfect, and we have identified two main issues for the protocol.} The first comprises the cases where the noiseless expectation values that we are measuring become close to 0. In this case, we observed that the RC result of the Trotter circuit sometimes gets the wrong sign (see e.g. Fig.~\ref{fig:mitigated evolution}-a, $1\leq t\leq 1.8$ and Fig.~\ref{fig:mitigated evolution}-e, $0.8\leq t\leq 1.4$). As a consequence, dividing by the fidelity estimate of NEC (which is positive by definition), only enhances this sign problem and makes the result deviate strongly from the noiseless curve, even at intermediate times. In spite of this, the oscillations of the exact evolution are still well reproduced by the mitigated curve. The second category relates to the strong deviations occurring at longer times (see e.g. Fig.~\ref{fig:mitigated evolution}-c,e $t\geq 2$). We identified the two problems as having different origins. The first one stems from the --- expected --- effect of a local, $2$-qubit depolarizing noise (cf. Eq.~\eqref{eq:3qubitchannel})\old{on our result (see Appendix~\ref{ap:Trotter noise})}, while the second issue is caused by the fact that the noise channel is not exactly depolarizing after RC, which negatively affects the output of the estimation circuits. \newest{This deviation becomes more important as the time increases since the estimated fidelity becomes very small}. \newest{The next two subsections are devoted to our analysis of these deviations and their origin.}

\subsection{Error mitigation enhances the wrong sign}

\newest{In this subsection, we discuss why the mitigated result may display a wrong sign as the expectation value approaches 0, inducing unwanted deviations from the perfect result, which are in turn enhanced by the NEC technique.}\old{First, let us address the issue of a local noise.} In the case of a pure global depolarising noise, Eq.~\eqref{eq:noise obs real} (or Eq.~\eqref{eq:noise obs global}) guarantees that the noisy result stays of the same sign as the noiseless one. However, when the noise is local (but still depolarising) this may not hold anymore. The equation becomes more involved, and the different terms in the Trotter expansion do not experience the same multiplicative error (see Eq.~\eqref{eq:formula1} of Appendix~\ref{ap:Trotter noise}). In Fig.~\ref{fig:mitigated evolution}, we simulate the Trotter circuit using the local noise channel of Eq.~\eqref{eq:3qubitchannel} for each CNOT and the depolarising parameter from the fit of Fig.~\ref{fig:randomized_evolution} (magenta and cyan curves on Fig.~\ref{fig:randomized_evolution}). We then apply the same noise channel to the estimation circuits and divide both results to obtain the mitigated one (green and blue solid curves in Fig.~\ref{fig:mitigated evolution}). In Fig.~\ref{fig:mitigated evolution}, we show classical simulations using the noise channel of Eq.~\eqref{eq:3qubitchannel}, which reproduce quite well the observed deviation of the QC results. Therefore, in this case, the deviation from the noiseless dynamics is explained by the local depolarising noise occurring on qubits. In this regime, the NEC protocol only works approximately, but other error mitigation protocols such as Zero-Noise Extrapolation (ZNE), are more efficient to tackle local depolarising noise, and can be combined with NEC to improve the results.
In Fig.~\ref{fig:mitigated evolution}-c,e, at longer time, we observe that mitigated fits (green solid curves) deviate slightly from the noiseless Trotter dynamics, but this deviation does not correspond to the one observed on the QC (green dotted curves) which indicates that it is not caused by the locality of the noise.

\subsection{Analysis of long-time deviations}
\label{subsec:analysis_long_time}

\par In Fig.~\ref{fig:mitigated evolution}, the error bars quantify the statistical fluctuations due to the finite sampling of the RC method and the finite number of shots (see Appendix~\ref{ap:uncertainty} for a detailed explanation of the method used to estimate statistical fluctuations). The deviation observed from the classical simulations performed using a quasi-local depolarising noise are not within these error bars. In Fig.~\ref{fig:evolution appendix XYZ}, we have increased $N_{RC}$ of NEC from $300$ to $600$ for $t\geq 1.8$ in order to reduce this uncertainty. If one compares Fig.~\ref{fig:mitigated evolution}-a,c,e together with Fig.~\ref{fig:evolution appendix XYZ}-a,c,e, data points have moved consistently within the range given by the error bars, there is no clear improvement of the results. Another source of error must \older{causes}\newest{be the cause of} this deviation.

\par
\begin{figure}[!!h]
\centering
\includegraphics[width=\linewidth]{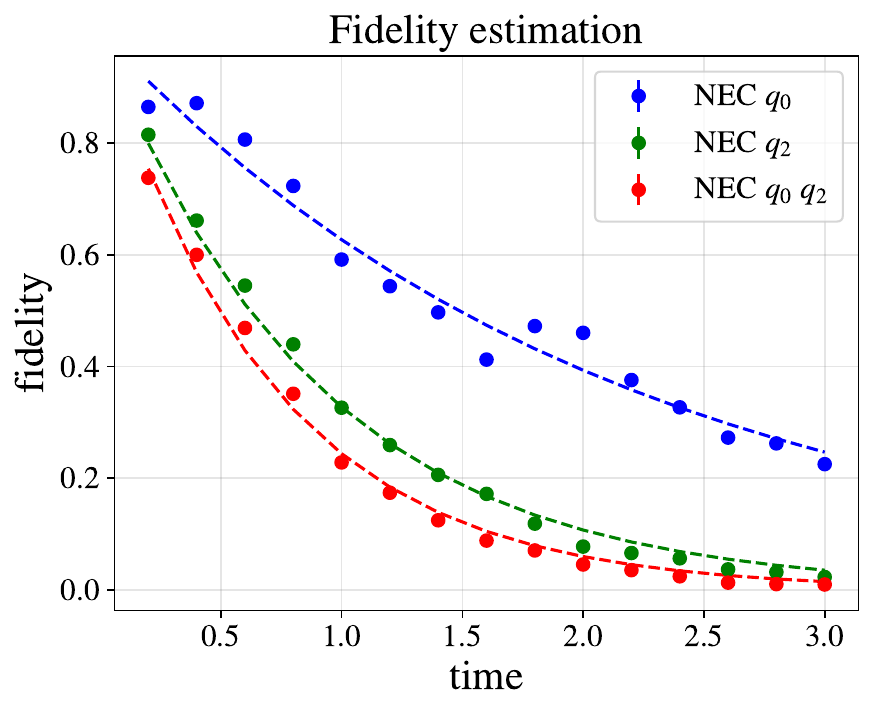}  
    \caption{Effective fidelity estimation for qubits $q_0$ (blue points), $q_2$  (green points) and $q_0q_2$ (red points), obtained by averaging over 300 randomly compiled noise estimation circuits. Solid lines are the results of the fit using the noise model of Eq.~\eqref{eq:3qubitchannel}. Error bars due to shot noise and finite sampling of RC configurations are smaller than the plot marker size.}
    \label{fig:fidelity estimation}
\end{figure}

 Since the deviations are not related to an insufficient number of RC samples, or to shot noise, it must originate from a poor estimation of the effective noise parameters on the Trotter cicuits after cRC by the NEC. To understand this, we compare the noise parameters of the Eq.~\eqref{eq:3qubitchannel} estimated using the Trotter circuits with cRC with the ones obtained from the fidelity estimations from the estimation circuits (see Fig.~\ref{fig:fidelity estimation}). We fit the 7 different estimation curves and find that resulting fit parameters obtained are: $\lambda_{\text{CNOT}}^{01}\simeq 0.007$, $\lambda_{\text{CNOT}}^{12}\simeq 0.016$, $\lambda_{\text{neigh.}}^{01}\simeq 0.05$, $\lambda_{\text{neigh.}}^{12}\simeq 0.009$ and $\lambda_{\text{glob.}}\simeq0.002$ and $\chi^2\simeq 4.10^{-4}$. These results only slightly differ from what we obtained from the randomized compiling of Trotter circuits (see Sec.~\ref{sec:twirlBCS}). These small deviations on the noise parameters estimated by the two types of circuits are not relevant at small times but become crucial at longer times. Indeed, the uncertainty on the mitigated result becomes large as the denominator (estimation circuit) approaches $0$. It can be estimated as (see Supplemental Material of~\cite{Urbanek2021}): 
\begin{eqnarray}
\sigma_m^2&=&\frac{\langle O\rangle^2_\text{noisy}\sigma_e^2+\langle E\rangle^2_\text{noisy}\sigma_o^2}{\langle E\rangle_\text{noisy}^4}
\label{eq:uncertainty}\\
&=&\frac{\langle O\rangle^2_\text{mitigated}\sigma_e^2+\sigma_o^2}{\langle E\rangle_\text{noisy}^2}
\label{eq:uncertainty2}
\end{eqnarray} where $\langle E\rangle_\text{noisy}$ is the mean value estimated from the NEC, $\langle O\rangle_\text{noisy}$ is the mean value estimated from the Trotter circuits, $\langle O\rangle_\text{mitigated}=\frac{\langle O\rangle_\text{noisy}}{\langle E\rangle _\text{noisy}}$, $\sigma_e$, $\sigma_o$ and $\sigma_m$ their respective standard deviation. Fig.~\ref{fig:fidelity estimation} shows that the fidelity is exponentially decreasing with the number of CNOT gates. Note, in particular, that the noise estimation method should be less stable for multiqubit observables because one expects their fidelity to be smaller compared to single-qubit observables (as they are more likely to be directly affected by errors). For an example, see Fig.~\ref{fig:evolution appendix XYZ}-g in Appendix~\ref{ap:evolution} where the error bars (uncertainty due to shot noise \new {and the finite sampling of the RC method}) become large for the observable $X_0 Y_1 Z_2$ at long time.\\

\par The noise parameter uncertainty stemming from the Trotter and the noise estimation circuits reflects the deviation of the effective noise from a pure quasi-local depolarising noise even after the application of cRC. This deviation can have two origins. On one hand, the assumption of a 2-qubit depolarising noise on the CNOT's active qubits is not exact, as the theorem only guarantees a Pauli noise \new{channel}. To avoid making this extra assumption, it will be either necessary to find a way to further simplify the Pauli noise \new{channel} on active qubits into a depolarising noise or to use more sophisticated error mitigation protocols \newest{(as compared to NEC)} which take into account the full effect of the Pauli noise. On the other hand, it may be possible that single-qubit gate's error cannot be neglected anymore. For the longest circuit using the crosstalk twirling protocol, we have on average (depending on the twirl configuration) $\newest{n_1 = 1187}$ single-qubit gates and $135$ CNOT gates. Single-qubit gate's error rate being estimated at $\newest{\epsilon_1}\sim 10^{-4}$, their total effect on the output of a circuit should be \older{ no more than $\sim10\%$} \newest{around $\sqrt{n_1}\times\epsilon_1\sim 0.3\%$}. However, as we have seen with Eq.~\eqref{eq:uncertainty} and  Eq.~\eqref{eq:uncertainty2}, small deviations can become important when the denominator approaches $0$.

\section{Conclusion}
\label{sec:conclusion}
\par
In this work, we have presented and analysed  simulations of the BCS hamiltonian for three energy levels on an IBM superconducting quantum computer. \older{Starting from \older{the}\newest{a quenched} mean-field ground state, we looked at the \newest{non-equilibrium} time evolution of different single-qubit and multiple-qubit observables. We used the Trotter algorithm in order to implement the hamiltonian evolution in terms of unitary gates. The inherent Trotter error remains small for the time evolution we are looking at (up to 15 Trotter steps).
\par On real IBM quantum computers, the primary source of errors is due to the application of noisy CNOT gates, the only 2-qubit native gates. They are needed to simulate the interaction between Cooper pairs and to apply SWAP gates that exchange the logical information contained in physical qubits, allowing all Cooper pairs to interact with each other, despite the linear layout of the physical qubits. To improve the \older{precision}\newest{accuracy} of our results, we used two quantum error mitigation protocols: readout error correction (REC) and noise estimation circuits (NEC). The NEC method estimates the amount of noise in a circuit by running extra circuits which are simplified versions of the original one. However, it performs better when the noise channel acting on the CNOT gates is essentially depolarising. In addition, we thus perform a standard RC protocol, which allows to effectively approach such a channel by averaging over different random instances of the original circuit. We showed that this approach improves the result somewhat (see blue bars in Fig.~\ref{fig:intro}b). 

However, the use of noisy CNOT gates makes the system  prone to crosstalk effects, which are known for being problematic in superconducting quantum computers~\cite{Ding2020,Sarovar2020,Murali2020,Kandala2021,Rudinger2021,Tripathi2022,Xie2021,Zhao2022}. In addition, the Pauli noise \new{channel} obtained after RC protocol is not exactly a depolarising one. To address these crucial issues we develop an extension of the RC protocol which aims to turn the coherent noise resulting from crosstalk effects into a depolarising channel. We show that this extended \textit{twirling on neighboring qubits qualitatively improves our results} simply by adding a few single-qubit gates on neighboring qubits of the CNOT gates (see green bars in Fig.~\ref{fig:intro}b). Most notably, we achieve this improvement without adding  extra circuits or qubits compared to the standard RC protocol.}\newest{To obtain reliable results from the noisy QC, we employ a simple error mitigation technique called NEC, whose efficiency is conditioned by the simplicity of the noise structure. In order to achieve the best possible results, we use the RC technique, which effectively turns the unknown and complex noise channels acting on the QC into a simpler channel.}

\newest{We use two versions of RC: the standard RC and the crosstalk RC. We find that the latter enables achieving much better accuracy, by further simplifying the noise caused by crosstalk effects, while not requiring extra circuit runs or qubits compared to the former. This demonstrates the significance of crosstalk on IBM quantum computers.
}

Our work closes a gap in the field of error mitigation on NISQs: while the specific structure of the noise on real devices is rather complex and usually not well understood, theoretical proposals for error mitigation techniques usually assume a simple depolarising error channel. We explicitly demonstrate that the use of crosstalk RC does indeed bring unknown noise closer to depolarising noise, allowing us to take advantage of existing error mitigation proposals. This is the main reason for the significant overall performance improvements stemming from the use of our crosstalk RC protocol, as shown in Fig.~\ref{fig:intro}. 

There is another appeal in our approach: crosstalk errors are notoriously hard to characterize. Being a systematic, coherent noise channel, it is difficult to predict their net effect on the output of a given quantum computation. Using crosstalk RC and our theoretical prediction for the structure of the effective error channel, we show that existing protocols significantly underestimate the real error in superconducting quantum computers. We find in particular that crosstalk error channels map to a depolarising error channel with \older{parameter} \newest{error rate per CNOT gate that can be} as high as 5\%, i.e. almost 5 times the values obtained \older{by randomized benchmarking for 2-qubit} \newest{by randomized benchmarking for the error rate on the active qubits of the CNOT}  gates\older{, which assumes depolarising noise from the start and neglects crosstalk errors}. This result is in good qualitative agreement with the recent papers \cite{Hines2023,Shirizly2023}, in which crosstalk is also found to be the main source of errors for quantum computation on IBM quantum computers. 


The \older{precision}\newest{accuracy} gain from using crosstalk RC, in comparison to the standard RC, in the explicit simulations of the dynamics of the BCS model on NISQs is significant. It allowed us to perform intermediate-time (up to 15 Trotter steps) quantum simulations of an \textit{all-to-all interacting model} with \older{a precision}\newest{an accuracy} beyond what can, to our knowledge, be found in the literature. This brings us one step closer to quantum simulations on NISQs matching \older{or even exceeding} the capabilities of classical computers\older{, thus becoming  relevant to a wide range of theoretical research in a number of fields}.\\


\acknowledgements
 We are indebted to Igor V. Gornyi for numerous crucial interactions and discussions throughout this project. Furthermore, we thank Emil Yuzbashyan, Jannis Ruh, Regina Finsterhoelzl, and Guido Burkard for fruitful discussions.
We acknowledge funding from the state of Baden-Württemberg through the Kompetenzzentrum Quantum
Computing (Project QC4BW).
J.S. is grateful to Departamento de Física, FCFM, University of Chile (Santiago) for hospitality during the final stage of this work and for support  by the European Commission’s Horizon 2020 RISE project Hydrotronics
(Grant No. 873028). 

\section*{Data and code availability}
Data and the code used to generate quantum circuits and analyze the results are available in \href{https://github.com/Perrin35/BCS-error-mitigation.git}{https://github.com/Perrin35/BCS-error-mitigation.git}

\appendix
\section{BCS hamiltonian in terms of Pauli matrices}
\label{ap:BCS_to_Pauli}
In this appendix, we show how to rewrite the standard hamiltonian for the BCS model for superconductivity in terms of spin operators. The BCS hamiltonian couples electrons pairs known as Cooper pairs~\cite{Bardeen1957} through a phonon-mediated interaction~\cite{Anderson1959}. The effective hamiltonian, obtained after eliminating the phonons can be written in the form~\cite{Yuzbashyan2005}:
\begin{equation}
    \mathcal{H}_\text{BCS}=\sum_{j=0}^{L-1}\epsilon_j (c^\dagger_{j \uparrow}c_{j \uparrow}+c^\dagger_{j \downarrow}c_{j \downarrow})-g\sum_{i,j=0}^{L-1} c^\dagger_{i\uparrow} c^\dagger_{i \downarrow} c_{j\downarrow} c_{j \uparrow}
\end{equation}
where $L$ is the number of energy levels, \old{$c^{(\dagger)}_{j,\sigma}$ are the usual electronic  creation (annihilation) operator}\new{$c^{(\dagger)}_{j,\sigma}$ are the usual electronic annihilation (creation) operator} acting on the energy level $j$ of energy $\epsilon_j$ and spin $\sigma\in\{\uparrow,\downarrow\}$. \new{The coupling constant $g$ describes the effective interaction between Cooper pairs.}\old{The first sum is the single-particle term while the second sum describes an effective coupling of strength $g$ between Cooper pairs. 
The hamiltonian can be rewritten in terms of Pauli matrices (see Appendix \ref{ap:pauli} for a detailed derivation):}
\new{ Single electrons are trivially decoupled from the dynamics. We therefore restrict the study to the Cooper pairs' subspace. The latter mimic the statistics of hardcore bosons~\cite{Llano2007}, their action in terms of creation and annihilation operators is easily related to the algebra of Pauli matrices:
\begin{align}
\sigma_j^z&=1-c_{j\downarrow}^\dagger c_{j\downarrow}-c^\dagger_{j\uparrow}c_{j\uparrow} \\
\sigma_j^+&=\frac{\sigma_j^x-i\sigma_j^y}{2}=c_{j\uparrow}^\dagger c_{j\downarrow}^\dagger\label{eq:sigplus}    \\
 \sigma_j^-&=\frac{\sigma_j^x+i\sigma_j^y}{2}=c_{j\downarrow}c_{j\uparrow}\label{eq:sigminus}    
\end{align} 

where $\sigma_j^a$, $a\in \{x,y,z\}$, refers to the operator such that the Pauli matrix $\sigma^a$ is applied on the energy level $j$ and the identity operator on the other energy levels. The  eigenstates $\ket{0},\ket{1}$ of $\sigma_j^z$ denote respectively the absence or presence of a Cooper pair on the energy level $j$. Using this transformation, the BCS hamiltonian can be recast as}
\begin{align}
\mathcal{H}_\text{BCS}&=-\sum_{j=0}^{L-1}(\epsilon_j -\frac{g}{2})\sigma_j^z
-\frac{g}{2}\sum_{0\leq i<j\leq L-1}\left(\sigma_i^x \sigma_j^x + \sigma_i^y \sigma_j^y\right),
\end{align}
\new{which is Eq.~(\ref{eq:BCS_pauli}). This rewriting in terms of Pauli matrices is closely related  to the Anderson $1/2$-pseudo-spin model~\cite{Anderson1958}. The only difference being that the $z$-component of the pseudo-spin is defined as $\smash{S_j^z=-\frac{\sigma_j^z}{2}}$ where $j$ is the energy level index. Note that the interaction term rewritten using Pauli matrices does not suffer from long Pauli string terms scaling with the system size as it is usually the case for electrons due to the Jordan-Wigner transformation~\cite{Jordan1928,Ovrum2007} which ensures the antisymmetrization of the wavefunction.}

\old{\subsection{Trotterization}
\label{sec:trotter}
\par In this article, we aim to simulate the dynamics resulting from $\mathcal{H}_\text{BCS}$. To this end, we need to implement the unitary evolution operator $\smash{U(t)=e^{-i\mathcal{H}_\text{BCS} t}}$ on the quantum computer i.e. to break down $U(t)$ into a series of 1- and 2-qubit unitary gates that can be directly applied on the QC (native gates). This operation, dubbed transpilation, is a very hard problem that usually cannot be tackled exactly, even for simple, integrable systems. 
\par To circumvent this issue, we will approximate the exact unitary evolution operator using the Trotter algorithm. It consists in slicing the full time evolution into small time steps $\Delta t$: 
\begin{equation}
e^{-i\mathcal{H}t}=\left(e^{-i\mathcal{H}\Delta t}\right)^r
\end{equation}
where $t$ is the time duration of the whole dynamics, $\smash{\Delta t=\frac{t}{r}}$ is the small time step  and $r$ the number of time steps. For each small time step evolution, we use the Suzuki-Trotter formula to expand the exponential \cite{Suzuki1976}, and neglect all the non-zero commutators between the  two terms of the hamiltonian such that we can write each of them in a separate exponential. For the BCS hamiltonian this yields:
\begin{equation}
e^{-i\mathcal{H}_\text{BCS}\Delta t}\simeq\prod_{j=0}^{L-1}e^{i(\epsilon_j-\frac{g}{2})\sigma_j^{z}\Delta t} \hspace{-0.5cm} \prod_{0\leq i<j\leq L-1}\hspace{-0.5cm} e^{i\frac{g}{2}(\sigma_i^x\sigma_j^x+\sigma_i^y\sigma_j^y)\Delta t}
\label{eq:Trotter}
\end{equation}
\par For each Trotter step, the error accumulated by this approximation has an upper bound of order $\mathcal{O}(\Delta t^2)$. This error scales  as $ \mathcal{O}(t \Delta t)$ for the whole dynamics (see e.g. ~\cite{Childs2018,Seetharam2021}). The Trotter algorithm can be extended to higher order in the commutator in order to lower this error but more gates are then required to simulate one Trotter step~\cite{Childs2018} . This is a major issue in the context of NISQ, as most of the error stem from the application of noisy gates. One then has to find a good trade-off between algorithmic and noise-induced errors. Fortunately, it turns out that the Trotter error's upper bound is usually not reached for the type of dynamics we are considering here. Consequently, even at the lowest order in the Trotter approximation, the noise coming from the gates remains the main source of error (see Sec.~\ref{sec:twirl} and ~\ref{sec:estimation}).}

\section{Circuit transpilation}
\label{ap:transpilation}
\par Each term of Eq.~\eqref{eq:Trotter} needs to be implemented with IBM QC's native gates. They are three single-qubit gates: $R_z(\theta)$, the Pauli matrix $\sigma^x$ (or simply denoted $X$) and $\sqrt{X}$ and one 2-qubit gate: the CNOT gate~\cite{Aleksandrowicz2019}. They form a universal gate set operations which allows to implement any unitary operator. An additional difficulty arises when taking into account the machine's layout: the unitary contains interaction terms that couple quibts which are not directly connected on the QC; we devote subsection~\ref{sec:layout} to this problem.
\subsection{Transpilation of 1-body operators}
\label{sec:transpilation_1}
\par The leftmost product of Eq.~\eqref{eq:Trotter} corresponds to on-site energies and can therefore be implemented as a product of 1-qubit gates. It is indeed a product of $\smash{R_z(\theta)=e^{-i\frac{\theta}{2}\sigma^z}}$ gates, which perform a rotation of angle $\theta$ around the Z-axis on the Bloch sphere. In our case $\theta_j=-\Delta t(2\epsilon_j-g)$ where $j$ is the index of the different qubits. The $R_z$ gate is a native gate for IBM quantum computers.
\subsection{Transpilation of 2-body operators}
\label{sec:transpilation_2}
\par The rightmost product of Eq.~\eqref{eq:Trotter} describes the interaction between Cooper pairs. To implement them, one needs to use 2-qubit gates for entangling both qubits. The CNOT gate is the native 2-qubit gate of most of IBM QC.
\par  Using $\left[\sigma_i^x\sigma_j^x,\sigma_i^y\sigma_j^y \right]=0$, we obtain:
\begin{equation}
    e^{-i\frac{\alpha}{2}(\sigma_i^x \sigma_j^x+\sigma_i^y \sigma_j^y)}= e^{-i\frac{\alpha}{2}\sigma_i^x \sigma_j^x} e^{-i\frac{\alpha}{2}\sigma_i^y \sigma_j^y},
\label{eq:interaction_term}
\end{equation}
\new{where $\alpha=-g\Delta t$ in our case.} To implement 2-body interaction terms of the form
\begin{equation}
    e^{-i\frac{\alpha}{2}\sigma_i^a \sigma_j^a}
    \label{eq:2qubitinteraction}
\end{equation} 
where $a\in\{x,y,z\}$, one needs 2 CNOT gates, a single $R_z(\alpha)$ gate, and a 1-qubit gate $O_a$:
\begin{equation}
O_z=\mathbb{I},\ O_x=H=\frac{1}{\sqrt{2}}
\begin{pmatrix}
1&1\\
1&-1\\
\end{pmatrix} \textrm{ and } O_y=\frac{1}{\sqrt{2}}
\begin{pmatrix}
1&i\\
i&1\\
\end{pmatrix}
\label{eq:1qubitgates_interaction}
\end{equation}

\begin{figure}[!!h]
\includegraphics[width=0.9\linewidth]{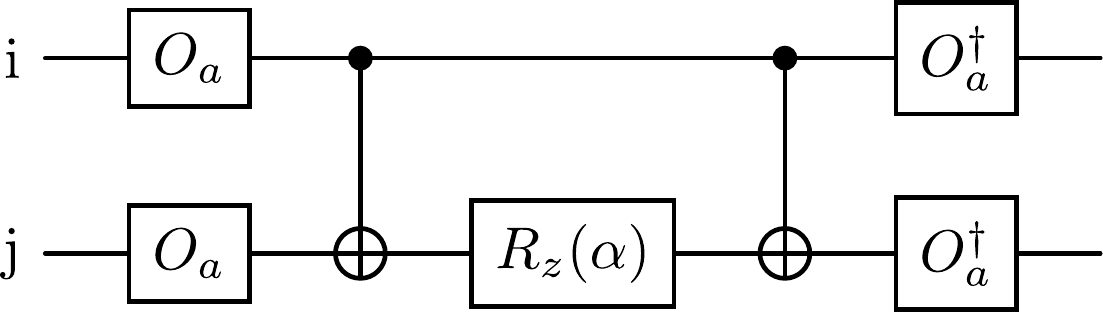}
\caption{Quantum circuit decomposition \new{of} the unitary operator $e^{-i\frac{\alpha}{2}\sigma_i^a \sigma_j^a}$ with $a\in\{x,y,z\}$. Here $i$ and $j$ denote qubit indices, the vertical lines represent the 2-qubit CNOT gates, for which the control qubit is indicated by the black dot. The other rectangles represent 1-qubit gates explicited in Eq.~\eqref{eq:1qubitgates_interaction}.}
\label{fig:interaction_circuit}
\end{figure}

 We represent in Fig.~\ref{fig:interaction_circuit} the 2-qubit quantum circuit equivalent to the exponential \eqref{eq:2qubitinteraction}.
 Therefore, implementing Eq.~\eqref{eq:interaction_term} as a quantum circuit requires 4 CNOT gates. Note however that after implementing such an interaction term using Qiskit \cite{Aleksandrowicz2019}, we use the \texttt{transpile} function with the option \texttt{optimization\_level=3} (i.e. the most optimized method), which outputs an equivalent quantum circuit using only 2 CNOT gates (with some single qubit native gates)

\subsection{Transpilation for the whole dynamics}
\label{sec:layout}

\par
\begin{figure}[!!h]
\centering
\includegraphics[width=.9\linewidth]{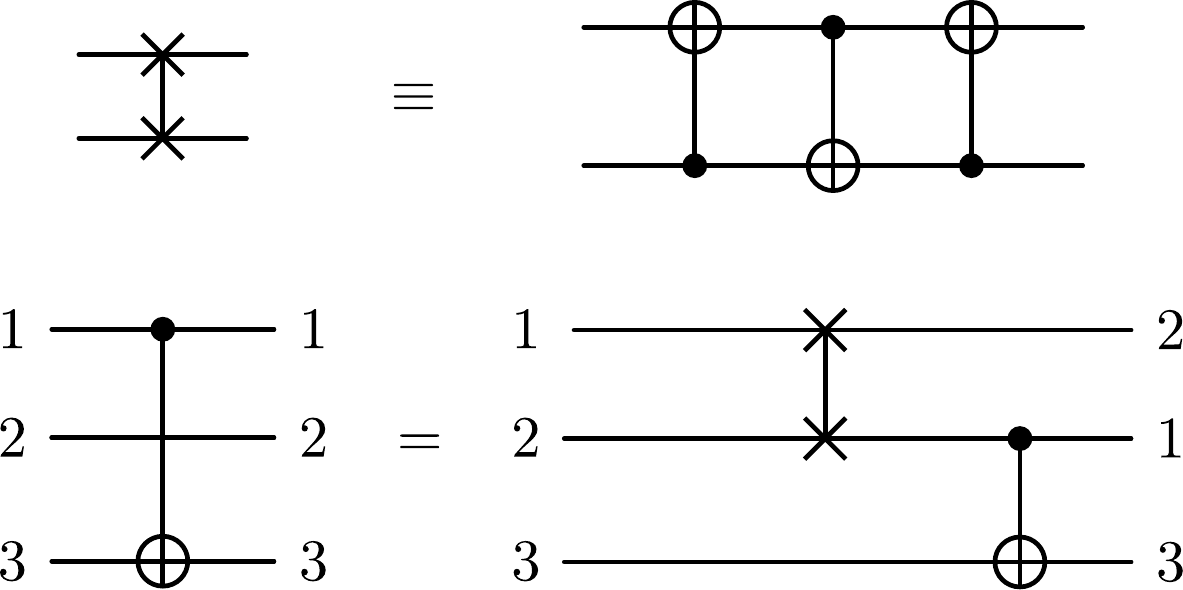}

\caption{Top: Logical decomposition of the SWAP gate (left) into 3 native CNOT gates (right). Bottom:  Example of usage of the SWAP gate (represented with crosses at each end). \new{The numbering refers to the position of logical qubits, while the solid horizontal lines are the physical qubits of the QC.}\old{The numbers denote the indices of logical qubits.} The leftmost circuit cannot be realised in a layout where qubits $1$, $2$ and $3$ are on a line, so we have to use the rightmost circuit instead. Because of this, the number of required CNOT increases by 3, and the positions of the logical qubits are modified.}
\label{fig:SWAP}
\end{figure}

 In this subsection, we use the 1-qubit and 2-qubits quantum circuits previously derived to build the circuit implementing the fully trotterized dynamics. Naively, we could simply concatenate the corresponding elementary circuits on the physical qubits $i$,$j\in\{0,L-1\}$ to reproduce the product of Eq.~\eqref{eq:Trotter}. We see then that the number of 2-qubits interaction circuits required grows as ${L\choose 2}=\frac{L(L-1)}{2}$ where $L$ is the number of energy levels, which translates into $L(L-1)$ noisy CNOT gates for each Trotter step. However, an additional difficulty stems from the architecture of the IBM quantum computers (and superconducting quantum computers in general). As shown in Fig.~\ref{fig:coupling}, QCs used (\texttt{ibm\_lagos} and \texttt{ibmq\_ehningen}) have a certain qubit coupling map, and CNOT gates can only be performed on neighboring physical qubits. As a consequence, it is not straightforward to implement an interaction between two distant qubits (in the sense of the coupling map of Fig.~\ref{fig:coupling}). To overcome this problem, one has to swap around the logical information contained in physical qubits, so that two interacting logical qubits are brought together on neighboring physical qubits. This swap of information is achieved using the so-called SWAP gate. A simple example of the use of SWAP gates is given in Fig.~\ref{fig:SWAP}. Note that the SWAP gate is not a native gate of the IBM machines. It requires three CNOT gates as shown in Fig.~\ref{fig:SWAP}.\\

\par To implement an interaction between two qubits initially separated by a distance $d$ (in the sense of the coupling map of Fig.~\ref{fig:coupling}), at least $d-1$ SWAP gates are required, i.e $3(d-1)$ CNOT gates. Consequently, the number of CNOT gates --and thus the number of errors-- grows very quickly with the size of the system we want to simulate (i.e. the number of qubits). 

\subsection{Initial state preparation}
\label{sec:initial_state}
\par In an effort to minimize the number of noisy CNOT gates in the circuit, we made the choice to start with an initial product (non entangled) state, which can be prepared using only single-qubit gates. A natural product state in the BCS problem is the mean-field ground state which is the ground state of the BCS hamiltonian in the thermodynamic limit but has a non trivial dynamic for finite system. To construct it, we use the gap equation derived from the mean field approximation~\cite{Tinkham1996}:
\begin{equation}
    1=g\sum_j\frac{1}{2\sqrt{\epsilon_j^2+|\Delta|^2}}
    \label{eq:gap}
\end{equation}
where $\Delta$ is the energy gap. The BCS mean-field ground state wave function can be written as:
\begin{align}
    \ket{\text{M.F.}}&=\prod_{i=0}^{L-1}(u_i+v_ic^\dagger_{i\uparrow}c^\dagger_{i\downarrow})\ket{0}_i\\\nonumber
    &=\prod_{i=0}^{L-1}\left(u_i\ket{0}_i+v_i\ket{1}_i\right)
\end{align} where 
\begin{equation}
    \frac{v_i}{u_i}=\frac{\sqrt{\epsilon_i^2+|\Delta|^2}-\epsilon_i}{\Delta^*}
\end{equation} and we also have  the normalization condition $|u_i|^2+|v_i|^2=1$. We rewrite $u_i$ and $v_i$ as: \begin{equation}
u_i=\cos\frac{\theta_i}{2}e^{i\varphi_i/2},\qquad v_i=\sin\frac{\theta_i}{2}e^{-i\varphi_i/2}
\end{equation}
 which yields:
\begin{align}
    \theta_i&=2\ \text{arctan}\left(\frac{\sqrt{\epsilon_i^2+|\Delta|^2}-\epsilon_i}{|\Delta|}\right)\\
    \varphi_i&=-\text{arg}(\Delta)\equiv \varphi
\end{align}
Thus, to prepare this initial state we simply apply on each qubit the gate $R_y(\theta_i)$ followed by the gate $R_z(\varphi)$. The first box on Fig.~\ref{fig:Trotter_circuit} shows an example of this preparation with the set of parameters specified in Sec.~\ref{sec:Params} \newest{(these correspond to $\Delta\approx 0.46$)}. 

\section{Evolution of all observables mitigated}
\label{ap:evolution}
\begin{figure}[!!h]
\centering
\begin{tabular}{cc}
\includegraphics[width=0.5\linewidth]{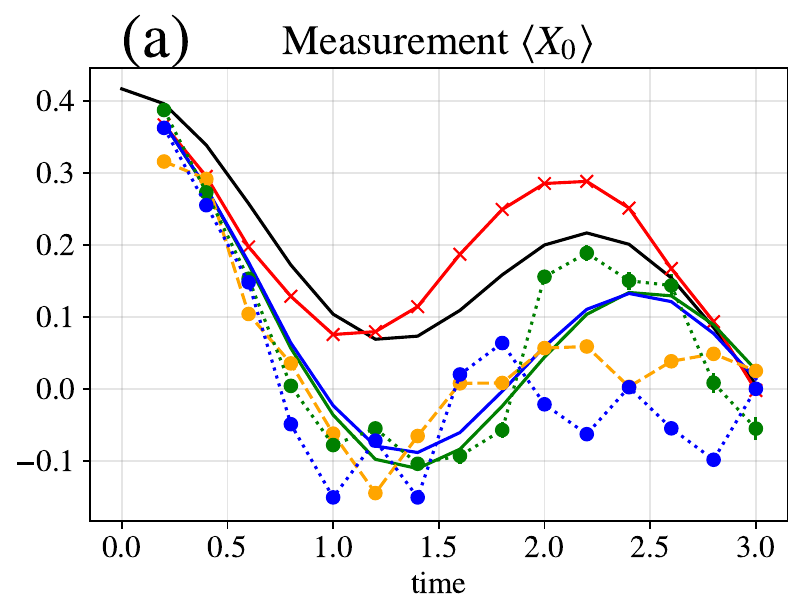}   &     \includegraphics[width=0.5\linewidth]{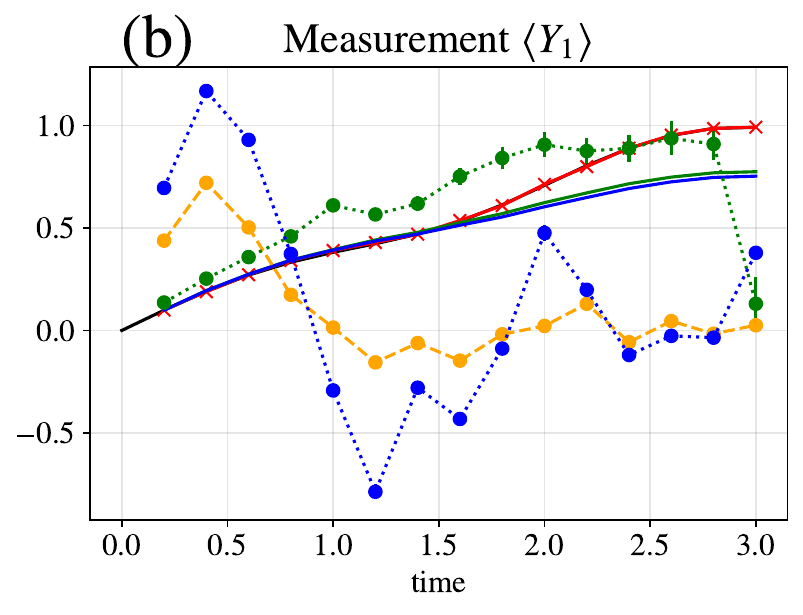} \\
\includegraphics[width=0.5\linewidth]{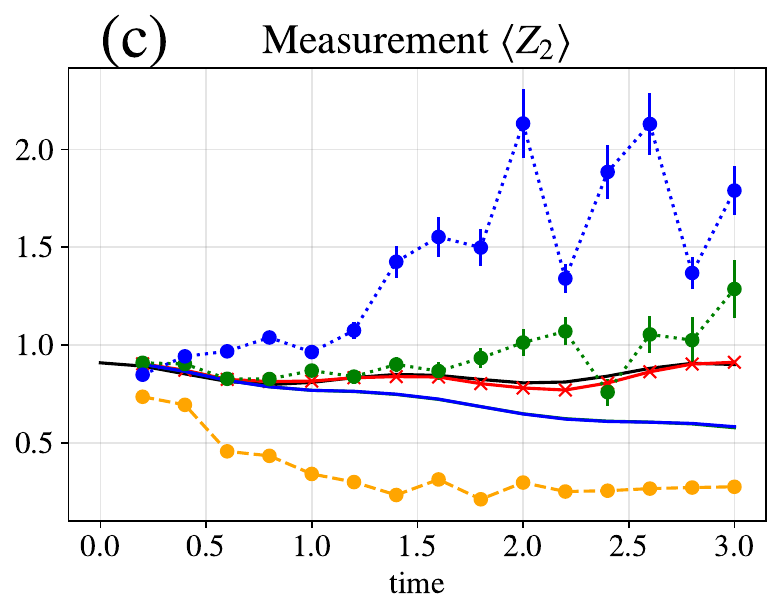}& \includegraphics[width=0.5\linewidth]{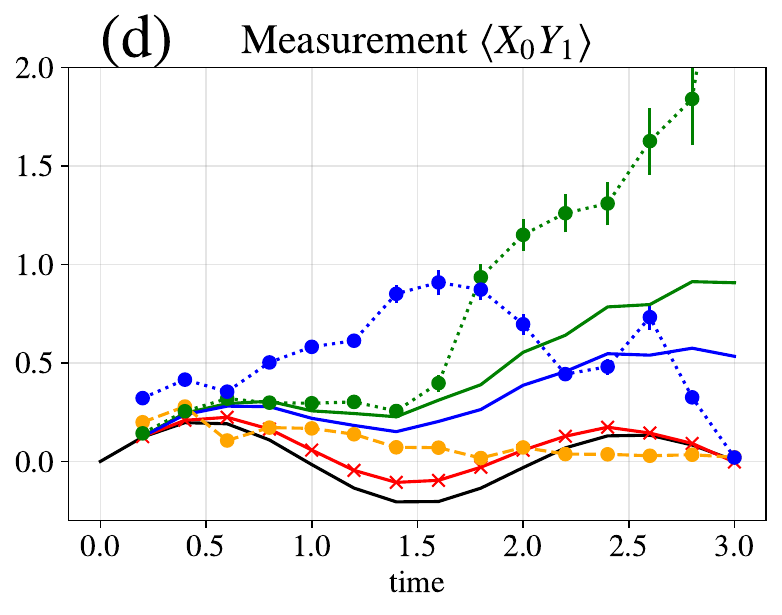}\\
\includegraphics[width=0.5\linewidth]{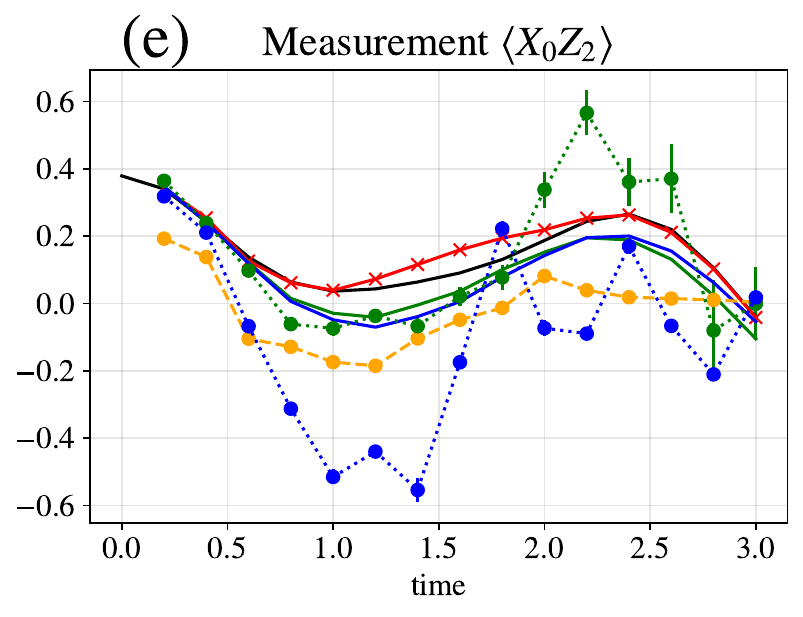}& \includegraphics[width=0.5\linewidth]{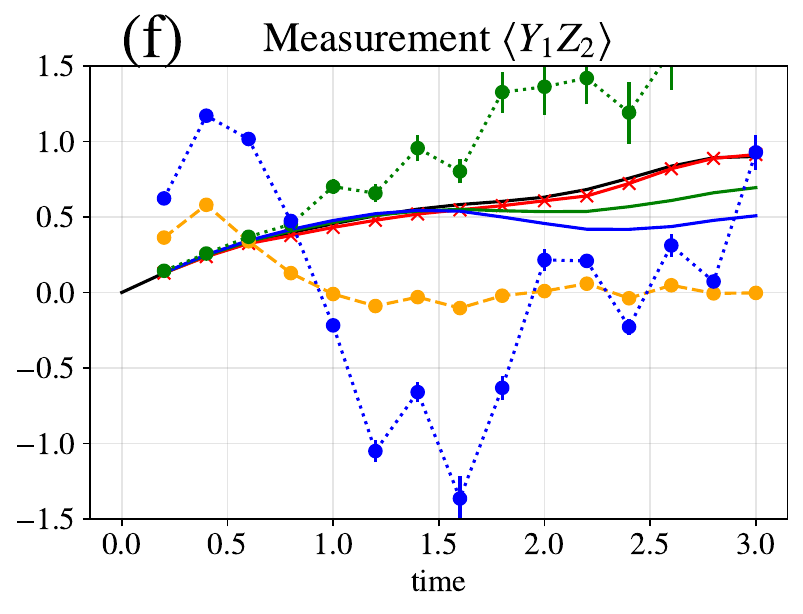}\\
      \multicolumn{2}{c}{\includegraphics[width=0.85\linewidth]{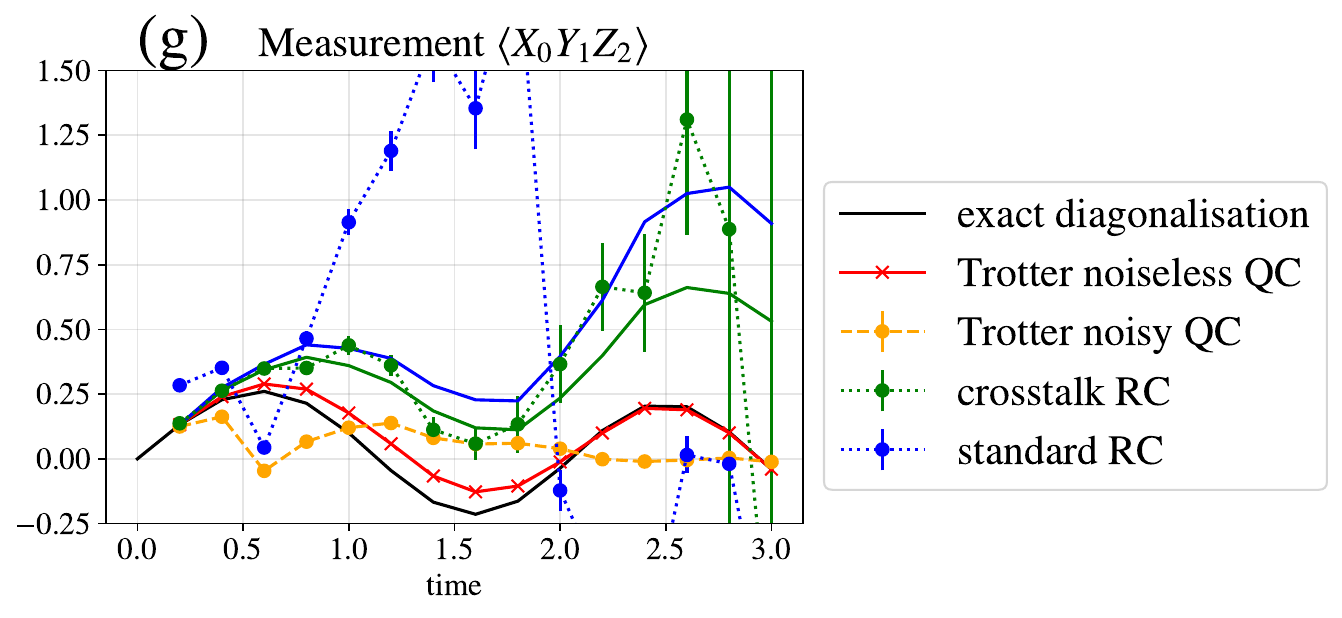}  }
\end{tabular}
    
    \caption{Evolution of the $7$ different observables derived from the measurement of of \texttt{ibm\_lagos}' qubits $q_0$, $q_1$ and $q_2$  respectively in the direction x, y, z after  application of mitigation protocols (NEC + RC + REC). Blue (resp. green) curves correspond to standard (resp. crosstalk) RC. Dashed curves are the result obtained on a QC and the solid line are the corresponding fit made with the noise model of Eq.~\eqref{eq:3qubitchannel}. Exact evolution (black lines), noiseless (resp. noisy) Trotter dynamics (red lines) (resp. dashed orange curves) are also plotted for each quantity. Error bars are due to the limited number of shots and the finite sampling over twirl configurations.}
    \label{fig:evolution appendix XYZ}
\end{figure}

\begin{figure}[!!h]
\centering
\begin{tabular}{ll}
\includegraphics[width=0.5\linewidth]{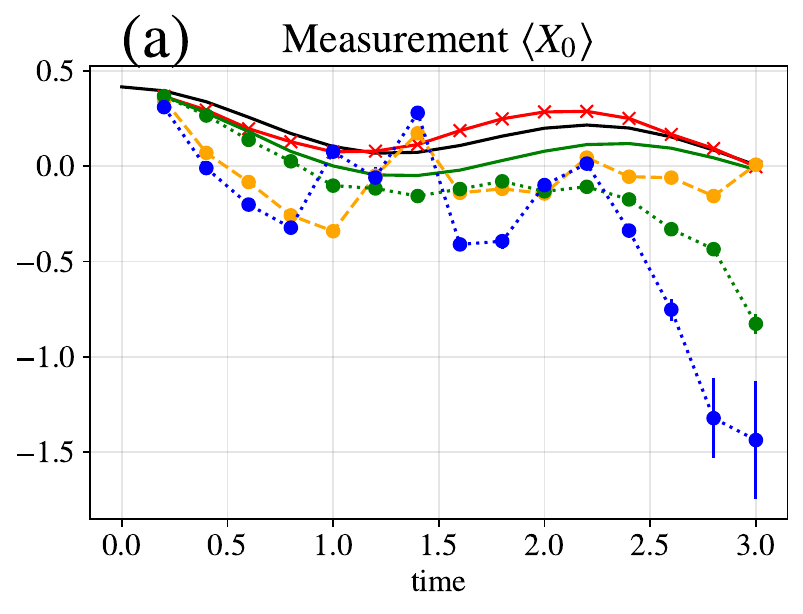}   &     \includegraphics[width=0.5\linewidth]{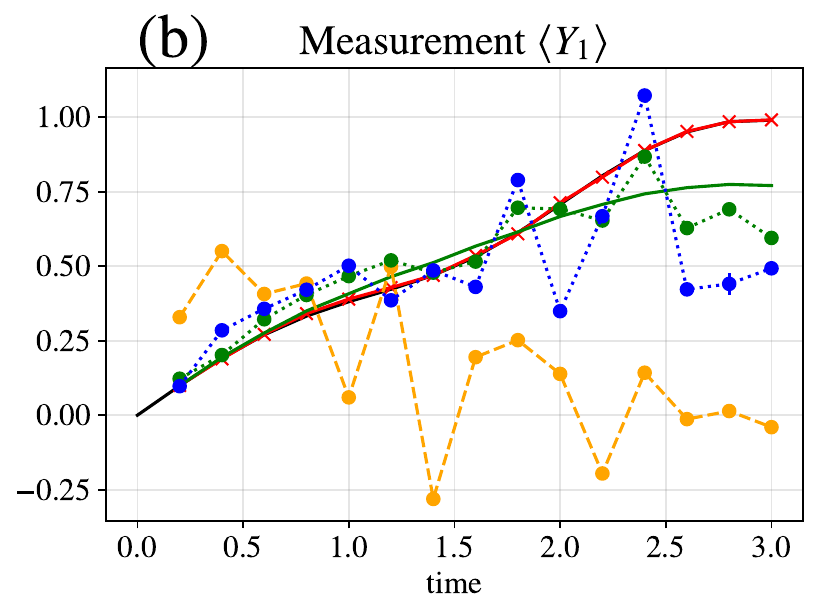} \\
\includegraphics[width=0.5\linewidth]{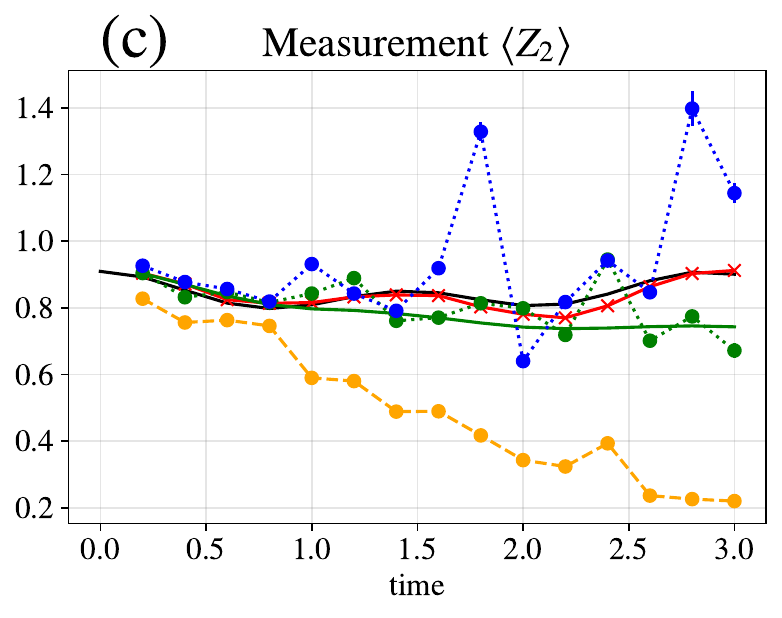}& \includegraphics[width=0.5\linewidth]{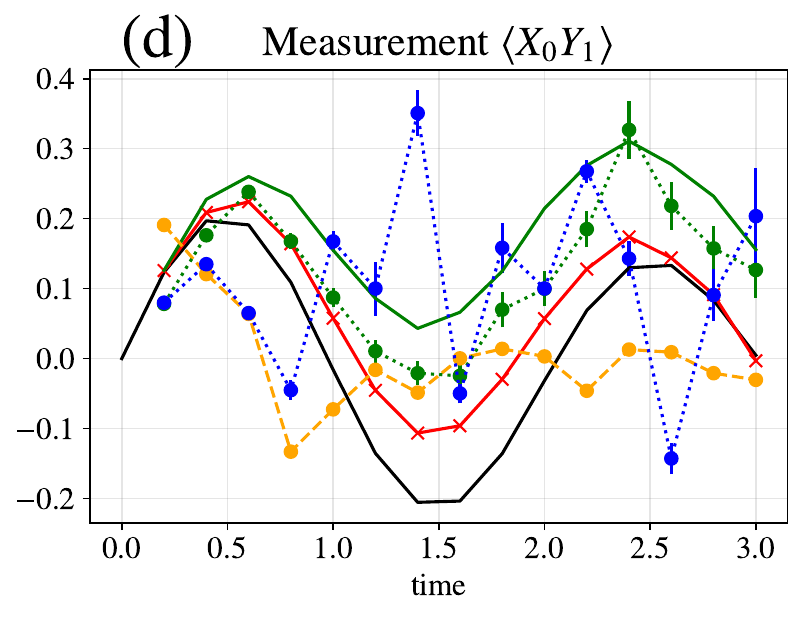}\\
\includegraphics[width=0.52\linewidth]{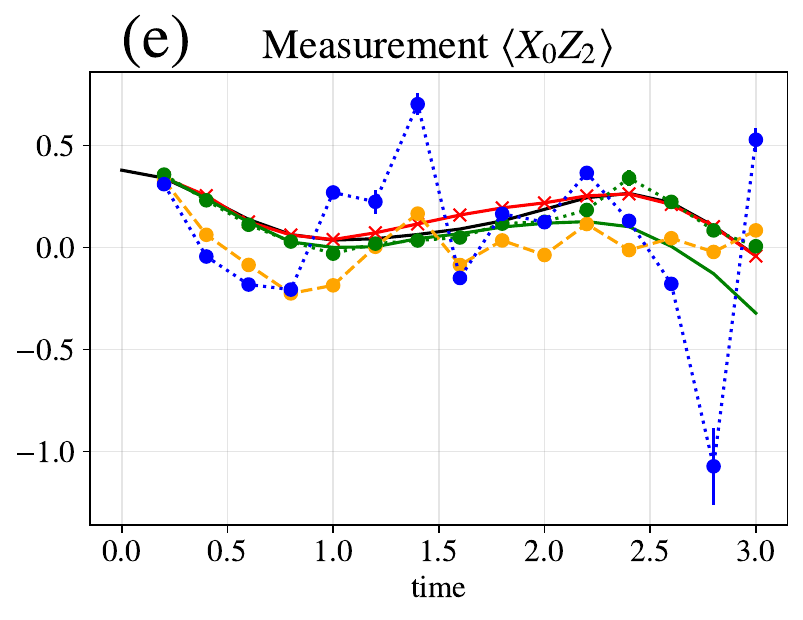}& \includegraphics[width=0.5\linewidth]{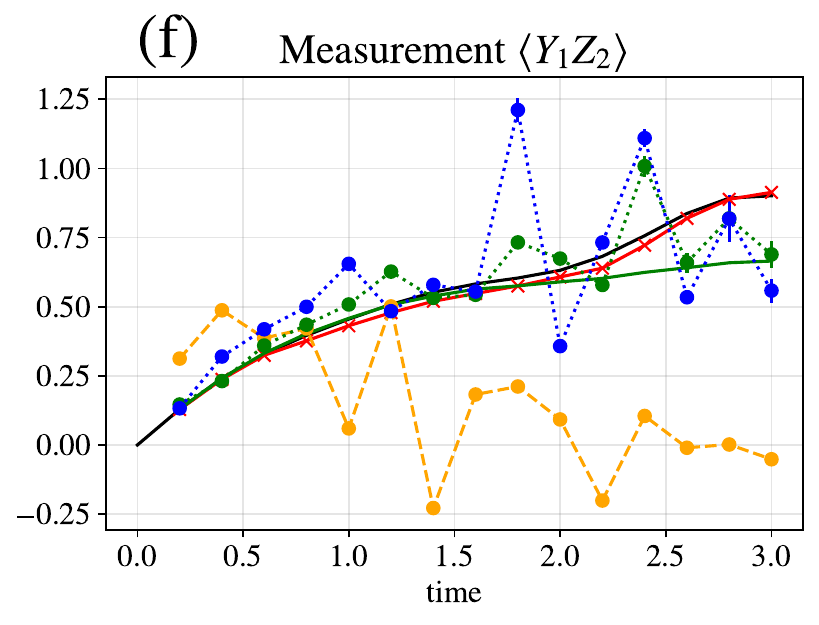}\\
      \multicolumn{2}{c}{\includegraphics[width=0.81\linewidth,trim={0 0.35cm 0 0},clip]{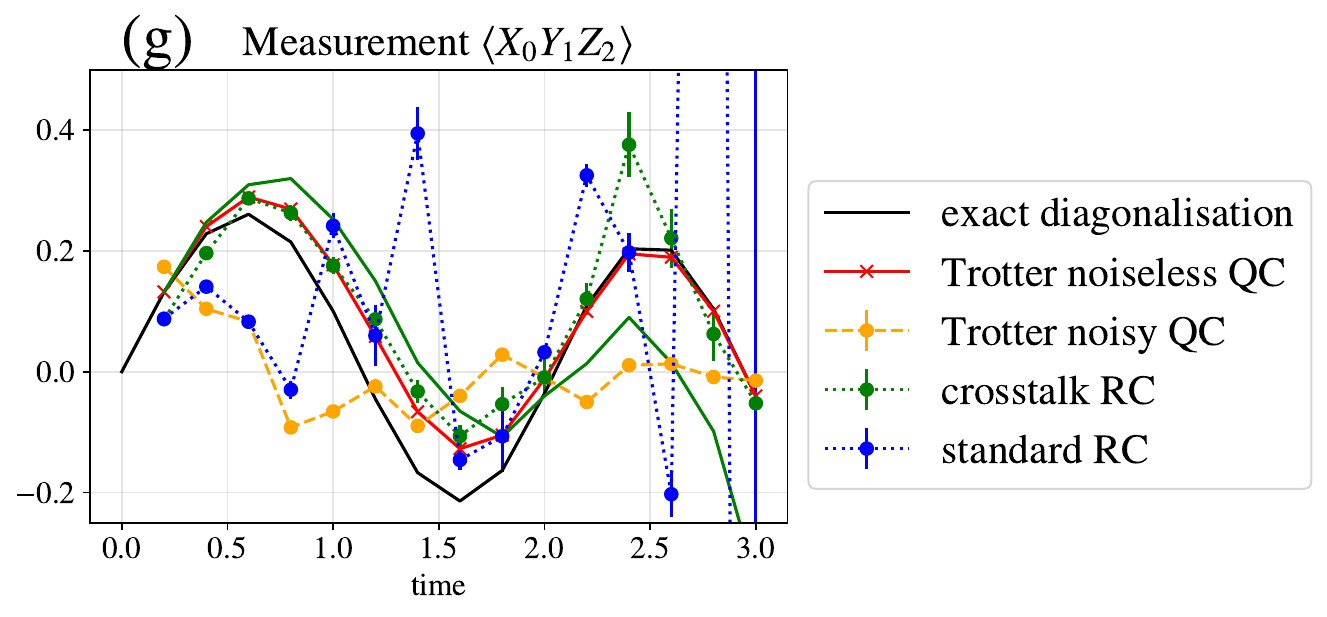}  }
\end{tabular}
    \caption{Evolution of the $7$ different observables derived from the measurement of \texttt{ibm\_ehningen}'s qubits $q_{18}$, $q_{21}$ and $q_{23}$ respectively in the direction x, y, z after  application of mitigation protocols (NEC + RC + REC). Blue (resp. green) curves correspond to standard (resp. crosstalk) RC. Dashed curves are the result obtained on a QC. The green solid line is the fit of crosstalk RC made with the noise model of Eq.~\eqref{eq:3qubitchannel} (fit for the standard RC did not converge). Exact evolution (black lines), noiseless (resp. noisy) Trotter dynamics (red lines) (resp. dashed orange curves) are also plotted for each quantity. Error bars are due to the limited number of shots and the finite sampling over twirl configurations.}
    \label{fig:evolution appendix XYZ ehningen}
\end{figure}

\begin{figure}[!!h]
\centering
\begin{tabular}{ll}
\includegraphics[width=0.5\linewidth]{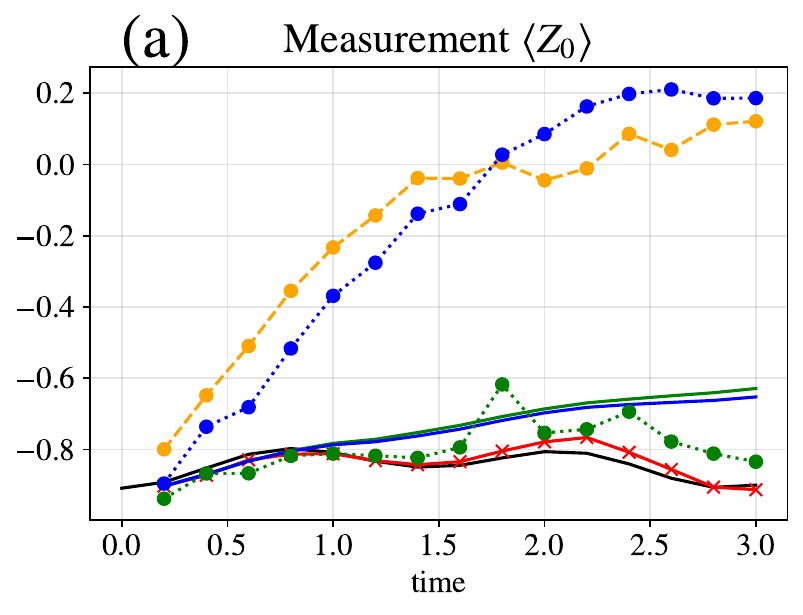}   &     \includegraphics[width=0.5\linewidth]{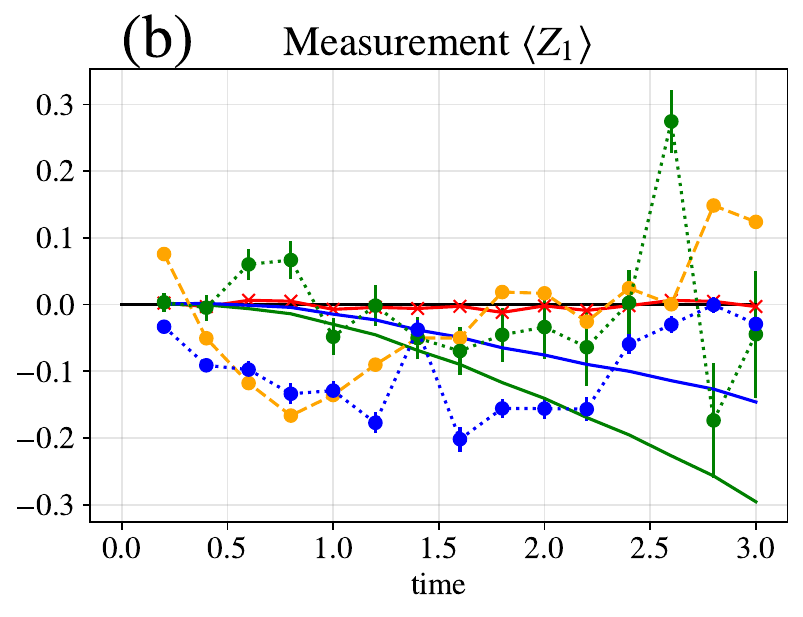} \\
\includegraphics[width=0.5\linewidth]{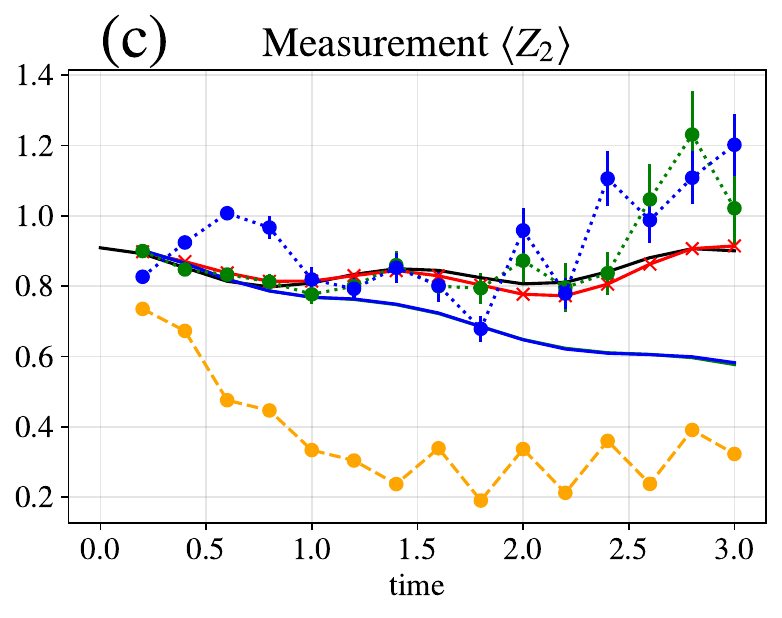}& \includegraphics[width=0.5\linewidth]{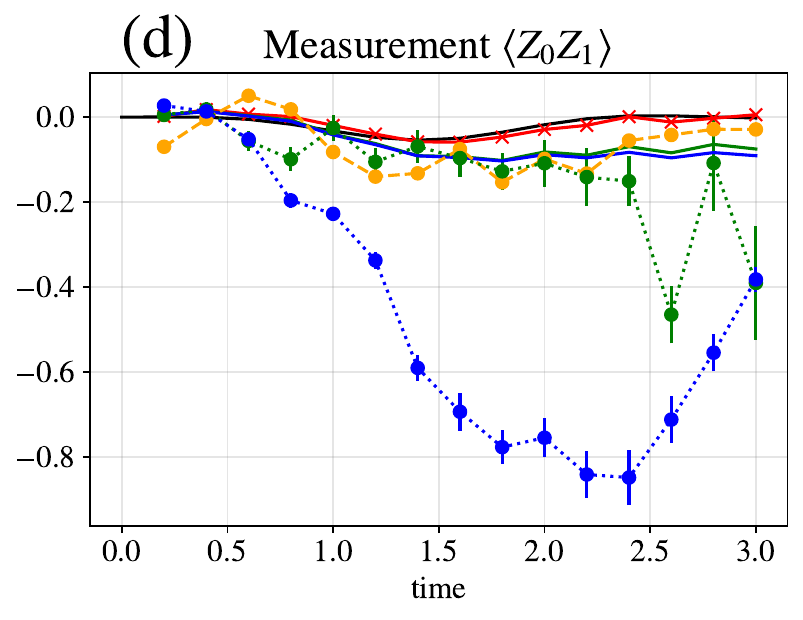}\\
\includegraphics[width=0.52\linewidth]{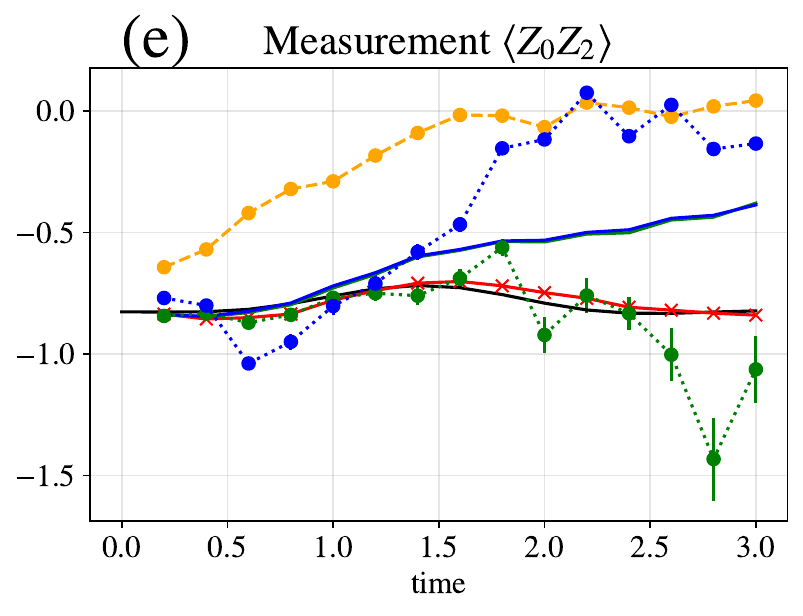}& \includegraphics[width=0.5\linewidth]{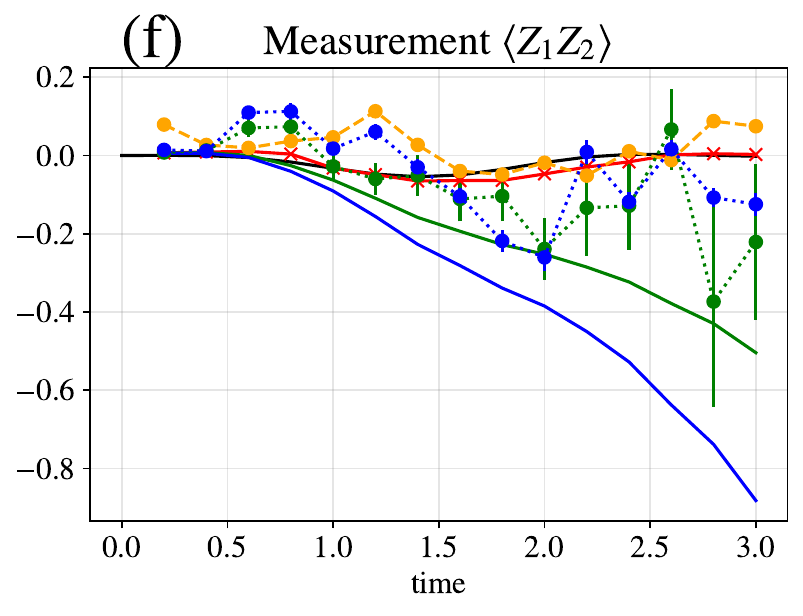}\\
      \multicolumn{2}{c}{\includegraphics[width=0.81\linewidth,trim={0 0.35cm 0 0},clip]{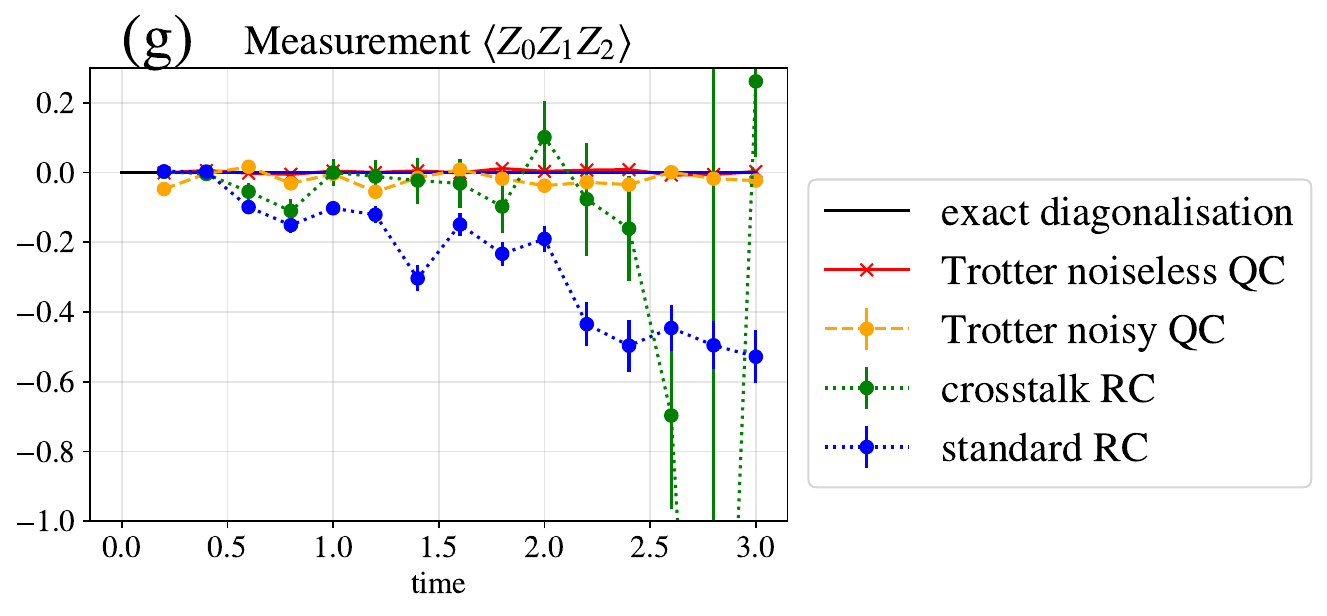}  }
\end{tabular}
    
    \caption{Evolution of the $7$ different observables derived from the measurement of \texttt{ibm\_lagos}' qubits $q_0$, $q_1$ and $q_2$ in the direction z after application of mitigation protocols (NEC + RC + REC). Blue (resp. green) curves correspond to standard (resp. crosstalk) RC. Dashed curves are the result obtained on a QC and the solid line are the corresponding fit made with the noise model of Eq.~\eqref{eq:3qubitchannel}. Exact evolution (black lines), noiseless (resp. noisy) Trotter dynamics (red lines) (resp. dashed orange curves) are also plotted for each quantity. Error bars correspond to the uncertainty due to shot noise and the finite sampling over twirl configurations.}
    \label{fig:evolution appendix ZZZ}
\end{figure}

In this appendix, we present the entirety of the results of our runs on the quantum computer, without RC, with standard RC and with crosstalk RC. This data was used to obtain the average absolute error plotted in Fig.~\ref{fig:intro}, and only a portion of it is presented in the main text. In Figs.~\ref{fig:evolution appendix XYZ} and \ref{fig:evolution appendix XYZ ehningen} we show the full time evolution for a first set of seven observables, executed respectively on \texttt{ibm\_lagos} and \texttt{ibmq\_ehningen}. In Fig.~\ref{fig:evolution appendix ZZZ}, we show the same evolution for a different set of seven observables, also obtained from \texttt{ibm\_lagos}. RC averages for both the Trotter circuits and the NEC have been performed over 300 samples, with the exception of the crosstalk RC curves at times $t \geq 1.8$ where we averaged over 600 NEC to check that the deviations do not come from a lack of
RC circuits. One can see that over different runs, the noise may change and the efficiency of the mitigated protocol varies.

\old{
\label{ap:evolution}
In this Appendix, we gather all the observables we have access in post-analysis: ($\langle X_0\rangle$, $\langle Y_1\rangle$, $\langle Z_2\rangle$, $\langle X_0Y_1\rangle$, $\langle X_0Z_2\rangle$, $\langle Y_1Z_2\rangle$ and $\langle X_0Y_1Z_2\rangle$) in Fig.~\ref{fig:evolution appendix XYZ}. We plot the final result after mitigation (green and blue dotted curves corresponding respectively to the crosstalk and standard RC protocol). We also plot the exact evolution of each observable (solid black lines) obtained after exact diagonalisation of the hamiltonian, the noiseless Trotter dynamics (solid red curves) and the one on a real QC (noisy) (dashed orange curves). Finally, one plots for comparison the mitigated result of the noise model Eq.~\eqref{eq:3qubitchannel} where the parameters are obtained from the fit of the Trotter circuits using crosstalk RC (green solid lines) and standard RC (blue solid lines). Expectation value of Trotter circuits have been averaged over $300$ RC circuits. It is also the case for noise estimation value, except for crosstalk RC at time $t\geq 1.8$ where we averaged over $600$ NEC to check that the deviation does not come from a lack of RC circuits. Green dotted curves of Fig.~\ref{fig:randomized_evolution} do not change quantitatively when compared to green dotted curves of Fig.~\ref{fig:evolution appendix XYZ}-a,c,e. It confirms that deviations should come from elsewhere (e.g Pauli noise instead of depolarising noise or single-qubit gate's errors).

\par We have also measured all qubits in direction z which give us access to seven other observables $\langle Z_0 \rangle$, $\langle Z_1 \rangle$, $\langle Z_2 \rangle$, $\langle Z_0 Z_1 \rangle$, $\langle Z_0Z_2\rangle$,  $\langle Z_1Z_2\rangle$ and  $\langle Z_0Z_1Z_2\rangle$. One observable is common to both measurements: $\langle Z_2 \rangle$. One can see that over different runs, the noise can change and the efficiency of the mitigated protocol varies depending if the approximation we made (depolarising noise instead of Pauli noise on active qubits or neglect single-qubit gate's error) is more or less verified. $300$ different RC circuits have been used for Trotter circuits as well as for noise estimation circuits at each time step. Results are displayed in Fig.~\ref{fig:evolution appendix ZZZ}. Fig.~\ref{fig:evolution appendix XYZ} and Fig.~\ref{fig:evolution appendix ZZZ} are simulations performed on \texttt{ibm\_lagos's} qubits $q_{0}$, $q_1$ and $q_2$.

\par Eventually, for comparison, we have also benchmarked the error mitigation protocols on another IBM device: \texttt{ibmq\_ehningen}. We used qubits $q_{18}$, $q_{21}$ and $q_{23}$ and measured them respectively in direction $x$, $y$ and $z$. RC for Trotter circuits and NEC has been performed on $300$ circuits.

\par In Fig.~\ref{fig:intro}-b, we define the relative error as $\epsilon_r=\left|\frac{n-p}{n}\right|$ where $p$ is the noiseless (perfect) result and $n$ the noisy one. We average it over all the data points available in this article (all gathered in Appendix~\ref{ap:evolution}) for both IBMQ devices. It is plotted for the Trotter bare results as well as both error mitigation protocols (standard and crosstalk RC +NEC +REC). This quantity captures the general trend of Fig.~\ref{fig:evolution appendix XYZ},~\ref{fig:evolution appendix ZZZ}, and~\ref{fig:evolution appendix XYZ ehningen} by showing that error mitigation protocols improve the accuracy of quantum simulations and crosstalk RC overcomes standard RC without any need of running additional circuits or qubits. This quantity has been built to strongly penalized noisy curves when they decay toward $0$ while the noiseless result is of order $1$ (which is the case for the Trotter bare results).}

\section{Noise channel after randomized compiling}
\label{ap:noise}
\par First, we review the detailed computation of a general n-qubit noise channel twirled over the set $\mathcal{T}_p^{\otimes n}=\{\mathbb{I}_2,\sigma_x,\sigma_y,\sigma_z\}^{\otimes n}$. Then, we argue that method can be applied to a 2-qubit noise channel induced by the application of a noisy CNOT gate. The main difference being that the first twirl can only be applied before the CNOT. Finally, we study the crosstalk RC protocol defined in Sec.~\ref{sec:twirl} and derived the n-qubit channel. We also show that when traced over the neighboring qubits this channel reduced to a 2-qubit Pauli noise \new{channel} and when traced over all the qubits except one neigbouring qubit of the CNOT it gives a 1-qubit depolarizing noise channel.

\subsection{Pauli twirling of a n-qubit noise channel}
\label{ap:nqubit noise}
\par Here, we follow the step of computation detailed in~\cite{Cai2019}. A general noise channel applied on a density matrix can be written as:
\begin{equation}
    \mathcal{E}^\text{gen.}_n(\rho)=\sum_M M.\rho. M^\dagger \quad\text{such that} \sum_M M.M^\dagger=\mathbb{I}
\end{equation}
$M$ is the so-called Kraus operator and we should stress that for a given noise channel this decomposition in terms of Kraus operators is not unique. Each Kraus operator can be decomposed over 
the Pauli basis:
\begin{equation}
    M=\frac{1}{2^n}\sum_{p\in P}\text{Tr}(M.p) p
\end{equation} where $P=\{\mathbb{I}_2,\sigma_x,\sigma_y,\sigma_z\}^{\otimes n}$. The general noise channel reads:
\begin{equation}
    \mathcal{E}^\text{gen.}_n(\rho)=\frac{1}{4^n}\sum_M \sum_{p_1,p_2\in P} \text{Tr}(M.p_1)  \text{Tr}(M.p_2)^* p_1.\rho .p_2
\end{equation} The twirling of the noise channel 
 over $\mathcal{T}_p^{\otimes n}=P$ consists \old{to apply}\new{of applying} Pauli matrices ($w$ in Eq.~\eqref{eq:twirl}) before and after the noise channel and average the result \new{over the twirling set}:
 \begin{eqnarray}
    \label{eq:twirl}
\mathcal{T}\left[\mathcal{E}^\text{gen.}_n\right](\rho) &=&\frac{1}{4^n}\sum_M \sum_{p_1,p_2\in P} \text{Tr}(M.p_1)  \text{Tr}(M .p_2)^*\nonumber\\
    &\times&\frac{1}{4^{n}} \sum_{w\in \mathcal{T}_p^{\otimes n }} w.p_1.w. \rho .w.p_2.w \\
    &=&\frac{1}{4^{2n}} \sum_M \sum_{p_1,p_2,w\in P}\text{Tr}(M.p_1)  \text{Tr}(M.p_2)^*\nonumber\\ 
    &\times&  p_1.\rho .p_2 \zeta(w,p_1) \zeta(w,p_2) \label{eq:twirl2}
\end{eqnarray} where $\zeta(w,p)$ is the sign of the commutation between $w$ and $p$: $+1$ if both commute, $-1$ if they anticommute. In particular, one has the relation:
\begin{equation}
 \zeta(w,p_1.p_2)=\zeta(w,p_1) \zeta(w,p_2) \label{eq:commute}
\end{equation}
 One inserts Eq.~\eqref{eq:commute} in Eq.~\eqref{eq:twirl2},  and splits the sums over $p_1$ and $p_2$ in two parts: the first one when $p_1=p_2$ and the second one  when $p_1\neq p_2$. The first sum evaluates $\zeta(w,p_1^2)=\zeta(w,\mathbb{I}_n)=+1$. For the second sum when $p_1\neq p_2$, we decompose the Pauli matrices over the different qubits $w=\bigotimes_{a=0}^{n-1} w^a$  and $p=p_1p_2=\bigotimes_{a=0}^{n-1} p^a$.  The sum over $w$ rewrites:
 \begin{equation}
     \sum_{w\in P} \zeta(w,p_1.p_2)=\prod_{a=0}^{n-1} \sum_{w^a \in \mathcal{T}_p}\zeta(w^a,p^a)
     \label{eq:commute2}
 \end{equation} When $p_1\neq p_2$, it exists at least one qubit $a$ such that $p^a\neq\mathbb{I}_2$. For that qubit, terms $w^a= \mathbb{I}_2$ and $w^a=p^a$ evaluate to $\zeta(w^a,p^a)=+1$ while the two other terms give $-1$. As a consequence, Eq.~\eqref{eq:commute2} vanishes. Only the diagonal term ($p_1=p_2$) is non zero in Eq.~\eqref{eq:twirl2} and gives:
 \begin{equation}
 \mathcal{T}\left[\mathcal{E}^\text{gen.}_n\right](\rho)= \frac{1}{4^{n}}  \sum_{p\in P} \left(\sum_M|\text{Tr}(Mp)|^2\right)  p.\rho .p 
 \label{eq:pauli twirl}
 \end{equation}
 \par Any noise channel is mapped, after twirling, to a Pauli noise \new{channel} with coefficients $\frac{1}{4^n}\sum_M|\text{Tr}(Mp)|^2$ for each Pauli matrix $p$ where the $M$'s are the Kraus operators of the original noise channel.

\subsection{Randomized compiling for CNOT gates}
\label{ap:random comp}
 \par In practice, a noise channel emerges from the application of a noisy gate. Consequently, RC cannot be directly applied to the noise channel; rather, it is performed on the entire noisy gate, modeled by its noiseless counterpart Eq.~\eqref{eq:twirl} rewrites:
 \begin{eqnarray}
 \label{eq:noisy gate twirl}
\mathcal{T}\left[\mathcal{E}_U\right](\rho) &=&\frac{1}{4^n}\sum_M \sum_{p_1,p_2\in P} \text{Tr}(M.p_1)  \text{Tr}(M .p_2)^*\nonumber\\
    &\times&\frac{1}{4^{n}} \sum_{w\in P} w.p_1.U.w'. \rho .w'U^\dagger.p_2.w
 \end{eqnarray}
 where $w'\in P $ and such that $w.U=U.w'$. In general, for any $U$ it is not \old{guarantee}\new{guaranteed} to find such $w'$. Following the main text, here the noisy gate $U_{\text{noisy}}$ will be the 2-qubit CNOT gate and for all $w$ it exists such $w'$. Let us denote $\text{CNOT}_{ij}$ the CNOT gate acting on qubit  $i$ and $j$ and $w=w^i\otimes w^j\bigotimes_{a\neq i,j} w^a$. For any $(w^i,w^j)\in \mathcal{T}_p^{\otimes 2}$ Table I of ~\cite{Urbanek2021} gives the corresponding $(w'^i,w'^j)$. For $a\neq i,j$, the action of the perfect CNOT is trivial and then $w'^a=w^a$. The rest of the computation follows the above derivation. 
 
 \par This twirling of noisy gates is called randomized compiling because, as it is explained in the main text, the space of twirling gates grows exponentially with the number of  CNOT gates in the circuit and one usually samples with a finite number of different twirl configuration this space by choosing randomly gates in the twirling set.
 
\subsection{Rotation twirling}
\label{ap:rotation twirl}
 \par In this section, we derive the noise channel obtained after Pauli twirling and apply the rotation twirling on neighboring qubits such as described in Sec.~\ref{sec:twirlcrosstalk} (see Fig.~\ref{fig:twirling}). First, we start by a simpler case, we consider a 1-qubit noise channel and apply this double RC procedure. The noise channel of Eq.~\eqref{eq:twirl} is modified as follows:
 \begin{eqnarray}
     \mathcal{T}^*\left[\mathcal{E}^{\text{gen.}}_{1}\right](\rho) =\frac{1}{16} \sum_M \sum_{p_1,p_2,w\in \mathcal{T}_p} \text{Tr}(M.p_1)  \text{Tr}(M .p_2)^*   \nonumber\\
    \times \frac{1}{3} \sum_{R\in\mathcal{T}_R} w.R^\dagger. p_1.R.w. \rho .w.R^\dagger.p_2.R.w \nonumber\\
    \label{eq:rotation 1 qubit}
 \end{eqnarray}
where $\mathcal{T}_R=\{R_x(\pi/2), R_y(\pi/2), R_z(\pi/2)\}$ as defined in Sec.~\ref{sec:twirlcrosstalk}. For each rotation $R$, the Pauli matrices $p_1$ and $p_2$ are mapped to the Pauli matrices $ p'_1= R^\dagger. p_1.R$ and $ p'_2= R^\dagger. p_2.R$  and we rewrite Eq.~\eqref{eq:rotation 1 qubit}:

 \begin{eqnarray}
\mathcal{T}^*\left[\mathcal{E}^{\text{gen.}}_{1}\right](\rho)=\frac{1}{48} \sum_{R\in\mathcal{T}_R}   \sum_M \sum_{p'_1,p'_2,w\in \mathcal{T}_p} \text{Tr}(M.R.p'_1.R^\dagger)  \nonumber\\
    \times \text{Tr}(M .R.p'_2.R^\dagger)^*   w. p'_1.w. \rho .w.p'_2.w \nonumber\\
    \label{eq:rotation 1 qubit 2}
 \end{eqnarray}
 \par The same derivation as in the above subsection can be done for $p'_1$ and $p'_2$ and therefore only the diagonal term $p'_1=p'_2$ does not vanish. We get:
  \begin{eqnarray}
\mathcal{T}^*\left[\mathcal{E}^{\text{gen.}}_{1}\right](\rho)=\frac{1}{12} \sum_{p\in P} \sum_M  \sum_{R}  |\text{Tr}(M.R.p.R^\dagger) |^2 p. \rho' p \nonumber\\
    \label{eq:rotation 1 qubit 3}
 \end{eqnarray}
 \par  The effect of the rotation matrices when the identity matrix is applied  ($p=\mathbb{I}_2$) does not change the matrix: 
 \begin{equation}
     R.p.R^\dagger=p=\mathbb{I}_2 \quad \forall R\in \mathcal{T}_R
     \label{eq:rotation identity}
 \end{equation} When $p=\sigma^x,\sigma^y$ or $\sigma^z$, rotation matrices $R\in \{R_x(\pi/2), R_y(\pi/2), R_z(\pi/2)\}$ equally mix the different directions:
 \begin{equation}
      R_k(\pi/2). \sigma^l. R_k(-\pi/2)=\sum_{m=x,y,z}\epsilon_{klm}\sigma^m+\delta_{kl}\sigma^l
      \label{eq:rotation pauli}
 \end{equation} where $k,l=x,y,z$ and $\epsilon_{klm}$ is the Levi-Civita tensor. Eq.~\eqref{eq:rotation 1 qubit 3} reduces to:
 \begin{eqnarray}
&\mathcal{T}^*&\left[\mathcal{E}^{\text{gen.}}_{1}\right](\rho)=\frac{1}{4}  \sum_M  |\text{Tr}(M) |^2  \rho \nonumber\\
&+&\frac{1}{3}  \left(\sum_{i=x,y,z}\frac{1}{4}\sum_M  |\text{Tr}(M.\sigma^i) |^2\right) \sum_{j=x,y,z}\sigma^j. \rho .\sigma^j\nonumber\\
\label{eq:rotation 1 qubit 4}
 \end{eqnarray}
\par Eq.~\eqref{eq:rotation 1 qubit 4} is a 1-qubit depolarising channel. Adding a rotation twirling set equally mixes the different coefficients of the Pauli noise \new{channel} to obtain a depolarising channel.\\

 Let us now consider the real case where a CNOT is applied and the rotation twirling is performed over the neighboring qubits. The twirled channel reads:
 \begin{eqnarray}
\mathcal{T}^*\left[\mathcal{E}_{\text{CNOT}}\right](\rho) =\frac{1}{4^{2n}} \sum_M \sum_{p_1,p_2,w\in P} \text{Tr}(M.p_1)  \text{Tr}(M .p_2)^*   \nonumber\\
    \times \frac{1}{3^{n-2}} \sum_{R\in\mathcal{T}_R^{\otimes n-2}} w.R^\dagger. p_1.R.w. \rho' .w.R^\dagger.p_2.R.w \nonumber\\
    \label{eq:rotation twirl}
 \end{eqnarray}
 where $\rho'=\text{CNOT}.\rho.\text{CNOT}$. Here $n$ denotes the number of qubits on which the noise channel applies. When considering crosstalk, $n$ is equal to 2 (the qubits on which the CNOT is applied) $+$ the number of direct neighbors.  The same trick can be applied as in Eq.~\eqref{eq:rotation 1 qubit 2} and after a similar derivation we end up with:
  \begin{eqnarray}
\mathcal{T}^*\left[\mathcal{E}_{\text{CNOT}}\right](\rho) =\frac{1}{4^{n}} \sum_{p\in P} \sum_M   \frac{1}{3^{n-2}} \sum_{R}  |\text{Tr}(M.R.p.R^\dagger) |^2 p. \rho' p \nonumber\\
    \label{eq:rotation twirl 2}
 \end{eqnarray}
 \par The probability to apply to the density matrix the Pauli gate $p$ is $\frac{1}{4^{n}} \sum_M   \frac{1}{3^{n-2}} \sum_{R}  |\text{Tr}(M.R.p.R^\dagger) |^2$. 
 \par Let us consider for the rest of the derivation that 1 and 2 denotes the index of the control and target qubits of the CNOT gate, hence, there is no rotation matrix applied on these qubits. We decompose the whole Pauli matrix and rotation matrix on each individual qubits: $P=\bigotimes_{a=0}^{n-1}p^a $ and $R=\bigotimes_{a\neq 1,2}r^a $. Using Eqs.~\eqref{eq:rotation identity} and ~\eqref{eq:rotation pauli}, the coefficients in Eq.~\eqref{eq:rotation twirl 2} only differs depending when $p^b=\mathbb{I}_2$ or $p^b\neq\mathbb{I}_2$ where $b$ is the index of a neighboring qubit. Therefore, the number of free parameter decreases thanks to this additional twirling. It goes from $4^n-1$ for a Pauli noise \new{channel} on $n$ qubits to $16\times 2^{n-2}-1=2^{n+2}-1$.
 \par One can also trace over all the qubits except one neighboring qubit $b$. The probability to apply the gate $p^b$ is:
 \begin{equation}
     \frac{1}{4^n\times 3}\prod_{a\neq b}\sum_{p^a\in\mathcal{T}_p}\sum_M\sum_{r^b\in\mathcal{T}_R}\left|\text{Tr}\left(M.\bigotimes_{c\neq b}p^c\otimes r^b.p^b.(r^b)^\dagger\right)\right|^2
 \end{equation}
 It reduces, when $p^b=\mathbb{I}_2$, to:
 \begin{equation}
     1-q_b=\frac{1}{4^n}\prod_{a\neq b}\sum_{p^a\in\mathcal{T}_p}\sum_M\left|\text{Tr}\left(M.\bigotimes_{c\neq b}p^c\otimes \mathbb{I}_2\right)\right|^2
     \label{eq:dep_noise3}
 \end{equation}
 When $p^b=\sigma^x,\sigma^y$ or $\sigma^z$, the probabilities  are the same:
 \begin{equation}
      \frac{q_b}{3}=\frac{1}{4^n\times 3}\prod_{a\neq b}\sum_{p^a\in\mathcal{T}_p}\sum_M\sum_{p^b\in\{\sigma^x,\sigma^y, \sigma^z\}}\left|\text{Tr}\left(M.\bigotimes_{c=0}^{n-1}p^c\right)\right|^2
      \label{eq:dep_noise4}
 \end{equation} where $q_b$ relates to the depolarizing parameter $\lambda_b=\frac{4 q_b}{3}$.
 \par Therefore, it mimics a depolarizing channel when looking \new{at} only one neighboring qubit. When tracing over all the neighboring qubits and looking at the noise channel on qubits 1 and 2, one has still a Pauli noise \new{channel} where the probability to apply the gate $p^1\otimes p^2$ is:
 \begin{equation}
     \frac{1}{4^n}\prod_{a\neq 1,2}\sum_{p^a\in\mathcal{T}_p}\sum_M\left|\text{Tr}\left(M.\bigotimes_{c=0}^{n-1}p^c\right)\right|^2
     \label{eq:pauli_noise2}
 \end{equation}

\par \newest{The general form of the noise channel obtained after the double RC protocol is:
\begin{align}
    &\mathcal{E}_n^{\text{crosstalk RC}}(\rho)=\sum_{i,j=0}^3\sum_{\vec{s}}\frac{p_{i,j,\vec{s}}}{3}\sum_{\vec{a}}P^{i,j,\vec{s}.\vec{a}}.\rho.P^{i,j,\vec{s},\vec{a}},\\
\label{eq:crosstalkRC noise}
&P^{i,j,\vec{s},\vec{a}}=\sigma_1^i\otimes\sigma_2^j\bigotimes_{k=3}^n\left(\sigma_k^{a_k}\right)^{s_k} ,\\
    & \textrm{with } \vec{s}\in\{0,1\}^{n-2} , \vec{a}\in\{1,2,3\}^{n-2}    
\end{align}
where $\vec{s}=(s_3,\cdots,s_n)$ is a binary vector indicating if an error occurred ($s_k=1$) on the neighboring qubit of index $k$ and $\vec{a}$ indicates if it was an X, Y or Z error ($a_k=1,2$ or $3$).}
\section{Statistical fluctuations due to finite sampling}
 \label{ap:uncertainty}
\par In this Appendix, we estimate the uncertainty due to the shot noise and the finite sampling of the twirl configurations.
The quantity we are interested to estimate is 
\begin{equation}
    O^\text{exp.}=\frac{1}{N_t}\sum_{r=1}^{N_t} O_r^{\text{exp}}
\end{equation}
where $N_t$ is the number of twirl configurations (in the main text $N_t=300$, in the Appendix~\ref{ap:evolution} for the NEC of the crosstalk RC and $t\geq 1.8$, $N_t=600$). $O_{r}^\text{exp.}$ is the experimental estimation of the expectation value of an observable $O$ for the twirled configuration circuit $C_r$. It is estimated by running $N_s$ times the circuit $C_r$:
\begin{equation}
    O_r^\text{exp}=\frac{1}{N_s}\sum_{m=1}^{N_s}O_{rm}^\text{exp.}
\end{equation}
where $N_s$ is the number of shots ($N_s=32000$ in the main text).
$O_{rm}^\text{exp.}$ is a bimodal random variable which is equal to $+1$ with probability $p(r)$ and $-1$ with probability $1- p(r)$.
The total expression is:
\begin{equation}
   O^\text{exp.}=\frac{1}{N_tN_s}\sum_{r=1}^{N_t} \sum_{m=1}^{N_s} O_{rm}^{\text{exp}}
\end{equation}
\par There are two different averages: the first one to estimate the expectation value of an observable $O_r$ corresponding to the circuit $C_r$ denoted as $\langle\cdots\rangle$:
\begin{eqnarray}
\langle O_r^\text{exp}\rangle&=&\frac{1}{N_s}\sum_{m=1}^{N_s}\langle O_{rm}^\text{exp}\rangle=2p(r)-1\\
&=&\underset{N_s\to\infty}{\text{lim}}O_r^\text{exp}
\end{eqnarray}
\par The second average is over the different twirl configurations and denoted $\overline{\cdots}$.  The probability that the expectation value of an observable corresponding to a certain circuit $C_r$ outputs $O_r^\text{exp}=o\in[-1,1]$ is given by the unknown probability distribution $f(o)$:
\begin{eqnarray}
    \overline{O^\text{exp.}}&=&\frac{1}{N_t}\sum_{r=1}^{N_t}\overline{ O_{r}^\text{exp}}=\int_{-1}^1o\times f(o)\mathrm{d}o\\
    &=&\underset{N_t\to\infty}{\text{lim}}O^\text{exp.}
\end{eqnarray}

\par To estimate the deviation from both averages due to the finite sampling it is useful to compute:
\begin{eqnarray}
\langle \left(O_r^\text{exp}\right)^2\rangle  - \langle O_r^\text{exp}\rangle^2 &=&\frac{1}{N_s}\text{Var}_{\langle\cdots\rangle}(O_r^\text{exp})\\
&=&\frac{1-\langle O_r^\text{exp}\rangle^2}{N_s}\\
\overline{\langle O^\text{exp}\rangle^2}  - \overline{\langle O^\text{exp}\rangle}^2 &=&\frac{1}{N_t}\text{Var}_{\overline{\cdots}}(\langle O^\text{exp}\rangle)
\end{eqnarray}
where $\text{Var}_{\langle\cdots\rangle}$ denotes the variance corresponding to the average over the different shots and $\text{Var}_{\overline{\cdots}}$ denotes the variance corresponding to the average over the different twirled configurations.
The total deviation is the sum of both quantities where we average the deviation due to shot noise over the different twirl configurations:
\begin{eqnarray}
\hspace{-1cm}
\overline{\langle \left(O^\text{exp.}\right)^2\rangle}-\overline{\langle O^\text{exp.}\rangle}^2&=&\overline{\langle\left(O^\text{exp.}\right)^2\rangle-\langle O^\text{exp.}\rangle^2}\nonumber\\
&+&\overline{\langle O^\text{exp.}\rangle^2}-\overline{\langle O^\text{exp.}\rangle}^2
\label{eq:samp_noise}\\
&=& \frac{1}{N_sN_t}\overline{\text{Var}_{\langle\cdots\rangle}(O^\text{exp})}\nonumber\\
&+&\frac{1}{N_t}\text{Var}_{\overline{\cdots}}(\langle O^\text{exp}\rangle )\\
&\simeq& \frac{1}{N_sN_t^2}\sum_{r=1}^{N_t}1-\langle O_r^\text{exp}\rangle^2\nonumber\\
&+&\frac{1}{N_t}\text{Var}_{\overline{\cdots}}(\langle O^\text{exp} \rangle)
\label{eq:total_unc}
\end{eqnarray} To obtain the last equation, we estimate the average over the twirl configurations empirically ($\overline{\cdots}\simeq\frac{1}{N_t}\sum_{r=1}^{N_t}\cdots$. $\langle O_r^\text{exp}\rangle\simeq O_r^\text{exp}$ and $\text{Var}_{\overline{\cdots}}$ are also estimated empirically from the data. Error bars in the plots correspond to the square root of Eq.~\eqref{eq:total_unc}. For the mitigated results, we combine this result with Eq.~\eqref{eq:uncertainty} to derive its uncertainty.
 \section{Mitigation for global and local depolarising channel}
 \label{ap:mitigation}
 \par In this appendix, we recall the result obtained for the noise estimation circuit in~\cite{Urbanek2021} in the case of a global and local depolarising noise channel. First, to understand how the noise estimation circuit protocol works, it is interesting to express the noisy expectation value in terms of the noiseless one.
 \subsection{Relation between noisy and noiseless outcome}
 \par In the simple case of a global depolarising channel, this can be done exactly. For local depolarising channel, the relation can be expanded in terms of number of errors occuring in the circuit and one keeps only the first order term. After considering these two noisy channels, the result generalizes to the case of Eq.~\eqref{eq:3qubitchannel} and yields Eq.~\eqref{eq:noise obs real} which includes these two cases plus a noise on the neighboring qubit.
\par A global depolarising channel on each CNOT reads:
\begin{equation}
\mathcal{E}_n^{\text{dep.}}(\rho)=(1-\lambda_\text{glob.})\rho+\lambda_\text{glob.}\frac{\mathbb{I}_{2^n}}{2^n}
\label{eq:global depo}
\end{equation}
where $n$ is the number of qubits in the circuit and $\smash{0\leq\lambda\leq\frac{4^n}{4^n-1}}$. If an error occurs in the circuit, it outputs the maximally mixed state. The noisy result of a traceless observable is therefore expressed as~\cite{Urbanek2021}:
\begin{equation}
    \langle O \rangle_{\text{noisy}}^\text{glob.}=(1-\lambda_\text{glob.})^{n_\text{CNOT}}\langle O \rangle_{\text{noiseless}}
    \label{eq:noise obs global}
\end{equation}
 where $n_{\text{CNOT}}$ is the number of CNOT gates in the circuit. 
 \par In the case of a depolarising channel acting only on the qubits $i,j$ where the CNOT is applied (called local depolarising channel), the error channel is written as:
 \begin{equation}
\mathcal{E}_2^{\text{dep.}}(\rho)=(1-\lambda_\text{CNOT})\rho+\lambda_\text{CNOT}\frac{\mathbb{I}_4}{4}\otimes \text{Tr}_{ij}(\rho)
\label{eq:local depo}
\end{equation}
where $\text{Tr}_{ij}$ is the partial trace over the active qubits $i,j$. An exact expression between the noisy result and the noiseless cannot be derived but one can expand it in terms of the number of errors:
\begin{eqnarray}
    \langle O \rangle_{\text{noisy}}^\text{local}&=&\sum_{i=0}^{\infty}(1-\lambda_\text{CNOT})^{n_\text{CNOT}-i}\lambda_\text{CNOT}^i \langle O\rangle_i\nonumber\\
    &=&(1-\lambda_\text{CNOT})^{n_\text{CNOT}}\sum_{i=0}^{\infty}\left(\frac{\lambda_\text{CNOT}}{1-\lambda_\text{CNOT}}\right)^{i}\langle O\rangle_i\nonumber\\
    &\simeq& \langle O \rangle_{0}\left[1+\lambda_\text{CNOT}\left(\frac{\langle O\rangle _1}{\langle O \rangle_{0}}-n_\text{CNOT}\right)\right]\nonumber\\&&+\mathcal{O}(\lambda_\text{CNOT}^2)
    \label{eq:noise obs local}
\end{eqnarray}
where $\langle O \rangle_i$ is the outcome such that $i$ errors occurred in the circuit, it is a sum of expectation values of all circuit where $i$ CNOT gates output the maximally mixed state on the active qubits. In the last line we expand the expression up to one error occuring in the circuit.
\subsection{Noisy expectation value of the noise estimation circuit}
\par In the case of a global depolarising noise, applying Eq.~\eqref{eq:noise obs global} to the NEC leads to:
\begin{equation}
\langle E \rangle_{\text{noisy}}^\text{glob.}=(1-\lambda_\text{glob.})^{n_{\text{CNOT}}}
    \label{eq:noise estimation global}
\end{equation} where $\langle E \rangle_{\text{noisy}}^\text{glob.}$ is the outcome of the NEC (the noiseless outcome being $1$). In this simple case, the NEC exactly reproduces the noise of the Trotter circuit. Dividing the Trotter circuits by the NEC yields the noiseless result:
\begin{equation}
    \frac{\langle O\rangle_{\text{noisy}}^\text{glob.}}{\langle E \rangle_{\text{noisy}}^\text{glob.}}=\langle O\rangle_{\text{noiseless}}^\text{glob.}
    \label{eq:noise mitigated global}
\end{equation}

\par For the local noise, we apply Eq.\eqref{eq:noise obs local} in the case of NEC and divide both results, which yields:
\begin{eqnarray}
\frac{\langle O \rangle_{\text{noisy}}^\text{local}}{\langle E \rangle_{\text{noisy}}^\text{local}}&\simeq&\langle O \rangle_{0}\left[1+\lambda_\text{CNOT}\left(\frac{\langle O\rangle _1}{\langle O \rangle_{0}}-\langle E\rangle_1\right)\right]\nonumber\\
&+&\mathcal{O}(\lambda_\text{CNOT}^2)
    \label{eq:noise mitigated local}
\end{eqnarray}
We will now show in the next appendix that the term in front of $\lambda_\text{CNOT}$ is small.
 
\section{First order Trotter expansion of an observable under a local depolarising noise}
\label{ap:Trotter noise}
\par In this Appendix, we show that assuming a local depolarising noise on each CNOT gate, the noise of the NEC reproduces part of the noise of the Trotter circuit up to the first Trotter order.
\par
\begin{figure*}
    \centering
 \includegraphics[width=\linewidth]{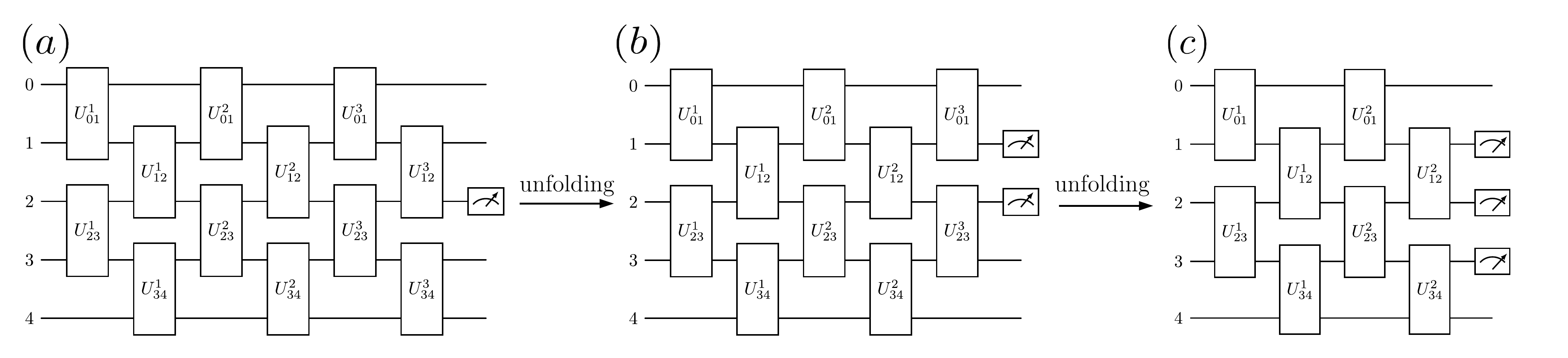}  
\caption{Schematic representation of the expansion performed. We start from a general circuit (a) and expand the different terms with respect to the last layer of gate and keep terms only up to first order in the Trotter time step$\Delta t$. (a): Example of a general circuit with a linear connection between qubits. We have 5 qubits and a 6 layers depth. (b): Circuit representation of the terms $\text{Tr}(\rho_1 \sigma_2^d)$, $\text{Tr}(\rho_1.A^{213}_2)$ and $\text{Tr}(\rho_1 B^{213}_2)$ (see Eq.~\eqref{eq:z'2}) as a circuit. For the two first terms, measurement is performed solely on qubit 2 while the last term measures both qubits 1 and 2. (c): Circuit representation of the terms in Eq.~\eqref{eq:z'24}. Measurement is performed only on qubit 2 for the three first term of this equation. The fourth term measures qubits 1 and 2 and the last term qubits 2 and 3.}
    \label{fig:Trotter_expansion}
\end{figure*}

 Every linear quantum circuits can be written as a sequence of a 2-body unitary gates (see Fig.~\ref{fig:Trotter_expansion}-a for an example).

\par We start by considering only one qubit measured 
 (qubit $m$) along the direction $d=x,y,z$ (in Fig.~\ref{fig:Trotter_expansion}-a $m=2$). The density matrix just before the measurement is denoted as $\rho_0$. Thus, one computes the quantity $\text{Tr}(\rho_0 \sigma^d_m)$. Each $U_{ij}^k$ unitary gates contains an arbitrary number of CNOT gates $r_{ij}^k$. Here $i$ and $j$ are the qubit indices on which the unitary gate is applied and k corresponds to the $2k-1$ layer of gates if $i$ is odd and $2k$ layer if $i$ is even. We suppose a local depolarizing noise after each CNOT gates with parameter $\varepsilon$:
$$\Lambda_{ij}(\rho)=(1-\varepsilon)\rho+\epsilon\frac{\mathbb{I}_{ij}}{4}\otimes\text{Tr}_{ij}(\rho)$$ where $\text{Tr}_{ij}$ denotes the partial trace over the qubits $i$ and $j$. No matter on which CNOT gates the noise occurs inside a $U_{ij}^k$ unitary gate the noisy outcome will be $\frac{\mathbb{I}_{ij}}{4}\otimes\text{Tr}_{ij}(\rho)$. Then the noisy channel after a $U_{ij}^k$ unitary gate is:
$$\text{E}_{ij}^k(\rho)=(1-\varepsilon)^{r_{ij}^k}\rho+(1-(1-\epsilon)^{r_{ij}^k})\frac{\mathbb{I}_{ij}}{4}\otimes\text{Tr}_{ij}(\rho)$$.
\par Suppose we have noise on the last layer of unitary gates (in the above example on $U_{34}^3$ and $U_{12}^3$). If the noise is on a unitary gate which is not on the measured qubit, this gate is not in the lightcone of the measured qubit and therefore its noise cannot   affect the result. 
If the noise is on a unitary gate which directly acts on the measured qubit, the resulting output vanishes because the Pauli matrix $\sigma_m^d$ is traceless $\text{Tr}(\left(\frac{\mathbb{I}_{ij}}{4}\otimes\text{Tr}_{ij}(\rho)\right).\sigma_m^d)= 0$ with $i=m$ or $j=m$. One can then write the noisy result of the circuit with respect to the noiseless one. For the example shown in Fig.~\ref{fig:Trotter_expansion}-a, it gives:
\begin{equation}
    \text{Tr}(\rho_0 .\sigma_2^d)=(1-\varepsilon)^{r_{12}^3} \text{Tr}(\rho_0^p .\sigma_2^d)
    \label{eq:1}
\end{equation}
where $\rho_0^p$ denotes the ``perfect'' result without noise i.e the unitary gates of the last layer have indeed been applied perfectly, $\rho^p_0=U_{12}^3U_{34}^3.\rho_1. (U_{34}^3)^\dagger(U_{12}^3)^\dagger$.
\par Let us denote $\rho_1$ as the density matrix just before the application of the last layer of unitary gates. Every unitary gates which do not act on the measured qubit $m=2$ commute with the Pauli matrix $\sigma_2^d$ then
\begin{equation}
\text{Tr}(\rho_0 .\sigma_2^d)=(1-\varepsilon)^{r_{12}^3} \text{Tr}(\rho_1.(U_{12}^3)^\dagger. \sigma_2^d. U_{12}^3)
\label{eq:z'20}
\end{equation}
\par In the following, we will make the assumption that every unitary gates are parametrized with a small time step $\Delta t$ as in the Trotter algorithm such that: 
\begin{equation}
U_{ij}^k=e^{-i \Delta t\sum_{a,b=0}^3\alpha_{ab}^{ijk}\sigma_i^a\otimes \sigma_j^b}
\end{equation} where  $\sigma_i^0$ denotes the identity matrix  on qubit $i$ and $\sigma_i^a$ are the the Pauli matrices where $a=1,2,3$ are identified respectively to the directions x,y,z.
\par Computation at first order in $\Delta t$ the quantity $(U_{12}^3)^\dagger.\sigma_2^d. U_{12}^3$ yields three terms:
\begin{eqnarray}(U_{12}^3)^\dagger.\sigma_2^d. U_{12}^3
    &\simeq&\sigma_2^d-2\Delta t (\mathbb{I}_1\otimes A^{213}_2+B^{213}_2)
    \label{eq:z'2}
\end{eqnarray} where $A^{213}_2=\sum_{b,c=1}^3\alpha_{0b}^{123}\epsilon_{bdc}\sigma_2^c$ and $B^{213}_2=\sum_{a,b,c=1}^3 \alpha_{ab}^{123} \epsilon_{bdc}\sigma_1^{a}\otimes\sigma_2^{b}$ and $\epsilon_{bdc}$ is the Levi-civita tensor. Substituting Eq.~\eqref{eq:z'2} in Eq.~\eqref{eq:z'20} one obtains:
\begin{eqnarray}
\text{Tr}(\rho_0.\sigma^d_2)
&\simeq&(1-\varepsilon)^{r_{12}^3}\left[\text{Tr}(\rho_1.\sigma_2^d)-2\Delta t\left(\text{Tr}(\rho_1.A^{123}_2)\right.\right.\nonumber\\
&&\left.\left.+\text{Tr}(\rho^p_1.B^{123}_2)\right)\right]
\label{eq:z'21}
\end{eqnarray}

The two first terms on the right hand side  of Eq.\eqref{eq:z'21} i.e. $\text{Tr}(\rho_1.\sigma_2^d)$ and $\text{Tr}(\rho_1.A^{213}_2)$ correspond to the circuit when only the qubit 2 is measured. The last term of Eq.~\eqref{eq:z'21} ($\text{Tr}(\rho^p_1.B^{213}_2)$) corresponds to the case where  qubits 1 and 2 are measured. A graphical representation of these terms as a quantum circuit is given in Fig.~\ref{fig:Trotter_expansion}-b.\\




\par Then, such as in Eq.~\eqref{eq:1}, when considering the last layer of gates  only the noise of the gates which acts directly on the qubits measured will matter. This leads us to:
\begin{eqnarray}
\text{Tr}(\rho_0.\sigma^d_2)
&\simeq&(1-\varepsilon)^{r_{12}^3+r_{23}^3}\left[\text{Tr}(\rho^p_1.\sigma_2^d)-2\Delta t\left(\text{Tr}(\rho^p_1.A^{213}_2)\right.\right.\nonumber\\
&&\left.\left.+(1-\varepsilon)^{r_{01}^3}\text{Tr}(\rho^p_1.B^{213}_2)\right)\right]
\label{eq:z'22}
\end{eqnarray}
\par Therefore, one can factorize the noise coming from the gate $U^3_{23}$ because it acts on qubit $2$ which is a common measured qubit for all the terms of Eq.~\eqref{eq:z'21}. For the last term there is one extra noise term coming from the CNOT acting on qubit 1 (gate $U^3_{01}$).
\par One continue the expansion of the expression until we reach the first layer of gate.
For the next step we get $\rho^p_1=U_{01}^3U_{23}^3.\rho_2. (U_{23}^3)^\dagger(U_{01}^3)^\dagger$. One keeps only the first order in $\Delta t$. Therefore, the evolution of the term $A^{123}_2$ and $B^{123}_2$ remains trivial for the rest of the computation (i.e identity operator):
\begin{eqnarray}(U_{23}^3)^\dagger(U_{01}^3)^\dagger.A^{123}_2.U_{01}^3U_{23}^3\simeq A^{213}_2\label{eq:A123}\\
(U_{23}^3)^\dagger(U_{01}^3)^\dagger.B^{123}_2.U_{01}^3U_{23}^3\simeq B^{213}_2 \label{eq:B123}
\end{eqnarray}

\par Only the main term  proportional to $\Delta t^0$ can produce a non trivial term. For the next layers of gate it couples to the qubit $3$ thanks to the gate $U^3_{23}$ (see Fig.~\ref{fig:Trotter_expansion}-c). 
We obtain a similar equation as in Eq.~\eqref{eq:z'2}:

\begin{eqnarray}
(U_{23}^3)^\dagger(U_{01}^3)^\dagger.\sigma^d_2.U_{01}^3U_{23}^3=\sigma_2^d-2\Delta t(A^{233}_2+B^{233}_2)\nonumber\\
\label{eq:z'23}
\end{eqnarray}

\par Fig.~\ref{fig:Trotter_expansion}-c represents the different term of this expansion.

\par After application of the noise of the unitary $U_{12}^2$ and $U_{34}^2$ we obtain:
\begin{eqnarray}
\text{Tr}(\rho_0.\sigma_2^d)&\simeq&(1-\varepsilon)^{r_{12}^3+r_{23}^3+r_{12}^2}\left[\text{Tr}(\rho^p_2.\sigma_2^d)\right.\nonumber\\
&&\left.-2\Delta t\left(\text{Tr}(\rho^p_2.A_2^{213})+\text{Tr}(\rho^p_2.A_2^{233})\right.\right.\nonumber\\
&&\left.\left.+(1-\varepsilon)^{r_{01}^3}\text{Tr}(\rho^p_2.B_2^{213})+(1-\varepsilon)^{r_{34}^2}\text{Tr}(\rho^p_2.B_2^{233})\right)\right]\nonumber \\
\label{eq:z'24}
\end{eqnarray}
where the general definitions of $A_k^{mm\pm 1 j}$ and $B_k^{mm\pm 1j}$ are given Eqs.~\eqref{eq:Adef},~\eqref{eq:Bdef}.





One can then generalize the above formula for a circuit of 2L layers and measured qubit $m$:
\begin{widetext}

\begin{eqnarray}
\text{Tr}(\rho_0 \sigma_m^d)&=&\prod_{j=1}^L (1-\epsilon)^{r_{m-1m}^j} (1-\epsilon)^{r_{mm+1}^j}\left[\text{Tr}(\rho_\text{ini}.\sigma_m^d)-2\Delta t\sum_{j=1}^L \text{Tr}(\rho_\text{ini}.A^{mm-1j}_m)+\text{Tr}(\rho_\text{ini}.A^{mm+1j}_m)\right.\nonumber\\
&&\left.-2\Delta t\sum_{j=1}^L(1-\varepsilon)^{\sum_{k=1}^j r_{m-2 m-1}^k}\text{Tr}(\rho_\text{ini}.B^{mm-1j}_m)+(1-\varepsilon)^{\sum_{k=1}^j r_{m+1 m+2}^k}\text{Tr}(\rho_\text{ini}.B^{mm+1j}_m)\right]+\mathcal{O}(\Delta t^2)\nonumber \\
\label{eq:formula1}
\end{eqnarray} 

\end{widetext}
where $\rho_{ini}$ is the initial density matrix before the first layer of unitary gates and
\begin{equation}
    A^{mm\pm 1j}_k=\sum_{b,c=1}^3 \alpha_{0b}^{mm\pm 1j}\epsilon_{bdc}\sigma_k^c
    \label{eq:Adef}
\end{equation}
\begin{equation}
    B^{mm\pm1j}_k=\sum_{a,b,c=1}^3 \alpha_{ab}^{mm\pm1j}
\epsilon_{bdc} \sigma_{m\pm1}^{a}\otimes \sigma_k^{b}
\label{eq:Bdef}
\end{equation}

\par To implement interaction between two non neighboring qubits, one need to use SWAP gates which are succession of three CNOT gates. These SWAP gates need to be taken into account at order 0 because they do not depend on any small parameter. This modifies the above formula: with SWAP gates the measured qubit does not remain in the same physical qubit  we need to follow it during the course of the circuit. In the formula of Eq.~\eqref{eq:formula1}, the qubit measured $m$ becomes a function of the layers: $m(j)$ which indicate the location of the measured qubit at layer $j$.
\par Eq.~\eqref{eq:formula1} can be applied to estimate the noise measured by the noise estimation circuit. In this case, the general $U^{a}_{bc}$ gates reduce to the identity operator up to some SWAP operations and $\Delta t=0$. Therefore, these circuits estimate only
the noise term in factor of the whole expression $\prod_{j=1}^L (1-\epsilon)^{r_{m-1m}^j} (1-\epsilon)^{r_{mm+1}^j}$ i.e the noise coming from the CNOT acting on the measured qubit. Then, it shows that if one divides the noise of the Trotter circuit by the noise of the estimation circuit, one can remove all the noise of the main term and a part of the first order Trotter term.
\par The number of higher order terms in $\Delta t$ scale exponentially with the total time evolution and therefore nothing guarantee that these terms will remain small. Their noise will deviate strongly as time goes on from the noise estimation circuit (see Fig.~\ref{fig:mitigated evolution} for a simulation of the noise estimation method using the error channel of Eq.~\eqref{eq:3qubitchannel} (quasi-local depolarizing noise)). Therefore, the noise estimation circuit is a method which delays the effect of the noise and allows simulation to longer evolution time.
\par Eq.~\eqref{eq:formula1} can be generalized to multiple measured qubits.  For each term containing the index $m$ for the measured qubit, one should add a product running over the different measured qubits. The formula for the noise estimation circuit also gets this modification and is still valid in that case.
\par This result generalizes to an error channel of the form of Eq.~\eqref{eq:3qubitchannel}. As long as a depolarising channel is applied on qubits whenever an error occurs, the main term of the Trotter expansion will be reproduced by the estimation circuit.


\begin{thebibliography}{65}%
\makeatletter
\providecommand \@ifxundefined [1]{%
 \@ifx{#1\undefined}
}%
\providecommand \@ifnum [1]{%
 \ifnum #1\expandafter \@firstoftwo
 \else \expandafter \@secondoftwo
 \fi
}%
\providecommand \@ifx [1]{%
 \ifx #1\expandafter \@firstoftwo
 \else \expandafter \@secondoftwo
 \fi
}%
\providecommand \natexlab [1]{#1}%
\providecommand \enquote  [1]{``#1''}%
\providecommand \bibnamefont  [1]{#1}%
\providecommand \bibfnamefont [1]{#1}%
\providecommand \citenamefont [1]{#1}%
\providecommand \href@noop [0]{\@secondoftwo}%
\providecommand \href [0]{\begingroup \@sanitize@url \@href}%
\providecommand \@href[1]{\@@startlink{#1}\@@href}%
\providecommand \@@href[1]{\endgroup#1\@@endlink}%
\providecommand \@sanitize@url [0]{\catcode `\\12\catcode `\$12\catcode
  `\&12\catcode `\#12\catcode `\^12\catcode `\_12\catcode `\%12\relax}%
\providecommand \@@startlink[1]{}%
\providecommand \@@endlink[0]{}%
\providecommand \url  [0]{\begingroup\@sanitize@url \@url }%
\providecommand \@url [1]{\endgroup\@href {#1}{\urlprefix }}%
\providecommand \urlprefix  [0]{URL }%
\providecommand \Eprint [0]{\href }%
\providecommand \doibase [0]{http://dx.doi.org/}%
\providecommand \selectlanguage [0]{\@gobble}%
\providecommand \bibinfo  [0]{\@secondoftwo}%
\providecommand \bibfield  [0]{\@secondoftwo}%
\providecommand \translation [1]{[#1]}%
\providecommand \BibitemOpen [0]{}%
\providecommand \bibitemStop [0]{}%
\providecommand \bibitemNoStop [0]{.\EOS\space}%
\providecommand \EOS [0]{\spacefactor3000\relax}%
\providecommand \BibitemShut  [1]{\csname bibitem#1\endcsname}%
\let\auto@bib@innerbib\@empty

\bibitem [{\citenamefont {Preskill}(2012)}]{Preskill2012}%
  \BibitemOpen
  \bibfield  {author} {\bibinfo {author} {\bibfnamefont {J.}\ \bibnamefont
  {Preskill}},\ } {\enquote {\bibinfo
  {title} {Quantum computing and the entanglement frontier}}\ }
  \href {\doibase 10.48550/ARXIV.1203.5813}
  {\bibfield  {journal} {\bibinfo  {journal}
  {{arXiv}}}:\ \bibinfo {pages} {1203.5813}
  }\BibitemShut {NoStop}%
  \bibitem [{\citenamefont {Preskill}(2018)}]{Preskill2018}%
  \BibitemOpen
  \bibfield  {author} {\bibinfo {author} {\bibfnamefont {J.}\ \bibnamefont
  {Preskill}},\ }\bibfield  {title} {\enquote {\bibinfo {title} {Quantum
  {C}omputing in the {NISQ} era and beyond},}\ }\href {\doibase
  10.22331/q-2018-08-06-79} {\bibfield  {journal} {\bibinfo  {journal}
  {{Quantum}}\ }\textbf {\bibinfo {volume} {2}},\ \bibinfo {pages} {79}
  (\bibinfo {year} {2018})}\BibitemShut {NoStop}%
\bibitem [{\citenamefont {Knill}\ \emph {et~al.}(1998)\citenamefont {Knill},
  \citenamefont {Laflamme},\ and\ \citenamefont {Zurek}}]{Knill1998}%
  \BibitemOpen
  \bibfield  {author} {\bibinfo {author} {\bibfnamefont {E.}\ \bibnamefont
  {Knill}}, \bibinfo {author} {\bibfnamefont {R.}\ \bibnamefont
  {Laflamme}}, \ and\ \bibinfo {author} {\bibfnamefont {W.~H.}\
  \bibnamefont {Zurek}},\ }\bibfield  {title} {\enquote {\bibinfo {title}
  {Resilient quantum computation},}\ }\href {\doibase
  10.1126/science.279.5349.342} {\bibfield  {journal} {\bibinfo  {journal}
  {Science}\ }\textbf {\bibinfo {volume} {279}},\ \bibinfo {pages} {342--345}
  (\bibinfo {year} {1998})} \BibitemShut
  {NoStop}%
\bibitem [{\citenamefont {Lidar}\ and\ \citenamefont {Brun}(2013)}]{Lidar2013}%
  \BibitemOpen
  \bibfield  {author} {\bibinfo {author} {\bibfnamefont {D.A.}\ \bibnamefont
  {Lidar}}\ and\ \bibinfo {author} {\bibfnamefont {T.A.}\ \bibnamefont
  {Brun}},\ }\href {\doibase 10.1017/CBO9781139034807} {\bibinfo {title}
  {Quantum Error Correction}}\ (\bibinfo  {publisher} {Cambridge University
  Press},\ \bibinfo {year} {2013})\BibitemShut {NoStop}%
\bibitem [{\citenamefont {Nielsen}\ and\ \citenamefont
  {Chuang}(2012)}]{Nielsen2012}%
  \BibitemOpen
  \bibfield  {author} {\bibinfo {author} {\bibfnamefont {M.~A.}\
  \bibnamefont {Nielsen}}\ and\ \bibinfo {author} {\bibfnamefont {I.~L.}\
  \bibnamefont {Chuang}},\ }\href {\doibase 10.1017/cbo9780511976667} {\bibinfo {title} {Quantum Computation and Quantum Information}}\ (\bibinfo
  {publisher} {Cambridge University Press},\ \bibinfo {year}
  {2012})\BibitemShut {NoStop}%
\bibitem [{\citenamefont {Quantum}(2021)}]{IBMQ2021}%
  \BibitemOpen
  \bibfield  {author} {\bibinfo {author}  \href {https://quantum-computing.ibm.com/}{\bibfnamefont {IBM}\ \bibnamefont
  {Quantum}}}\BibitemShut {NoStop}%
\bibitem [{\citenamefont {Cai}\ \emph {et~al.}(2022)\citenamefont {Cai},
  \citenamefont {Babbush}, \citenamefont {Benjamin}, \citenamefont {Endo},
  \citenamefont {Huggins}, \citenamefont {Li}, \citenamefont {McClean},\ and\
  \citenamefont {O'Brien}}]{Cai2022}%
  \BibitemOpen
  \bibfield  {author} {\bibinfo {author} {\bibfnamefont {Z.}\ \bibnamefont
  {Cai}}, \bibinfo {author} {\bibfnamefont {R.}\ \bibnamefont {Babbush}},
  \bibinfo {author} {\bibfnamefont {S.~C.}\ \bibnamefont {Benjamin}},
  \bibinfo {author} {\bibfnamefont {S.}\ \bibnamefont {Endo}}, \bibinfo
  {author} {\bibfnamefont {W.~J.}\ \bibnamefont {Huggins}}, \bibinfo
  {author} {\bibfnamefont {Y.}\ \bibnamefont {Li}}, \bibinfo {author}
  {\bibfnamefont {J.~R.}\ \bibnamefont {McClean}}, \ and\ \bibinfo {author}
  {\bibfnamefont {T.~E.}\ \bibnamefont {O'Brien}},\ } {\enquote {\bibinfo {title} {Quantum error
  mitigation}}\ } \href {\doibase
  10.48550/ARXIV.2210.00921} {\bibfield  {journal} {\bibinfo  {journal} {arXiv:}}\ \bibinfo {pages} {2210.00921}}
  \BibitemShut {NoStop}%
\bibitem [{\citenamefont {Endo}\ \emph {et~al.}(2021)\citenamefont {Endo},
  \citenamefont {Cai}, \citenamefont {Benjamin},\ and\ \citenamefont
  {Yuan}}]{Endo2021}%
  \BibitemOpen
  \bibfield  {author} {\bibinfo {author} {\bibfnamefont {Suguru}\ \bibnamefont
  {Endo}}, \bibinfo {author} {\bibfnamefont {Z.}\ \bibnamefont {Cai}},
  \bibinfo {author} {\bibfnamefont {S.~C.}\ \bibnamefont {Benjamin}}, \ and\
  \bibinfo {author} {\bibfnamefont {X.}\ \bibnamefont {Yuan}},\ }\bibfield
  {title} {\enquote {\bibinfo {title} {Hybrid quantum-classical algorithms and
  quantum error mitigation}}\ }\href {\doibase 10.7566/JPSJ.90.032001}
  {\bibfield  {journal} {\bibinfo  {journal} {Journal of the Physical Society
  of Japan}\ }\textbf {\bibinfo {volume} {90}},\ \bibinfo {pages} {032001}
  (\bibinfo {year} {2021})}
 \BibitemShut {NoStop}%
\bibitem [{\citenamefont {Czarnik}\ \emph {et~al.}(2021)\citenamefont
  {Czarnik}, \citenamefont {Arrasmith}, \citenamefont {Coles},\ and\
  \citenamefont {Cincio}}]{Czarnik2021}%
  \BibitemOpen
  \bibfield  {author} {\bibinfo {author} {\bibfnamefont {P.}\ \bibnamefont
  {Czarnik}}, \bibinfo {author} {\bibfnamefont {A.}\ \bibnamefont
  {Arrasmith}}, \bibinfo {author} {\bibfnamefont {P.~J.}\ \bibnamefont
  {Coles}}, \ and\ \bibinfo {author} {\bibfnamefont {L.}\ \bibnamefont
  {Cincio}},\ }\bibfield  {title} {\enquote {\bibinfo {title} {Error mitigation
  with {C}lifford quantum-circuit data}}\ }\href {\doibase
  10.22331/q-2021-11-26-592} {\bibfield  {journal} {\bibinfo  {journal}
  {{Quantum}}\ }\textbf {\bibinfo {volume} {5}},\ \bibinfo {pages} {592}
  (\bibinfo {year} {2021})}\BibitemShut {NoStop}%
\bibitem [{\citenamefont {McArdle}\ \emph {et~al.}(2019)\citenamefont
  {McArdle}, \citenamefont {Yuan},\ and\ \citenamefont
  {Benjamin}}]{McArdle2019}%
  \BibitemOpen
  \bibfield  {author} {\bibinfo {author} {\bibfnamefont {S.}\ \bibnamefont
  {McArdle}}, \bibinfo {author} {\bibfnamefont {X.}\ \bibnamefont {Yuan}}, \
  and\ \bibinfo {author} {\bibfnamefont {S.~C.}\ \bibnamefont {Benjamin}},\
  }\bibfield  {title} {\enquote {\bibinfo {title} {Error-mitigated digital
  quantum simulation}}\ }\href {\doibase 10.1103/PhysRevLett.122.180501}
  {\bibfield  {journal} {\bibinfo  {journal} {Phys. Rev. Lett.}\ }\textbf
  {\bibinfo {volume} {122}},\ \bibinfo {pages} {180501} (\bibinfo {year}
  {2019})}\BibitemShut {NoStop}%
\bibitem [{\citenamefont {Temme}\ \emph {et~al.}(2017)\citenamefont {Temme},
  \citenamefont {Bravyi},\ and\ \citenamefont {Gambetta}}]{Temme2017}%
  \BibitemOpen
  \bibfield  {author} {\bibinfo {author} {\bibfnamefont {K.}\ \bibnamefont
  {Temme}}, \bibinfo {author} {\bibfnamefont {S.}\ \bibnamefont {Bravyi}},
  \ and\ \bibinfo {author} {\bibfnamefont {J.~M.}\ \bibnamefont {Gambetta}},\
  }\bibfield  {title} {\enquote {\bibinfo {title} {Error mitigation for
  short-depth quantum circuits}}\ }\href {\doibase
  10.1103/PhysRevLett.119.180509} {\bibfield  {journal} {\bibinfo  {journal}
  {Phys. Rev. Lett.}\ }\textbf {\bibinfo {volume} {119}},\ \bibinfo {pages}
  {180509} (\bibinfo {year} {2017})}\BibitemShut {NoStop}%
\bibitem [{\citenamefont {Li}\ and\ \citenamefont {Benjamin}(2017)}]{Li2017}%
  \BibitemOpen
  \bibfield  {author} {\bibinfo {author} {\bibfnamefont {Y.}\ \bibnamefont
  {Li}}\ and\ \bibinfo {author} {\bibfnamefont {S.~C.}\ \bibnamefont
  {Benjamin}},\ }\bibfield  {title} {\enquote {\bibinfo {title} {Efficient
  variational quantum simulator incorporating active error minimization}}\
  }\href {\doibase 10.1103/PhysRevX.7.021050} {\bibfield  {journal} {\bibinfo
  {journal} {Phys. Rev. X}\ }\textbf {\bibinfo {volume} {7}},\ \bibinfo {pages}
  {021050} (\bibinfo {year} {2017})}\BibitemShut {NoStop}%
\bibitem [{\citenamefont {Bharti}\ \emph {et~al.}(2022)\citenamefont {Bharti},
  \citenamefont {Cervera-Lierta}, \citenamefont {Kyaw}, \citenamefont {Haug},
  \citenamefont {Alperin-Lea}, \citenamefont {Anand}, \citenamefont {Degroote},
  \citenamefont {Heimonen}, \citenamefont {Kottmann}, \citenamefont {Menke},
  \citenamefont {Mok}, \citenamefont {Sim}, \citenamefont {Kwek},\ and\
  \citenamefont {Aspuru-Guzik}}]{Bharti2022}%
  \BibitemOpen
  \bibfield  {author} {\bibinfo {author} {\bibfnamefont {K.}\ \bibnamefont
  {Bharti}}, \bibinfo {author} {\bibfnamefont {A.}\ \bibnamefont
  {Cervera-Lierta}}, \bibinfo {author} {\bibfnamefont {T.~H.}\ \bibnamefont
  {Kyaw}}, \bibinfo {author} {\bibfnamefont {T.}\ \bibnamefont {Haug}},
  \bibinfo {author} {\bibfnamefont {S.}\ \bibnamefont {Alperin-Lea}},
  \bibinfo {author} {\bibfnamefont {A.}\ \bibnamefont {Anand}}, \bibinfo
  {author} {\bibfnamefont {M.}\ \bibnamefont {Degroote}}, \bibinfo
  {author} {\bibfnamefont {H.}\ \bibnamefont {Heimonen}}, \bibinfo
  {author} {\bibfnamefont {J.~S.}\ \bibnamefont {Kottmann}}, \bibinfo
  {author} {\bibfnamefont {T.}\ \bibnamefont {Menke}}, \bibinfo {author}
  {\bibfnamefont {W.-K.}\ \bibnamefont {Mok}}, \bibinfo {author}
  {\bibfnamefont {S.}\ \bibnamefont {Sim}}, \bibinfo {author} {\bibfnamefont
  {L.-C.}\ \bibnamefont {Kwek}}, \ and\ \bibinfo {author} {\bibfnamefont
  {A.}\ \bibnamefont {Aspuru-Guzik}},\ }\bibfield  {title} {\enquote
  {\bibinfo {title} {Noisy intermediate-scale quantum algorithms}}\ }\href
  {\doibase 10.1103/RevModPhys.94.015004} {\bibfield  {journal} {\bibinfo
  {journal} {Rev. Mod. Phys.}\ }\textbf {\bibinfo {volume} {94}},\ \bibinfo
  {pages} {015004} (\bibinfo {year} {2022})}\BibitemShut {NoStop}%
\bibitem [{\citenamefont {Leymann}\ and\ \citenamefont
  {Barzen}(2020)}]{Leymann2020}%
  \BibitemOpen
  \bibfield  {author} {\bibinfo {author} {\bibfnamefont {F.}\ \bibnamefont
  {Leymann}}\ and\ \bibinfo {author} {\bibfnamefont {J.}\ \bibnamefont
  {Barzen}},\ }\bibfield  {title} {\enquote {\bibinfo {title} {The bitter truth
  about gate-based quantum algorithms in the nisq era}}\ }\href {\doibase
  10.1088/2058-9565/abae7d} {\bibfield  {journal} {\bibinfo  {journal} {Quantum
  Science and Technology}\ }\textbf {\bibinfo {volume} {5}},\ \bibinfo {pages}
  {044007} (\bibinfo {year} {2020})}\BibitemShut {NoStop}%
\bibitem [{\citenamefont {Zhou}\ \emph {et~al.}(2020)\citenamefont {Zhou},
  \citenamefont {Stoudenmire},\ and\ \citenamefont {Waintal}}]{Zhou2020}%
  \BibitemOpen
  \bibfield  {author} {\bibinfo {author} {\bibfnamefont {Y.}\ \bibnamefont
  {Zhou}}, \bibinfo {author} {\bibfnamefont {E.~M.}\ \bibnamefont
  {Stoudenmire}}, \ and\ \bibinfo {author} {\bibfnamefont {X.}\
  \bibnamefont {Waintal}},\ }\bibfield  {title} {\enquote {\bibinfo {title}
  {{What Limits the Simulation of Quantum Computers?}}}\ }\href {\doibase
  10.1103/PhysRevX.10.041038} {\bibfield  {journal} {\bibinfo  {journal}
  {Physical Review X}\ }\textbf {\bibinfo {volume} {10}},\ \bibinfo {pages}
  {041038} (\bibinfo {year} {2020})},\ \Eprint
  {http://arxiv.org/abs/2002.07730} {arXiv:2002.07730} \BibitemShut {NoStop}%
\bibitem [{\citenamefont {Ayral}\ \emph {et~al.}(2022)\citenamefont {Ayral},
  \citenamefont {Louvet}, \citenamefont {Zhou}, \citenamefont {Lambert},
  \citenamefont {Stoudenmire},\ and\ \citenamefont {Waintal}}]{Ayral2022}%
  \BibitemOpen
  \bibfield  {author} {\bibinfo {author} {\bibfnamefont {T.}\ \bibnamefont
  {Ayral}}, \bibinfo {author} {\bibfnamefont {Thibaud}\ \bibnamefont {Louvet}},
  \bibinfo {author} {\bibfnamefont {Y.}\ \bibnamefont {Zhou}}, \bibinfo
  {author} {\bibfnamefont {C.}\ \bibnamefont {Lambert}}, \bibinfo {author}
  {\bibfnamefont {E.~Miles}\ \bibnamefont {Stoudenmire}}, \ and\ \bibinfo
  {author} {\bibfnamefont {X.}\ \bibnamefont {Waintal}},\ }\enquote {\bibinfo {title} {{A
  density-matrix renormalization group algorithm for simulating quantum
  circuits with a finite fidelity}}},\ 
  \href
  {http://arxiv.org/abs/2207.05612}
{arXiv:2207.05612} \BibitemShut {NoStop}%
\bibitem [{\citenamefont {Kandala}\ \emph {et~al.}(2021)\citenamefont
  {Kandala}, \citenamefont {Wei}, \citenamefont {Srinivasan}, \citenamefont
  {Magesan}, \citenamefont {Carnevale}, \citenamefont {Keefe}, \citenamefont
  {Klaus}, \citenamefont {Dial},\ and\ \citenamefont {McKay}}]{Kandala2021}%
  \BibitemOpen
  \bibfield  {author} {\bibinfo {author} {\bibfnamefont {A.}~\bibnamefont
  {Kandala}}, \bibinfo {author} {\bibfnamefont {K.~X.}\ \bibnamefont {Wei}},
  \bibinfo {author} {\bibfnamefont {S.}~\bibnamefont {Srinivasan}}, \bibinfo
  {author} {\bibfnamefont {E.}~\bibnamefont {Magesan}}, \bibinfo {author}
  {\bibfnamefont {S.}~\bibnamefont {Carnevale}}, \bibinfo {author}
  {\bibfnamefont {G.~A.}\ \bibnamefont {Keefe}}, \bibinfo {author}
  {\bibfnamefont {D.}~\bibnamefont {Klaus}}, \bibinfo {author} {\bibfnamefont
  {O.}~\bibnamefont {Dial}}, \ and\ \bibinfo {author} {\bibfnamefont {D.~C.}\
  \bibnamefont {McKay}},\ }\bibfield  {title} {\enquote {\bibinfo {title}
  {Demonstration of a high-fidelity cnot gate for fixed-frequency transmons
  with engineered $zz$ suppression}}\ }\href {\doibase
  10.1103/PhysRevLett.127.130501} {\bibfield  {journal} {\bibinfo  {journal}
  {Phys. Rev. Lett.}\ }\textbf {\bibinfo {volume} {127}},\ \bibinfo {pages}
  {130501} (\bibinfo {year} {2021})}\BibitemShut {NoStop}%
\bibitem [{\citenamefont {Magesan}\ \emph {et~al.}(2011)\citenamefont
  {Magesan}, \citenamefont {Gambetta},\ and\ \citenamefont
  {Emerson}}]{Magesan2011}%
  \BibitemOpen
  \bibfield  {author} {\bibinfo {author} {\bibfnamefont {E.}\ \bibnamefont
  {Magesan}}, \bibinfo {author} {\bibfnamefont {J.~M.}\ \bibnamefont
  {Gambetta}}, \ and\ \bibinfo {author} {\bibfnamefont {J.}\ \bibnamefont
  {Emerson}},\ }\bibfield  {title} {\enquote {\bibinfo {title} {Scalable and
  robust randomized benchmarking of quantum processes}}\ }\href {\doibase
  10.1103/PhysRevLett.106.180504} {\bibfield  {journal} {\bibinfo  {journal}
  {Phys. Rev. Lett.}\ }\textbf {\bibinfo {volume} {106}},\ \bibinfo {pages}
  {180504} (\bibinfo {year} {2011})}\BibitemShut {NoStop}%
\bibitem [{\citenamefont {Magesan}\ \emph {et~al.}(2012)\citenamefont
  {Magesan}, \citenamefont {Gambetta},\ and\ \citenamefont
  {Emerson}}]{Magesan2012}%
  \BibitemOpen
  \bibfield  {author} {\bibinfo {author} {\bibfnamefont {E.}\ \bibnamefont
  {Magesan}}, \bibinfo {author} {\bibfnamefont {J.~M.}\ \bibnamefont
  {Gambetta}}, \ and\ \bibinfo {author} {\bibfnamefont {J.}\ \bibnamefont
  {Emerson}},\ }\bibfield  {title} {\enquote {\bibinfo {title} {Characterizing
  quantum gates via randomized benchmarking}}\ }\href {\doibase
  10.1103/PhysRevA.85.042311} {\bibfield  {journal} {\bibinfo  {journal} {Phys.
  Rev. A}\ }\textbf {\bibinfo {volume} {85}},\ \bibinfo {pages} {042311}
  (\bibinfo {year} {2012})}\BibitemShut {NoStop}%
\bibitem [{\citenamefont {Georgopoulos}\ \emph
  {et~al.}(2021{\natexlab{a}})\citenamefont {Georgopoulos}, \citenamefont
  {Emary},\ and\ \citenamefont {Zuliani}}]{Georgopoulos2021}%
  \BibitemOpen
  \bibfield  {author} {\bibinfo {author} {\bibfnamefont {K.}\
  \bibnamefont {Georgopoulos}}, \bibinfo {author} {\bibfnamefont {C.}\
  \bibnamefont {Emary}}, \ and\ \bibinfo {author} {\bibfnamefont {P.}\
  \bibnamefont {Zuliani}},\ }\bibfield  {title} {\enquote {\bibinfo {title}
  {Modeling and simulating the noisy behavior of near-term quantum
  computers}}\ }\href {\doibase 10.1103/PhysRevA.104.062432} {\bibfield
  {journal} {\bibinfo  {journal} {Phys. Rev. A}\ }\textbf {\bibinfo {volume}
  {104}},\ \bibinfo {pages} {062432} (\bibinfo {year}
  {2021}{\natexlab{a}})}\BibitemShut {NoStop}%
\bibitem [{\citenamefont {Johnstun}\ and\ \citenamefont
  {Van~Huele}(2021)}]{Johnstun2021}%
  \BibitemOpen
  \bibfield  {author} {\bibinfo {author} {\bibfnamefont {S.}\ \bibnamefont
  {Johnstun}}\ and\ \bibinfo {author} {\bibfnamefont {J.-F.}\
  \bibnamefont {Van~Huele}},\ }\bibfield  {title} {\enquote {\bibinfo {title}
  {Understanding and compensating for noise on ibm quantum computers}}\ }\href
  {\doibase 10.1119/10.0006204} {\bibfield  {journal} {\bibinfo  {journal}
  {American Journal of Physics}\ }\textbf {\bibinfo {volume} {89}},\ \bibinfo
  {pages} {935--942} (\bibinfo {year} {2021})} \BibitemShut {NoStop}%
\bibitem [{\citenamefont {Georgopoulos}\ \emph
  {et~al.}(2021{\natexlab{b}})\citenamefont {Georgopoulos}, \citenamefont
  {Emary},\ and\ \citenamefont {Zuliani}}]{Georgopoulos2021b}%
  \BibitemOpen
  \bibfield  {author} {\bibinfo {author} {\bibfnamefont {K.}\
  \bibnamefont {Georgopoulos}}, \bibinfo {author} {\bibfnamefont {C.}\
  \bibnamefont {Emary}}, \ and\ \bibinfo {author} {\bibfnamefont {P.}\
  \bibnamefont {Zuliani}},\ }
  {\enquote {\bibinfo {title} {Quantum computer benchmarking via quantum
  algorithms}}\ } \href {\doibase 10.48550/ARXIV.2112.09457}
  {\bibfield  {journal} {\bibinfo  {journal} {arXiv:}},\ \bibinfo {pages}
  {2112.09457}}
  \BibitemShut {NoStop}%
\bibitem [{\citenamefont {Iverson}\ and\ \citenamefont
  {Preskill}(2020)}]{Iverson2020}%
  \BibitemOpen
  \bibfield  {author} {\bibinfo {author} {\bibfnamefont {J.~K.}\
  \bibnamefont {Iverson}}\ and\ \bibinfo {author} {\bibfnamefont {J.}\
  \bibnamefont {Preskill}},\ }\bibfield  {title} {\enquote {\bibinfo {title}
  {Coherence in logical quantum channels}}\ }\href {\doibase
  10.1088/1367-2630/ab8e5c} {\bibfield  {journal} {\bibinfo  {journal} {New
  Journal of Physics}\ }\textbf {\bibinfo {volume} {22}},\ \bibinfo {pages}
  {073066} (\bibinfo {year} {2020})}\BibitemShut {NoStop}%
\bibitem [{\citenamefont {Sarovar}\ \emph {et~al.}(2020)\citenamefont
  {Sarovar}, \citenamefont {Proctor}, \citenamefont {Rudinger}, \citenamefont
  {Young}, \citenamefont {Nielsen},\ and\ \citenamefont
  {Blume-Kohout}}]{Sarovar2020}%
  \BibitemOpen
  \bibfield  {author} {\bibinfo {author} {\bibfnamefont {M.}\ \bibnamefont
  {Sarovar}}, \bibinfo {author} {\bibfnamefont {T.}\ \bibnamefont
  {Proctor}}, \bibinfo {author} {\bibfnamefont {K.}\ \bibnamefont
  {Rudinger}}, \bibinfo {author} {\bibfnamefont {K.}\ \bibnamefont {Young}},
  \bibinfo {author} {\bibfnamefont {E.}\ \bibnamefont {Nielsen}}, \ and\
  \bibinfo {author} {\bibfnamefont {R.}\ \bibnamefont {Blume-Kohout}},\
  }\bibfield  {title} {\enquote {\bibinfo {title} {Detecting crosstalk errors
  in quantum information processors}}\ }\href {\doibase
  10.22331/q-2020-09-11-321} {\bibfield  {journal} {\bibinfo  {journal}
  {Quantum}\ }\textbf {\bibinfo {volume} {4}},\ \bibinfo {pages} {321}
  (\bibinfo {year} {2020})}\BibitemShut {NoStop}%
\bibitem {Note1}%
  \BibitemOpen
  \bibinfo {note} {\new{ Strictly speaking,
  the term ``Crosstalk'' encapsulates a wide variety of effects where qubits
  are affected by their neighbors, as described extensively in \cite
  {Sarovar2020}. In this work, we designate by ``crosstalk'' the result of the
  application of an imperfect quantum gate on the neighboring qubits, usually
  refered to as ``spillover crosstalk''.}}\BibitemShut {Stop}%
\bibitem [{\citenamefont {Zhao}\ \emph {et~al.}(2022)\citenamefont {Zhao},
  \citenamefont {Linghu}, \citenamefont {Li}, \citenamefont {Xu}, \citenamefont
  {Wang}, \citenamefont {Xue}, \citenamefont {Jin},\ and\ \citenamefont
  {Yu}}]{Zhao2022}%
  \BibitemOpen
  \bibfield  {author} {\bibinfo {author} {\bibfnamefont {P.}\ \bibnamefont
  {Zhao}}, \bibinfo {author} {\bibfnamefont {K.}\ \bibnamefont {Linghu}},
  \bibinfo {author} {\bibfnamefont {Z.}\ \bibnamefont {Li}}, \bibinfo
  {author} {\bibfnamefont {P.}\ \bibnamefont {Xu}}, \bibinfo {author}
  {\bibfnamefont {R.}\ \bibnamefont {Wang}}, \bibinfo {author}
  {\bibfnamefont {G.}\ \bibnamefont {Xue}}, \bibinfo {author}
  {\bibfnamefont {Y.}\ \bibnamefont {Jin}}, \ and\ \bibinfo {author}
  {\bibfnamefont {H.}\ \bibnamefont {Yu}},\ }\bibfield  {title} {\enquote
  {\bibinfo {title} {Quantum crosstalk analysis for simultaneous gate
  operations on superconducting qubits}}\ }\href {\doibase
  10.1103/prxquantum.3.020301} {\bibfield  {journal} {\bibinfo  {journal}
  {{PRX} Quantum}\ }\textbf {\bibinfo {volume} {3} } {\bibinfo {page} {020301} } (\bibinfo {year} {2022})}\BibitemShut {NoStop}%
\bibitem [{\citenamefont {Ketterer}\ \emph {et~al.}(2023)\citenamefont {Ketterer},
  \citenamefont {Wellens}}]{Ketterer2023}%
  \BibitemOpen
  \bibfield  {author} {\bibinfo {author} {\bibfnamefont {A.}\ \bibnamefont
  {Ketterer}}, \bibinfo {author} {\bibfnamefont {T.}\ \bibnamefont {Wellens}},\ }\bibfield  {title} {\enquote
  {\bibinfo {title} {Characterizing crosstalk of superconducting transmon processors}}\ }\href {\doibase 10.48550/ARXIV.2303.14103}{\bibfield  {journal} {\bibinfo  {journal}
  {arXiv:}}\bibinfo {pages} {2303.14103}}\BibitemShut {NoStop}%
\bibitem [{\citenamefont {Rudinger}\ \emph {et~al.}(2021)\citenamefont
  {Rudinger}, \citenamefont {Hogle}, \citenamefont {Naik}, \citenamefont
  {Hashim}, \citenamefont {Lobser}, \citenamefont {Santiago}, \citenamefont
  {Grace}, \citenamefont {Nielsen}, \citenamefont {Proctor}, \citenamefont
  {Seritan}, \citenamefont {Clark}, \citenamefont {Blume-Kohout}, \citenamefont
  {Siddiqi},\ and\ \citenamefont {Young}}]{Rudinger2021}%
  \BibitemOpen
  \bibfield  {author} {\bibinfo {author} {\bibfnamefont {K.}\ \bibnamefont
  {Rudinger}}, \bibinfo {author} {\bibfnamefont {C.~W.}\ \bibnamefont
  {Hogle}}, \bibinfo {author} {\bibfnamefont {R.~K.}\ \bibnamefont {Naik}},
  \bibinfo {author} {\bibfnamefont {A.}\ \bibnamefont {Hashim}}, \bibinfo
  {author} {\bibfnamefont {D.}\ \bibnamefont {Lobser}}, \bibinfo {author}
  {\bibfnamefont {D.~I.}\ \bibnamefont {Santiago}}, \bibinfo {author}
  {\bibfnamefont {M.~D.}\ \bibnamefont {Grace}}, \bibinfo {author}
  {\bibfnamefont {E.}\ \bibnamefont {Nielsen}}, \bibinfo {author}
  {\bibfnamefont {T.}\ \bibnamefont {Proctor}}, \bibinfo {author}
  {\bibfnamefont {S.}\ \bibnamefont {Seritan}}, \bibinfo {author}
  {\bibfnamefont {S.~M.}\ \bibnamefont {Clark}}, \bibinfo {author}
  {\bibfnamefont {R.}\ \bibnamefont {Blume-Kohout}}, \bibinfo {author}
  {\bibfnamefont {I.}\ \bibnamefont {Siddiqi}}, \ and\ \bibinfo {author}
  {\bibfnamefont {K.~C.}\ \bibnamefont {Young}},\ }\bibfield  {title}
  {\enquote {\bibinfo {title} {Experimental characterization of crosstalk
  errors with simultaneous gate set tomography}}\ }\href {\doibase
  10.1103/prxquantum.2.040338} {\bibfield  {journal} {\bibinfo  {journal}
  {{PRX} Quantum}\ }\textbf {\bibinfo {volume} {2}} {\bibinfo {page} {040338} } (\bibinfo {year} {2021})}\BibitemShut {NoStop}%
\bibitem [{\citenamefont {Tripathi}\ \emph {et~al.}(2022)\citenamefont
  {Tripathi}, \citenamefont {Chen}, \citenamefont {Khezri}, \citenamefont
  {Yip}, \citenamefont {Levenson-Falk},\ and\ \citenamefont
  {Lidar}}]{Tripathi2022}%
  \BibitemOpen
  \bibfield  {author} {\bibinfo {author} {\bibfnamefont {V.}\ \bibnamefont
  {Tripathi}}, \bibinfo {author} {\bibfnamefont {H.}\ \bibnamefont {Chen}},
  \bibinfo {author} {\bibfnamefont {M.}\ \bibnamefont {Khezri}}, \bibinfo
  {author} {\bibfnamefont {K.-W.}\ \bibnamefont {Yip}}, \bibinfo {author}
  {\bibfnamefont {E.M.}\ \bibnamefont {Levenson-Falk}}, \ and\ \bibinfo
  {author} {\bibfnamefont {D.~A.}\ \bibnamefont {Lidar}},\ }\bibfield
  {title} {\enquote {\bibinfo {title} {Suppression of crosstalk in
  superconducting qubits using dynamical decoupling}}\ }\href {\doibase
  10.1103/PhysRevApplied.18.024068} {\bibfield  {journal} {\bibinfo  {journal}
  {Phys. Rev. Appl.}\ }\textbf {\bibinfo {volume} {18}},\ \bibinfo {pages}
  {024068} (\bibinfo {year} {2022})}\BibitemShut {NoStop}%
\bibitem [{\citenamefont {Ding}\ \emph {et~al.}(2020)\citenamefont {Ding},
  \citenamefont {Gokhale}, \citenamefont {Lin}, \citenamefont {Rines},
  \citenamefont {Propson},\ and\ \citenamefont {Chong}}]{Ding2020}%
  \BibitemOpen
  \bibfield  {author} {\bibinfo {author} {\bibfnamefont {Y.}\
  \bibnamefont {Ding}}, \bibinfo {author} {\bibfnamefont {P.}\ \bibnamefont
  {Gokhale}}, \bibinfo {author} {\bibfnamefont {S.~F.}\ \bibnamefont
  {Lin}}, \bibinfo {author} {\bibfnamefont {R.}\ \bibnamefont {Rines}},
  \bibinfo {author} {\bibfnamefont {T.}\ \bibnamefont {Propson}}, \ and\
  \bibinfo {author} {\bibfnamefont {F.~T.}\ \bibnamefont {Chong}},\
  }\bibfield  {title} {\enquote {\bibinfo {title} {Systematic crosstalk
  mitigation for superconducting qubits via frequency-aware compilation}}\
  }in\ \href {\doibase 10.1109/micro50266.2020.00028} { {\bibinfo
  {booktitle} { International Symposium on
  Microarchitecture ({MICRO})}},\ \bibinfo  {publisher} {53rd Annual {IEEE/ACM}}\ (\bibinfo
  {year} {2020})}\BibitemShut {NoStop}%
\bibitem [{\citenamefont {Murali}\ \emph {et~al.}(2020)\citenamefont {Murali},
  \citenamefont {Mckay}, \citenamefont {Martonosi},\ and\ \citenamefont
  {Javadi-Abhari}}]{Murali2020}%
  \BibitemOpen
  \bibfield  {author} {\bibinfo {author} {\bibfnamefont {P.}\ \bibnamefont
  {Murali}}, \bibinfo {author} {\bibfnamefont {D.~C.}\ \bibnamefont
  {Mckay}}, \bibinfo {author} {\bibfnamefont {M.}\ \bibnamefont
  {Martonosi}}, \ and\ \bibinfo {author} {\bibfnamefont {A.}\ \bibnamefont
  {Javadi-Abhari}},\ }\bibfield  {title} {\enquote {\bibinfo {title} {Software
  mitigation of crosstalk on noisy intermediate-scale quantum computers}}\
  }in\ \href {\doibase 10.1145/3373376.3378477} { \bibinfo {booktitle}{Proceedings of the 25th International Conference on Architectural Support for Programming Languages and Operating Systems},\ \bibinfo {publisher} {{ACM}}\ (\bibinfo {year} {2020})}\BibitemShut {NoStop}%
\bibitem [{\citenamefont {Xie}\ \emph {et~al.}(2021)\citenamefont {Xie},
  \citenamefont {Zhai},\ and\ \citenamefont {Zheng}}]{Xie2021}%
  \BibitemOpen
  \bibfield  {author} {\bibinfo {author} {\bibfnamefont {L.}\ \bibnamefont
  {Xie}}, \bibinfo {author} {\bibfnamefont {J.}\ \bibnamefont {Zhai}}, \
  and\ \bibinfo {author} {\bibfnamefont {W.}\ \bibnamefont {Zheng}},\
  }\bibfield  {title} {\enquote {\bibinfo {title} {Mitigating crosstalk in
  quantum computers through commutativity-based instruction reordering}}\ }in\
  \href {\doibase 10.1109/dac18074.2021.9586145} {\bibinfo {booktitle}
  { Design Automation Conference},\ \bibinfo
  {publisher} {{58th ACM/IEEE}}\ (\bibinfo {year} {2021})}\BibitemShut {NoStop}%
\bibitem [{\citenamefont {Woitzik}\ \emph {et~al.}(2023)\citenamefont
  {Woitzik}, \citenamefont {Hoffmann}, \citenamefont {Buchleitner},\ and\
  \citenamefont {Carnio}}]{Woitzik2023}%
  \BibitemOpen
  \bibfield  {author} {\bibinfo {author} {\bibfnamefont {A. J.~C.}\
  \bibnamefont {Woitzik}}, \bibinfo {author} {\bibfnamefont {L.}\
  \bibnamefont {Hoffmann}}, \bibinfo {author} {\bibfnamefont {A.}\
  \bibnamefont {Buchleitner}}, \ and\ \bibinfo {author} {\bibfnamefont
  {E.~G.}\ \bibnamefont {Carnio}},\ }{\enquote {\bibinfo {title} {An energy estimation
  benchmark for quantum computers}}\ }\href {\doibase
  10.48550/ARXIV.2302.04144}  
  {\bibfield  {journal} {\bibinfo  {journal}
  {arXiv:}}\bibinfo {pages} {2302.04144}}\BibitemShut
  {NoStop}%
\bibitem [{\citenamefont {Blume-Kohout}\ \emph {et~al.}(2013)\citenamefont
  {Blume-Kohout}, \citenamefont {Gamble}, \citenamefont {Nielsen},
  \citenamefont {Mizrahi}, \citenamefont {Sterk},\ and\ \citenamefont
  {Maunz}}]{BlumeKohout2013}%
  \BibitemOpen
  \bibfield  {author} {\bibinfo {author} {\bibfnamefont {R.}\ \bibnamefont
  {Blume-Kohout}}, \bibinfo {author} {\bibfnamefont {J.~K.}\ \bibnamefont
  {Gamble}}, \bibinfo {author} {\bibfnamefont {E.}\ \bibnamefont {Nielsen}},
  \bibinfo {author} {\bibfnamefont {J.}\ \bibnamefont {Mizrahi}},
  \bibinfo {author} {\bibfnamefont {J.~D.}\ \bibnamefont {Sterk}}, \ and\
  \bibinfo {author} {\bibfnamefont {P.}\ \bibnamefont {Maunz}},\ }{\enquote {\bibinfo {title} {Robust,
  self-consistent, closed-form tomography of quantum logic gates on a trapped
  ion qubit}}\ }\href
  {\doibase 10.48550/ARXIV.1310.4492}  {\bibfield  {journal} {\bibinfo  {journal}
  {arXiv:}}\bibinfo {pages} {1310.4492}}\BibitemShut {NoStop}%
\bibitem [{\citenamefont {Blume-Kohout}\ \emph {et~al.}(2017)\citenamefont
  {Blume-Kohout}, \citenamefont {Gamble}, \citenamefont {Nielsen},
  \citenamefont {Rudinger}, \citenamefont {Mizrahi}, \citenamefont {Fortier},\
  and\ \citenamefont {Maunz}}]{BlumeKohout2017}%
  \BibitemOpen
  \bibfield  {author} {\bibinfo {author} {\bibfnamefont {R.}\ \bibnamefont
  {Blume-Kohout}}, \bibinfo {author} {\bibfnamefont {J.~K.}\ \bibnamefont
  {Gamble}}, \bibinfo {author} {\bibfnamefont {E.}\ \bibnamefont {Nielsen}},
  \bibinfo {author} {\bibfnamefont {K.}\ \bibnamefont {Rudinger}},
  \bibinfo {author} {\bibfnamefont {J.}\ \bibnamefont {Mizrahi}},
  \bibinfo {author} {\bibfnamefont {K.}\ \bibnamefont {Fortier}}, \ and\
  \bibinfo {author} {\bibfnamefont {P.}\ \bibnamefont {Maunz}},\ }\bibfield
  {title} {\enquote {\bibinfo {title} {Demonstration of qubit operations below
  a rigorous fault tolerance threshold with gate set tomography}}\ }\href
  {\doibase 10.1038/ncomms14485} {\bibfield  {journal} {\bibinfo  {journal}
  {Nature Communications}\ }\textbf {\bibinfo {volume} {8}} (\bibinfo {year}
  {2017})}\BibitemShut {NoStop}%
\bibitem [{\citenamefont {Endo}\ \emph {et~al.}(2018)\citenamefont {Endo},
  \citenamefont {Benjamin},\ and\ \citenamefont {Li}}]{Endo2018}%
  \BibitemOpen
  \bibfield  {author} {\bibinfo {author} {\bibfnamefont {S.}\ \bibnamefont
  {Endo}}, \bibinfo {author} {\bibfnamefont {S.~C.}\ \bibnamefont
  {Benjamin}}, \ and\ \bibinfo {author} {\bibfnamefont {Y.}\ \bibnamefont
  {Li}},\ }\bibfield  {title} {\enquote {\bibinfo {title} {Practical quantum
  error mitigation for near-future applications}}\ }\href {\doibase
  10.1103/PhysRevX.8.031027} {\bibfield  {journal} {\bibinfo  {journal} {Phys.
  Rev. X}\ }\textbf {\bibinfo {volume} {8}},\ \bibinfo {pages} {031027}
  (\bibinfo {year} {2018})}\BibitemShut {NoStop}%
\bibitem [{\citenamefont {Nielsen}\ \emph {et~al.}(2021)\citenamefont
  {Nielsen}, \citenamefont {Gamble}, \citenamefont {Rudinger}, \citenamefont
  {Scholten}, \citenamefont {Young},\ and\ \citenamefont
  {Blume-Kohout}}]{Nielsen2021}%
  \BibitemOpen
  \bibfield  {author} {\bibinfo {author} {\bibfnamefont {E.}\ \bibnamefont
  {Nielsen}}, \bibinfo {author} {\bibfnamefont {J.~K.}\ \bibnamefont
  {Gamble}}, \bibinfo {author} {\bibfnamefont {K.}\ \bibnamefont
  {Rudinger}}, \bibinfo {author} {\bibfnamefont {T.}\ \bibnamefont
  {Scholten}}, \bibinfo {author} {\bibfnamefont {K.}\ \bibnamefont {Young}},
  \ and\ \bibinfo {author} {\bibfnamefont {R.}\ \bibnamefont
  {Blume-Kohout}},\ }\bibfield  {title} {\enquote {\bibinfo {title} {Gate set
  tomography}}\ }\href {\doibase 10.22331/q-2021-10-05-557} {\bibfield
  {journal} {\bibinfo  {journal} {Quantum}\ }\textbf {\bibinfo {volume} {5}},\
  \bibinfo {pages} {557} (\bibinfo {year} {2021})}\BibitemShut {NoStop}%
\bibitem [{\citenamefont {Dumitrescu}\ \emph {et~al.}(2018)\citenamefont
  {Dumitrescu}, \citenamefont {McCaskey}, \citenamefont {Hagen}, \citenamefont
  {Jansen}, \citenamefont {Morris}, \citenamefont {Papenbrock}, \citenamefont
  {Pooser}, \citenamefont {Dean},\ and\ \citenamefont
  {Lougovski}}]{Dumitrescu2018}%
  \BibitemOpen
  \bibfield  {author} {\bibinfo {author} {\bibfnamefont
  {E.{\hspace{0.167em}}F.}\ \bibnamefont {Dumitrescu}}, \bibinfo {author}
  {\bibfnamefont {A.{\hspace{0.167em}}J.}\ \bibnamefont {McCaskey}}, \bibinfo
  {author} {\bibfnamefont {G.}~\bibnamefont {Hagen}}, \bibinfo {author}
  {\bibfnamefont {G.{\hspace{0.167em}}R.}\ \bibnamefont {Jansen}}, \bibinfo
  {author} {\bibfnamefont {T.{\hspace{0.167em}}D.}\ \bibnamefont {Morris}},
  \bibinfo {author} {\bibfnamefont {T.}~\bibnamefont {Papenbrock}}, \bibinfo
  {author} {\bibfnamefont {R.{\hspace{0.167em}}C.}\ \bibnamefont {Pooser}},
  \bibinfo {author} {\bibfnamefont {D.{\hspace{0.167em}}J.}\ \bibnamefont
  {Dean}}, \ and\ \bibinfo {author} {\bibfnamefont {P.}~\bibnamefont
  {Lougovski}},\ }\bibfield  {title} {\enquote {\bibinfo {title} {Cloud quantum
  computing of an atomic nucleus},}\ }\href {\doibase
  10.1103/physrevlett.120.210501} {\bibfield  {journal} {\bibinfo  {journal}
  {Physical Review Letters}\ }\textbf {\bibinfo {volume} {120}} {\bibinfo {page} {210501}} (\bibinfo
  {year} {2018})}\BibitemShut {NoStop}%
\bibitem [{\citenamefont {He}\ \emph {et~al.}(2020)\citenamefont {He},
  \citenamefont {Nachman}, \citenamefont {de~Jong},\ and\ \citenamefont
  {Bauer}}]{He2020}%
  \BibitemOpen
  \bibfield  {author} {\bibinfo {author} {\bibfnamefont {A.}\ \bibnamefont
  {He}}, \bibinfo {author} {\bibfnamefont {B.}\ \bibnamefont {Nachman}},
  \bibinfo {author} {\bibfnamefont {W.~A.}\ \bibnamefont {de~Jong}}, \ and\
  \bibinfo {author} {\bibfnamefont {C.~W.}\ \bibnamefont {Bauer}},\
  }\bibfield  {title} {\enquote {\bibinfo {title} {Zero-noise extrapolation for
  quantum-gate error mitigation with identity insertions}}\ }\href {\doibase
  10.1103/physreva.102.012426} {\bibfield  {journal} {\bibinfo  {journal}
  {Physical Review A}\ }\textbf {\bibinfo {volume} {102}} {\bibinfo {page} {012426}} (\bibinfo {year}
  {2020})}\BibitemShut {NoStop}%
\bibitem [{\citenamefont {Giurgica-Tiron}\ \emph {et~al.}(2020)\citenamefont
  {Giurgica-Tiron}, \citenamefont {Hindy}, \citenamefont {LaRose},
  \citenamefont {Mari},\ and\ \citenamefont {Zeng}}]{GiurgicaTiron2020}%
  \BibitemOpen
  \bibfield  {author} {\bibinfo {author} {\bibfnamefont {T.}\ \bibnamefont
  {Giurgica-Tiron}}, \bibinfo {author} {\bibfnamefont {Y.}\ \bibnamefont
  {Hindy}}, \bibinfo {author} {\bibfnamefont {R.}\ \bibnamefont {LaRose}},
  \bibinfo {author} {\bibfnamefont {A.}\ \bibnamefont {Mari}}, \ and\
  \bibinfo {author} {\bibfnamefont {W.~J.}\ \bibnamefont {Zeng}},\
  }\bibfield  {title} {\enquote {\bibinfo {title} {Digital zero noise
  extrapolation for quantum error mitigation}}\ }in\ \href {\doibase
  10.1109/qce49297.2020.00045} {\bibinfo {booktitle} {
  International Conference on Quantum Computing and Engineering},\
  \bibinfo  {publisher} {{IEEE}}\ (\bibinfo {year} {2020})}\BibitemShut
  {NoStop}%
\bibitem [{\citenamefont {Urbanek}\ \emph {et~al.}(2021)\citenamefont
  {Urbanek}, \citenamefont {Nachman}, \citenamefont {Pascuzzi}, \citenamefont
  {He}, \citenamefont {Bauer},\ and\ \citenamefont {de~Jong}}]{Urbanek2021}%
  \BibitemOpen
  \bibfield  {author} {\bibinfo {author} {\bibfnamefont {M.}\
  \bibnamefont {Urbanek}}, \bibinfo {author} {\bibfnamefont {B.}\
  \bibnamefont {Nachman}}, \bibinfo {author} {\bibfnamefont {V.~R.}\
  \bibnamefont {Pascuzzi}}, \bibinfo {author} {\bibfnamefont {A.}\
  \bibnamefont {He}}, \bibinfo {author} {\bibfnamefont {C.~W.}\
  \bibnamefont {Bauer}},\ and\ \bibinfo {author} {\bibfnamefont {W.~A.}\
  \bibnamefont {de~Jong}},\ }\bibfield  {title} {\enquote {\bibinfo {title}
  {Mitigating depolarizing noise on quantum computers with noise-estimation
  circuits}}\ }\href {\doibase 10.1103/physrevlett.127.270502} {\bibfield
  {journal} {\bibinfo  {journal} {Physical Review Letters}\ }\textbf {\bibinfo
  {volume} {127}} {\bibinfo {page} {270502}} (\bibinfo {year} {2021})}\BibitemShut {NoStop}%
\bibitem [{\citenamefont {Quek}\ \emph {et~al.}(2022)\citenamefont {Quek},
  \citenamefont {França}, \citenamefont {Khatri}, \citenamefont {Meyer},\ and\
  \citenamefont {Eisert}}]{Quek2022}%
  \BibitemOpen
  \bibfield  {author} {\bibinfo {author} {\bibfnamefont {Y.}\ \bibnamefont
  {Quek}}, \bibinfo {author} {\bibfnamefont {D.~S.}\ \bibnamefont
  {França}}, \bibinfo {author} {\bibfnamefont {S.}\ \bibnamefont
  {Khatri}}, \bibinfo {author} {\bibfnamefont {J.~J.}\ \bibnamefont
  {Meyer}}, \ and\ \bibinfo {author} {\bibfnamefont {J.}\ \bibnamefont
  {Eisert}},\ }{\enquote {\bibinfo
  {title} {Exponentially tighter bounds on limitations of quantum error
  mitigation}}\ } \href {\doibase 10.48550/ARXIV.2210.11505}{\bibfield  {journal} {\bibinfo  {journal}
  {arXiv:}}\bibinfo {pages} {2210.11505}} \BibitemShut {NoStop}%
\bibitem [{\citenamefont {Kern}\ \emph {et~al.}(2005)\citenamefont {Kern},
  \citenamefont {Alber},\ and\ \citenamefont {Shepelyansky}}]{Kern2005}%
  \BibitemOpen
  \bibfield  {author} {\bibinfo {author} {\bibfnamefont {O.}~\bibnamefont
  {Kern}}, \bibinfo {author} {\bibfnamefont {G.}~\bibnamefont {Alber}}, \ and\
  \bibinfo {author} {\bibfnamefont {D.~L.}\ \bibnamefont {Shepelyansky}},\
  }\bibfield  {title} {\enquote {\bibinfo {title} {Quantum error correction of
  coherent errors by randomization}}\ }\href {\doibase
  10.1140/epjd/e2004-00196-9} {\bibfield  {journal} {\bibinfo  {journal} {The
  European Physical Journal D}\ }\textbf {\bibinfo {volume} {32}},\ \bibinfo
  {pages} {153--156} (\bibinfo {year} {2005})}\BibitemShut {NoStop}%
\bibitem [{\citenamefont {Wallman}\ and\ \citenamefont
  {Emerson}(2016)}]{Wallman2016}%
  \BibitemOpen
  \bibfield  {author} {\bibinfo {author} {\bibfnamefont {J.~J.}\ \bibnamefont
  {Wallman}}\ and\ \bibinfo {author} {\bibfnamefont {J.}\ \bibnamefont
  {Emerson}},\ }\bibfield  {title} {\enquote {\bibinfo {title} {Noise tailoring
  for scalable quantum computation via randomized compiling}}\ }\href
  {\doibase 10.1103/physreva.94.052325} {\bibfield  {journal} {\bibinfo
  {journal} {Physical Review A}\ }\textbf {\bibinfo {volume} {94}} {\bibinfo {page} {052325}} (\bibinfo
  {year} {2016})}\BibitemShut {NoStop}%
\bibitem [{\citenamefont {Hashim}\ \emph {et~al.}(2021)\citenamefont {Hashim},
  \citenamefont {Naik}, \citenamefont {Morvan}, \citenamefont {Ville},
  \citenamefont {Mitchell}, \citenamefont {Kreikebaum}, \citenamefont {Davis},
  \citenamefont {Smith}, \citenamefont {Iancu}, \citenamefont {O'Brien},
  \citenamefont {Hincks}, \citenamefont {Wallman}, \citenamefont {Emerson},\
  and\ \citenamefont {Siddiqi}}]{Hashim2021}%
  \BibitemOpen
  \bibfield  {author} {\bibinfo {author} {\bibfnamefont {A.}\ \bibnamefont
  {Hashim}}, \bibinfo {author} {\bibfnamefont {R.~K.}\ \bibnamefont {Naik}},
  \bibinfo {author} {\bibfnamefont {A.}\ \bibnamefont {Morvan}}, \bibinfo
  {author} {\bibfnamefont {J.-L.}\ \bibnamefont {Ville}}, \bibinfo {author}
  {\bibfnamefont {B.}\ \bibnamefont {Mitchell}}, \bibinfo {author}
  {\bibfnamefont {J.~M.}\ \bibnamefont {Kreikebaum}}, \bibinfo {author}
  {\bibfnamefont {M.}\ \bibnamefont {Davis}}, \bibinfo {author}
  {\bibfnamefont {E.}\ \bibnamefont {Smith}}, \bibinfo {author}
  {\bibfnamefont {C.}\ \bibnamefont {Iancu}}, \bibinfo {author}
  {\bibfnamefont {K.~P.}\ \bibnamefont {O'Brien}}, \bibinfo {author}
  {\bibfnamefont {I.}\ \bibnamefont {Hincks}}, \bibinfo {author}
  {\bibfnamefont {J.~J.}\ \bibnamefont {Wallman}}, \bibinfo {author}
  {\bibfnamefont {J.}\ \bibnamefont {Emerson}}, \ and\ \bibinfo {author}
  {\bibfnamefont {I.}\ \bibnamefont {Siddiqi}},\ }\bibfield  {title}
  {\enquote {\bibinfo {title} {Randomized compiling for scalable quantum
  computing on a noisy superconducting quantum processor}}\ }\href {\doibase
  10.1103/physrevx.11.041039} {\bibfield  {journal} {\bibinfo  {journal}
  {Physical Review X}\ }\textbf {\bibinfo {volume} {11}} {\bibinfo {page} {041039}} (\bibinfo {year}
  {2021})}\BibitemShut {NoStop}%
\bibitem [{\citenamefont {Cai}\ and\ \citenamefont {Benjamin}(2019)}]{Cai2019}%
  \BibitemOpen
  \bibfield  {author} {\bibinfo {author} {\bibfnamefont {Z.}\ \bibnamefont
  {Cai}}\ and\ \bibinfo {author} {\bibfnamefont {S.~C.}\ \bibnamefont
  {Benjamin}},\ }\bibfield  {title} {\enquote {\bibinfo {title} {Constructing
  smaller pauli twirling sets for arbitrary error channels}}\ }\href {\doibase
  10.1038/s41598-019-46722-7} {\bibfield  {journal} {\bibinfo  {journal}
  {Scientific Reports}\ }\textbf {\bibinfo {volume} {9}} (\bibinfo {year}
  {2019})}\BibitemShut {NoStop}%
\bibitem [{\citenamefont {Rosenberg}\ \emph {et~al.}(2022)\citenamefont
  {Rosenberg}, \citenamefont {Ginsparg},\ and\ \citenamefont
  {McMahon}}]{Rosenberg2022}%
  \BibitemOpen
  \bibfield  {author} {\bibinfo {author} {\bibfnamefont {E.}\ \bibnamefont
  {Rosenberg}}, \bibinfo {author} {\bibfnamefont {P.}\ \bibnamefont
  {Ginsparg}}, \ and\ \bibinfo {author} {\bibfnamefont {P.~L.}\ \bibnamefont
  {McMahon}},\ }\bibfield  {title} {\enquote {\bibinfo {title} {Experimental
  error mitigation using linear rescaling for variational quantum eigensolving
  with up to 20 qubits},}\ }\href {\doibase 10.1088/2058-9565/ac3b37}
  {\bibfield  {journal} {\bibinfo  {journal} {Quantum Science and Technology}\
  }\textbf {\bibinfo {volume} {7}},\ \bibinfo {pages} {015024} (\bibinfo {year}
  {2022})}\BibitemShut {NoStop}%
\bibitem [{\citenamefont {Kurita}\ \emph {et~al.}(2022)\citenamefont {Kurita},
  \citenamefont {Qassim}, \citenamefont {Ishii}, \citenamefont {Oshima},
  \citenamefont {Sato},\ and\ \citenamefont {Emerson}}]{Kurita2022}%
  \BibitemOpen
  \bibfield  {author} {\bibinfo {author} {\bibfnamefont {T.}\
  \bibnamefont {Kurita}}, \bibinfo {author} {\bibfnamefont {H.}\
  \bibnamefont {Qassim}}, \bibinfo {author} {\bibfnamefont {M.}\
  \bibnamefont {Ishii}}, \bibinfo {author} {\bibfnamefont {H.}\
  \bibnamefont {Oshima}}, \bibinfo {author} {\bibfnamefont {S.}\
  \bibnamefont {Sato}}, \ and\ \bibinfo {author} {\bibfnamefont {J.}\
  \bibnamefont {Emerson}},\ }
{\enquote {\bibinfo {title} {Synergetic quantum error mitigation by
  randomized compiling and zero-noise extrapolation for the variational quantum
  eigensolver},}\ } \href {\doibase 10.48550/ARXIV.2212.11198} {\bibfield  {journal} {\bibinfo  {journal}
  {arXiv:}}\bibinfo {pages} {2212.11198}} \BibitemShut {NoStop}%
\bibitem [{\citenamefont {Vazquez}\ \emph {et~al.}(2023)\citenamefont
  {Vazquez}, \citenamefont {Egger}, \citenamefont {Ochsner},\ and\
  \citenamefont {Woerner}}]{Vazquez2023}%
  \BibitemOpen
  \bibfield  {author} {\bibinfo {author} {\bibfnamefont {A.~C.}\
  \bibnamefont {Vazquez}}, \bibinfo {author} {\bibfnamefont {D.~J.}\
  \bibnamefont {Egger}}, \bibinfo {author} {\bibfnamefont {D.}\ \bibnamefont
  {Ochsner}}, \ and\ \bibinfo {author} {\bibfnamefont {S.}\ \bibnamefont
  {Woerner}},\ }\bibfield  {title} {\enquote {\bibinfo {title}
  {Well-conditioned multi-product formulas for hardware-friendly hamiltonian
  simulation},}\ }\href {\doibase 10.22331/q-2023-07-25-1067} {\bibfield
  {journal} {\bibinfo  {journal} {Quantum}\ }\textbf {\bibinfo {volume} {7}},\
  \bibinfo {pages} {1067} (\bibinfo {year} {2023})}\BibitemShut {NoStop}%
\bibitem [{\citenamefont {Kim}\ \emph {et~al.}(2023)\citenamefont {Kim},
  \citenamefont {Wood}, \citenamefont {Yoder}, \citenamefont {Merkel},
  \citenamefont {Gambetta}, \citenamefont {Temme},\ and\ \citenamefont
  {Kandala}}]{Kim2023}%
  \BibitemOpen
  \bibfield  {author} {\bibinfo {author} {\bibfnamefont {Y.}\
  \bibnamefont {Kim}}, \bibinfo {author} {\bibfnamefont {C.~J.}\
  \bibnamefont {Wood}}, \bibinfo {author} {\bibfnamefont {T.~J.}\
  \bibnamefont {Yoder}}, \bibinfo {author} {\bibfnamefont {S.~T.}\
  \bibnamefont {Merkel}}, \bibinfo {author} {\bibfnamefont {J.~M.}\
  \bibnamefont {Gambetta}}, \bibinfo {author} {\bibfnamefont {K.}\
  \bibnamefont {Temme}}, \ and\ \bibinfo {author} {\bibfnamefont {A.}\
  \bibnamefont {Kandala}},\ }\bibfield  {title} {\enquote {\bibinfo {title}
  {Scalable error mitigation for noisy quantum circuits produces competitive
  expectation values},}\ }\href {\doibase 10.1038/s41567-022-01914-3}
  {\bibfield  {journal} {\bibinfo  {journal} {Nature Physics}\ }\textbf
  {\bibinfo {volume} {19}},\ \bibinfo {pages} {752--759} (\bibinfo {year}
  {2023})}\BibitemShut {NoStop}%
\bibitem [{\citenamefont {Bardeen}\ \emph {et~al.}(1957)\citenamefont
  {Bardeen}, \citenamefont {Cooper},\ and\ \citenamefont
  {Schrieffer}}]{Bardeen1957}%
  \BibitemOpen
  \bibfield  {author} {\bibinfo {author} {\bibfnamefont {J.}~\bibnamefont
  {Bardeen}}, \bibinfo {author} {\bibfnamefont {L.~N.}\ \bibnamefont {Cooper}},
  \ and\ \bibinfo {author} {\bibfnamefont {J.~R.}\ \bibnamefont {Schrieffer}},\
  }\bibfield  {title} {\enquote {\bibinfo {title} {Theory of
  superconductivity}}\ }\href {\doibase 10.1103/physrev.108.1175} {\bibfield
  {journal} {\bibinfo  {journal} {Physical Review}\ }\textbf {\bibinfo {volume}
  {108}},\ \bibinfo {pages} {1175--1204} (\bibinfo {year} {1957})}\BibitemShut
  {NoStop}%
\bibitem [{\citenamefont {Richardson}(1963)}]{Richardson1963}%
  \BibitemOpen
  \bibfield  {author} {\bibinfo {author} {\bibfnamefont {R.W.}\ \bibnamefont
  {Richardson}},\ }\bibfield  {title} {\enquote {\bibinfo {title} {A restricted
  class of exact eigenstates of the pairing-force hamiltonian}}\ }\href
  {\doibase 10.1016/0031-9163(63)90259-2} {\bibfield  {journal} {\bibinfo
  {journal} {Physics Letters}\ }\textbf {\bibinfo {volume} {3}},\ \bibinfo
  {pages} {277--279} (\bibinfo {year} {1963})}\BibitemShut {NoStop}%
\bibitem [{\citenamefont {Richardson}\ and\ \citenamefont
  {Sherman}(1964)}]{Richardson1964}%
  \BibitemOpen
  \bibfield  {author} {\bibinfo {author} {\bibfnamefont {R.W.}\ \bibnamefont
  {Richardson}}\ and\ \bibinfo {author} {\bibfnamefont {N.}~\bibnamefont
  {Sherman}},\ }\bibfield  {title} {\enquote {\bibinfo {title} {Exact
  eigenstates of the pairing-force hamiltonian}}\ }\href {\doibase
  10.1016/0029-5582(64)90687-x} {\bibfield  {journal} {\bibinfo  {journal}
  {Nuclear Physics}\ }\textbf {\bibinfo {volume} {52}},\ \bibinfo {pages}
  {221--238} (\bibinfo {year} {1964})}\BibitemShut {NoStop}%
\bibitem [{\citenamefont {Dukelsky}\ \emph {et~al.}(2004)\citenamefont
  {Dukelsky}, \citenamefont {Pittel},\ and\ \citenamefont
  {Sierra}}]{Dukelsky2004}%
  \BibitemOpen
  \bibfield  {author} {\bibinfo {author} {\bibfnamefont {J.}~\bibnamefont
  {Dukelsky}}, \bibinfo {author} {\bibfnamefont {S.}~\bibnamefont {Pittel}}, \
  and\ \bibinfo {author} {\bibfnamefont {G.}~\bibnamefont {Sierra}},\
  }\bibfield  {title} {\enquote {\bibinfo {title} {Colloquium: Exactly solvable
  Richardson-Gaudin models for many-body quantum systems}}\ }\href {\doibase
  10.1103/revmodphys.76.643} {\bibfield  {journal} {\bibinfo  {journal}
  {Reviews of Modern Physics}\ }\textbf {\bibinfo {volume} {76}},\ \bibinfo
  {pages} {643--662} (\bibinfo {year} {2004})}\BibitemShut {NoStop}%
\bibitem [{\citenamefont {Rombouts}\ \emph {et~al.}(2004)\citenamefont
  {Rombouts}, \citenamefont {Neck},\ and\ \citenamefont
  {Dukelsky}}]{Rombouts2004}%
  \BibitemOpen
  \bibfield  {author} {\bibinfo {author} {\bibfnamefont {S.}~\bibnamefont
  {Rombouts}}, \bibinfo {author} {\bibfnamefont {D.~Van}\ \bibnamefont {Neck}},
  \ and\ \bibinfo {author} {\bibfnamefont {J.}~\bibnamefont {Dukelsky}},\
  }\bibfield  {title} {\enquote {\bibinfo {title} {Solving the Richardson
  equations for fermions}}\ }\href {\doibase 10.1103/physrevc.69.061303}
  {\bibfield  {journal} {\bibinfo  {journal} {Physical Review C}\ }\textbf
  {\bibinfo {volume} {69}} {\bibinfo {page} {061303(R)}} (\bibinfo {year} {2004})}\BibitemShut {NoStop}%
\bibitem [{\citenamefont {Faribault}\ \emph {et~al.}(2009)\citenamefont
  {Faribault}, \citenamefont {Calabrese},\ and\ \citenamefont
  {Caux}}]{Faribault2009}%
  \BibitemOpen
  \bibfield  {author} {\bibinfo {author} {\bibfnamefont {A.}\
  \bibnamefont {Faribault}}, \bibinfo {author} {\bibfnamefont {P.}\
  \bibnamefont {Calabrese}}, \ and\ \bibinfo {author} {\bibfnamefont
  {J.-S.}\ \bibnamefont {Caux}},\ }\bibfield  {title} {\enquote
  {\bibinfo {title} {Bethe ansatz approach to quench dynamics in the Richardson
  model}}\ }\href {\doibase 10.1063/1.3183720} {\bibfield  {journal} {\bibinfo
   {journal} {Journal of Mathematical Physics}\ }\textbf {\bibinfo {volume}
  {50}},\ \bibinfo {pages} {095212} (\bibinfo {year} {2009})}\BibitemShut
  {NoStop}%
\bibitem [{\citenamefont {Lacroix}(2020)}]{Lacroix2020}%
  \BibitemOpen
  \bibfield  {author} {\bibinfo {author} {\bibfnamefont {D.}\ \bibnamefont
  {Lacroix}},\ }\bibfield  {title} {\enquote {\bibinfo {title}
  {Symmetry-assisted preparation of entangled many-body states on a quantum
  computer}}\ }\href {\doibase 10.1103/physrevlett.125.230502} {\bibfield
  {journal} {\bibinfo  {journal} {Physical Review Letters}\ }\textbf {\bibinfo
  {volume} {125}} {\bibinfo {page} {230502}} (\bibinfo {year} {2020})}\BibitemShut {NoStop}%
\bibitem [{\citenamefont {Khamoshi}\ \emph {et~al.}(2020)\citenamefont
  {Khamoshi}, \citenamefont {Evangelista},\ and\ \citenamefont
  {Scuseria}}]{Khamoshi2020}%
  \BibitemOpen
  \bibfield  {author} {\bibinfo {author} {\bibfnamefont {A.}\ \bibnamefont
  {Khamoshi}}, \bibinfo {author} {\bibfnamefont {F.~A.}\ \bibnamefont
  {Evangelista}}, \ and\ \bibinfo {author} {\bibfnamefont {G.~E.}\
  \bibnamefont {Scuseria}},\ }\bibfield  {title} {\enquote {\bibinfo {title}
  {Correlating {AGP} on a quantum computer}}\ }\href {\doibase
  10.1088/2058-9565/abc1bb} {\bibfield  {journal} {\bibinfo  {journal} {Quantum
  Science and Technology}\ }\textbf {\bibinfo {volume} {6}},\ \bibinfo {pages}
  {014004} (\bibinfo {year} {2020})}\BibitemShut {NoStop}%
\bibitem [{\citenamefont {Guzman}\ and\ \citenamefont
  {Lacroix}(2022)}]{Guzman2022}%
  \BibitemOpen
  \bibfield  {author} {\bibinfo {author} {\bibfnamefont {E. A.}\
  \bibnamefont {Ruiz Guzman}}\ and\ \bibinfo {author} {\bibfnamefont {D.}\
  \bibnamefont {Lacroix}},\ }\bibfield  {title} {\enquote {\bibinfo {title}
  {Accessing ground-state and excited-state energies in a many-body system
  after symmetry restoration using quantum computers}}\ }\href {\doibase
  10.1103/physrevc.105.024324} {\bibfield  {journal} {\bibinfo  {journal}
  {Physical Review C}\ }\textbf {\bibinfo {volume} {105}} {\bibinfo {page} {024324}} (\bibinfo {year}
  {2022})}\BibitemShut {NoStop}%
  \bibitem [{\citenamefont {Ruh}\ \emph {et~al.}(2023)\citenamefont {Ruh},
  \citenamefont {Finsterhoelzl},\ and\ \citenamefont {Burkard}}]{Ruh2023}%
  \BibitemOpen
  \bibfield  {author} {\bibinfo {author} {\bibfnamefont {Jannis}\ \bibnamefont
  {Ruh}}, \bibinfo {author} {\bibfnamefont {Regina}\ \bibnamefont
  {Finsterhoelzl}}, \ and\ \bibinfo {author} {\bibfnamefont {Guido}\
  \bibnamefont {Burkard}},\ }\bibfield  {title} {\enquote {\bibinfo {title}
  {Digital quantum simulation of the {BCS} model with a central-spin-like
  quantum processor},}\ }\href {\doibase 10.1103/physreva.107.062604}
  {\bibfield  {journal} {\bibinfo  {journal} {Physical Review A}\ }\textbf
  {\bibinfo {volume} {107}} {\bibinfo {page} 062604 }(\bibinfo {year} {2023})}\BibitemShut {NoStop}%
\bibitem [{\citenamefont {Anderson}(1959)}]{Anderson1959}%
  \BibitemOpen
  \bibfield  {author} {\bibinfo {author} {\bibfnamefont {P.W.}\ \bibnamefont
  {Anderson}},\ }\bibfield  {title} {\enquote {\bibinfo {title} {Theory of
  dirty superconductors}}\ }\href {\doibase 10.1016/0022-3697(59)90036-8}
  {\bibfield  {journal} {\bibinfo  {journal} {Journal of Physics and Chemistry
  of Solids}\ }\textbf {\bibinfo {volume} {11}},\ \bibinfo {pages} {26--30}
  (\bibinfo {year} {1959})}\BibitemShut {NoStop}%
\bibitem [{\citenamefont {Yuzbashyan}\ \emph {et~al.}(2005)\citenamefont
  {Yuzbashyan}, \citenamefont {Altshuler}, \citenamefont {Kuznetsov},\ and\
  \citenamefont {Enolskii}}]{Yuzbashyan2005}%
  \BibitemOpen
  \bibfield  {author} {\bibinfo {author} {\bibfnamefont {E.~A.}\ \bibnamefont
  {Yuzbashyan}}, \bibinfo {author} {\bibfnamefont {B.~L.}\ \bibnamefont
  {Altshuler}}, \bibinfo {author} {\bibfnamefont {V.~B.}\ \bibnamefont
  {Kuznetsov}}, \ and\ \bibinfo {author} {\bibfnamefont {V.~Z.}\
  \bibnamefont {Enolskii}},\ }\bibfield  {title} {\enquote {\bibinfo {title}
  {Solution for the dynamics of the {BCS} and central spin problems}}\ }\href
  {\doibase 10.1088/0305-4470/38/36/003} {\bibfield  {journal} {\bibinfo
  {journal} {Journal of Physics A: Mathematical and General}\ }\textbf
  {\bibinfo {volume} {38}},\ \bibinfo {pages} {7831--7849} (\bibinfo {year}
  {2005})}\BibitemShut {NoStop}%
\bibitem [{\citenamefont {Jordan}\ and\ \citenamefont
  {Wigner}(1928)}]{Jordan1928}%
  \BibitemOpen
  \bibfield  {author} {\bibinfo {author} {\bibfnamefont {P.}~\bibnamefont
  {Jordan}}\ and\ \bibinfo {author} {\bibfnamefont {E.}~\bibnamefont
  {Wigner}},\ }\bibfield  {title} {\enquote {\bibinfo {title} {Über das
  paulische Equivalenzverbot}}\ }\href {\doibase 10.1007/bf01331938}
  {\bibfield  {journal} {\bibinfo  {journal} {Zeitschrift für Physik}\
  }\textbf {\bibinfo {volume} {47}},\ \bibinfo {pages} {631--651} (\bibinfo
  {year} {1928})}\BibitemShut {NoStop}%
\bibitem [{\citenamefont {Ovrum}\ and\ \citenamefont
  {Hjorth-Jensen}(2007)}]{Ovrum2007}%
  \BibitemOpen
  \bibfield  {author} {\bibinfo {author} {\bibfnamefont {E.}~\bibnamefont
  {Ovrum}}\ and\ \bibinfo {author} {\bibfnamefont {M.}~\bibnamefont
  {Hjorth-Jensen}},\ } {\enquote
  {\bibinfo {title} {Quantum computation algorithm for many-body studies}}\ }\href {\doibase 10.48550/ARXIV.0705.1928}
  {\bibfield  {journal} {\bibinfo  {journal}
  {arXiv:}}\bibinfo {pages} {0705.1928}}\BibitemShut {NoStop}%
\bibitem [{\citenamefont {Suzuki}(1976)}]{Suzuki1976}%
  \BibitemOpen
  \bibfield  {author} {\bibinfo {author} {\bibfnamefont {M.}\ \bibnamefont
  {Suzuki}},\ }\bibfield  {title} {\enquote {\bibinfo {title} {Generalized
  trotter{\textquotesingle}s formula and systematic approximants of exponential
  operators and inner derivations with applications to many-body problems}}\
  }\href {\doibase 10.1007/bf01609348} {\bibfield  {journal} {\bibinfo
  {journal} {Communications in Mathematical Physics}\ }\textbf {\bibinfo
  {volume} {51}},\ \bibinfo {pages} {183--190} (\bibinfo {year}
  {1976})}\BibitemShut {NoStop}%
\bibitem [{\citenamefont {Childs}\ \emph {et~al.}(2018)\citenamefont {Childs},
  \citenamefont {Maslov}, \citenamefont {Nam}, \citenamefont {Ross},\ and\
  \citenamefont {Su}}]{Childs2018}%
  \BibitemOpen
  \bibfield  {author} {\bibinfo {author} {\bibfnamefont {A.~M.}\
  \bibnamefont {Childs}}, \bibinfo {author} {\bibfnamefont {D.}\
  \bibnamefont {Maslov}}, \bibinfo {author} {\bibfnamefont {Y.}\
  \bibnamefont {Nam}}, \bibinfo {author} {\bibfnamefont {N.~J.}\ \bibnamefont
  {Ross}}, \ and\ \bibinfo {author} {\bibfnamefont {Y.}\ \bibnamefont {Su}},\
  }\bibfield  {title} {\enquote {\bibinfo {title} {Toward the first quantum
  simulation with quantum speedup}}\ }\href {\doibase 10.1073/pnas.1801723115}
  {\bibfield  {journal} {\bibinfo  {journal} {Proceedings of the National
  Academy of Sciences}\ }\textbf {\bibinfo {volume} {115}},\ \bibinfo {pages}
  {9456--9461} (\bibinfo {year} {2018})}\BibitemShut {NoStop}%
\bibitem [{\citenamefont {Seetharam}\ \emph {et~al.}(2021)\citenamefont
  {Seetharam}, \citenamefont {Biswas}, \citenamefont {Noel}, \citenamefont
  {Risinger}, \citenamefont {Zhu}, \citenamefont {Katz}, \citenamefont
  {Chattopadhyay}, \citenamefont {Cetina}, \citenamefont {Monroe},
  \citenamefont {Demler},\ and\ \citenamefont {Sels}}]{Seetharam2021}%
  \BibitemOpen
  \bibfield  {author} {\bibinfo {author} {\bibfnamefont {K.}\ \bibnamefont
  {Seetharam}}, \bibinfo {author} {\bibfnamefont {D.}\ \bibnamefont
  {Biswas}}, \bibinfo {author} {\bibfnamefont {C.}\ \bibnamefont {Noel}},
  \bibinfo {author} {\bibfnamefont {A.}\ \bibnamefont {Risinger}}, \bibinfo
  {author} {\bibfnamefont {D.}\ \bibnamefont {Zhu}}, \bibinfo {author}
  {\bibfnamefont {O.}~\bibnamefont {Katz}}, \bibinfo {author} {\bibfnamefont
  {S.}\ \bibnamefont {Chattopadhyay}}, \bibinfo {author} {\bibfnamefont
  {M.}\ \bibnamefont {Cetina}}, \bibinfo {author} {\bibfnamefont
  {C.}\ \bibnamefont {Monroe}}, \bibinfo {author} {\bibfnamefont
  {E.}\ \bibnamefont {Demler}}, and\ \bibinfo {author} {\bibfnamefont
  {D.}\ \bibnamefont {Sels}},\ }
  {\enquote {\bibinfo {title} {Digital quantum simulation of NMR
  experiments}}\ }\href{\doibase 10.48550/ARXIV.2109.13298}  {\bibfield  {journal} {\bibinfo  {journal}
  {arXiv:}}\bibinfo {pages} {2109.13298}}\BibitemShut {NoStop}%
\bibitem [{\citenamefont {Aleksandrowicz}\ \emph {et~al.}(2019)\citenamefont
  {Aleksandrowicz}, \citenamefont {Alexander}, \citenamefont {Barkoutsos},
  \citenamefont {Bello}, \citenamefont {Ben-Haim}, \citenamefont {Bucher},
  \citenamefont {Cabrera-Hernández}, \citenamefont {Carballo-Franquis},
  \citenamefont {Chen}, \citenamefont {Chen}, \citenamefont {Chow},
  \citenamefont {Córcoles-Gonzales}, \citenamefont {Cross}, \citenamefont
  {Cross}, \citenamefont {Cruz-Benito}, \citenamefont {Culver}, \citenamefont
  {González}, \citenamefont {Torre}, \citenamefont {Ding}, \citenamefont
  {Dumitrescu}, \citenamefont {Duran}, \citenamefont {Eendebak}, \citenamefont
  {Everitt}, \citenamefont {Sertage}, \citenamefont {Frisch}, \citenamefont
  {Fuhrer}, \citenamefont {Gambetta}, \citenamefont {Gago}, \citenamefont
  {Gomez-Mosquera}, \citenamefont {Greenberg}, \citenamefont {Hamamura},
  \citenamefont {Havlicek}, \citenamefont {Hellmers}, \citenamefont {{Łukasz
  Herok}}, \citenamefont {Horii}, \citenamefont {{Shaohan Hu}}, \citenamefont
  {Imamichi}, \citenamefont {{Toshinari Itoko}}, \citenamefont {Javadi-Abhari},
  \citenamefont {Kanazawa}, \citenamefont {Karazeev}, \citenamefont {Krsulich},
  \citenamefont {Liu}, \citenamefont {Luh}, \citenamefont {{Yunho Maeng}},
  \citenamefont {Marques}, \citenamefont {Martín-Fernández}, \citenamefont
  {McClure}, \citenamefont {McKay}, \citenamefont {{Srujan Meesala}},
  \citenamefont {Mezzacapo}, \citenamefont {Moll}, \citenamefont {Rodríguez},
  \citenamefont {Nannicini}, \citenamefont {Nation}, \citenamefont
  {Ollitrault}, \citenamefont {O'Riordan}, \citenamefont {{Hanhee Paik}},
  \citenamefont {Pérez}, \citenamefont {Phan}, \citenamefont {Pistoia},
  \citenamefont {Prutyanov}, \citenamefont {Reuter}, \citenamefont {Rice},
  \citenamefont {{Abdón Rodríguez Davila}}, \citenamefont {Rudy},
  \citenamefont {{Mingi Ryu}}, \citenamefont {{Ninad Sathaye}}, \citenamefont
  {Schnabel}, \citenamefont {Schoute}, \citenamefont {{Kanav Setia}},
  \citenamefont {{Yunong Shi}}, \citenamefont {{Adenilton Silva}},
  \citenamefont {Siraichi}, \citenamefont {{Seyon Sivarajah}}, \citenamefont
  {Smolin}, \citenamefont {Soeken}, \citenamefont {Takahashi}, \citenamefont
  {Tavernelli}, \citenamefont {Taylor}, \citenamefont {Taylour}, \citenamefont
  {{Kenso Trabing}}, \citenamefont {Treinish}, \citenamefont {Turner},
  \citenamefont {Vogt-Lee}, \citenamefont {Vuillot}, \citenamefont {Wildstrom},
  \citenamefont {Wilson}, \citenamefont {Winston}, \citenamefont {Wood},
  \citenamefont {Wood}, \citenamefont {Wörner}, \citenamefont {Akhalwaya},\
  and\ \citenamefont {Zoufal}}]{Aleksandrowicz2019}%
  \BibitemOpen
  \bibfield  {author} {\bibinfo {author} {\bibfnamefont {G.}\ \bibnamefont
  {Aleksandrowicz}} \bibinfo {author} {\bibfnamefont {and }\ \bibnamefont
  {al.}}} \href {\doibase
  10.5281/ZENODO.2562111} {\enquote {\bibinfo {title} {Qiskit: An open-source
  framework for quantum computing}}\ } (\bibinfo {year} {2019})\BibitemShut
  {NoStop}%
\bibitem [{\citenamefont {Tinkham}(1996)}]{Tinkham1996}%
  \BibitemOpen
  \bibfield  {author} {\bibinfo {author} {\bibfnamefont {M.}~\bibnamefont
  {Tinkham}},\ }\href@noop {} { \bibinfo {title} {Introduction to
  Superconductivity, 2nd Edition,}}\ (\bibinfo  {publisher} {McGraw-Hill, New
  York},\ \bibinfo {year} {1996})\BibitemShut {NoStop}%
\bibitem {Note2}%
  \BibitemOpen
  \bibinfo {note} {\new{After each Trotter step, logical qubits have been swapped
  around and are not at their initial location. To simulate strictly the
  operator of Eq.~\protect \eqref {eq:Trotter}, one should put them back to
  their original configuration at the end of each Trotter step again using
  additional SWAP gates. However, the order in which the interaction is applied
  does not matter, as we neglected the non-commuting part of the evolution
  operator within the Trotter approximation. To limit the number of CNOT gates,
  we keep track of the positions of the logical qubits and apply the following
  Trotter step accordingly.}}\BibitemShut {Stop}%
\bibitem [{\citenamefont {Jurcevic}\ \emph {et~al.}(2021)\citenamefont
  {Jurcevic}, \citenamefont {Javadi-Abhari}, \citenamefont {Bishop},
  \citenamefont {Lauer}, \citenamefont {Bogorin}, \citenamefont {Brink},
  \citenamefont {Capelluto}, \citenamefont {Günlük}, \citenamefont {Itoko},
  \citenamefont {Kanazawa}, \citenamefont {Kandala}, \citenamefont {Keefe},
  \citenamefont {Krsulich}, \citenamefont {Landers}, \citenamefont
  {Lewandowski}, \citenamefont {McClure}, \citenamefont {Nannicini},
  \citenamefont {Narasgond}, \citenamefont {Nayfeh}, \citenamefont {Pritchett},
  \citenamefont {Rothwell}, \citenamefont {Srinivasan}, \citenamefont
  {Sundaresan}, \citenamefont {Wang}, \citenamefont {Wei}, \citenamefont
  {Wood}, \citenamefont {Yau}, \citenamefont {Zhang}, \citenamefont {Dial},
  \citenamefont {Chow},\ and\ \citenamefont {Gambetta}}]{Jurcevic2021}%
  \BibitemOpen
  \bibfield  {author} {\bibinfo {author} {\bibfnamefont {P.}\ \bibnamefont
  {Jurcevic}}, \bibinfo {author} {\bibfnamefont {and}\ \bibnamefont
  {al.}} }\bibfield  {title} {\enquote {\bibinfo {title}
  {Demonstration of quantum volume 64 on a superconducting quantum computing
  system}}\ }\href {\doibase 10.1088/2058-9565/abe519} {\bibfield  {journal}
  {\bibinfo  {journal} {Quantum Science and Technology}\ }\textbf {\bibinfo
  {volume} {6}},\ \bibinfo {pages} {025020} (\bibinfo {year}
  {2021})}\BibitemShut {NoStop}%
\bibitem [{\citenamefont {Nachman}\ \emph {et~al.}(2020)\citenamefont
  {Nachman}, \citenamefont {Urbanek}, \citenamefont {de~Jong},\ and\
  \citenamefont {Bauer}}]{Nachman2020}%
  \BibitemOpen
  \bibfield  {author} {\bibinfo {author} {\bibfnamefont {B.}\
  \bibnamefont {Nachman}}, \bibinfo {author} {\bibfnamefont {M.}\
  \bibnamefont {Urbanek}}, \bibinfo {author} {\bibfnamefont {W.~A.}\
  \bibnamefont {de~Jong}}, \ and\ \bibinfo {author} {\bibfnamefont
  {C.~W.}\ \bibnamefont {Bauer}},\ }\bibfield  {title} {\enquote
  {\bibinfo {title} {Unfolding quantum computer readout noise}}\ }\href
  {\doibase 10.1038/s41534-020-00309-7} {\bibfield  {journal} {\bibinfo
  {journal} {npj Quantum Information}\ }\textbf {\bibinfo {volume} {6}}
  (\bibinfo {year} {2020})}\BibitemShut {NoStop}%
  \bibitem [{\citenamefont {Hines}\ \emph {et~al.}(2023)\citenamefont {Hines},
  \citenamefont {Lu}, \citenamefont {Naik}, \citenamefont {Hashim},
  \citenamefont {Ville}, \citenamefont {Mitchell}, \citenamefont {Kriekebaum},
  \citenamefont {Santiago}, \citenamefont {Seritan}, \citenamefont {Nielsen},
  \citenamefont {Blume-Kohout}, \citenamefont {Young}, \citenamefont {Siddiqi},
  \citenamefont {Whaley},\ and\ \citenamefont {Proctor}}]{Hines2023}%
  \BibitemOpen
  \bibfield  {author} {\bibinfo {author} {\bibfnamefont {J.}\ \bibnamefont
  {Hines}}, \bibinfo {author} {\bibfnamefont {M.}\ \bibnamefont {Lu}},
  \bibinfo {author} {\bibfnamefont {R.~K.}\ \bibnamefont {Naik}}, \bibinfo
  {author} {\bibfnamefont {A.}\ \bibnamefont {Hashim}}, \bibinfo {author}
  {\bibfnamefont {J.-L.}\ \bibnamefont {Ville}}, \bibinfo {author}
  {\bibfnamefont {B.}\ \bibnamefont {Mitchell}}, \bibinfo {author}
  {\bibfnamefont {J.~M.}\ \bibnamefont {Kriekebaum}}, \bibinfo {author}
  {\bibfnamefont {D.~I.}\ \bibnamefont {Santiago}}, \bibinfo {author}
  {\bibfnamefont {S.}\ \bibnamefont {Seritan}}, \bibinfo {author}
  {\bibfnamefont {E.}\ \bibnamefont {Nielsen}}, \bibinfo {author}
  {\bibfnamefont {R.}\ \bibnamefont {Blume-Kohout}}, \bibinfo {author}
  {\bibfnamefont {K.}\ \bibnamefont {Young}}, \bibinfo {author}
  {\bibfnamefont {I.}\ \bibnamefont {Siddiqi}}, \bibinfo {author}
  {\bibfnamefont {B.}\ \bibnamefont {Whaley}}, \ and\ \bibinfo {author}
  {\bibfnamefont {T.}\ \bibnamefont {Proctor}},\ }\bibfield  {title}
  {\enquote {\bibinfo {title} {Demonstrating scalable randomized benchmarking
  of universal gate sets},}\ }\href {\doibase 10.1103/PhysRevX.13.041030}
  {\bibfield  {journal} {\bibinfo  {journal} {Phys. Rev. X}\ }\textbf {\bibinfo
  {volume} {13}},\ \bibinfo {pages} {041030} (\bibinfo {year}
  {2023})}\BibitemShut {NoStop}%
\bibitem [{\citenamefont {L.~Shirizly}()}]{Shirizly2023}%
  \BibitemOpen
  \bibfield  {author} {\bibinfo {author} {\bibfnamefont {H.~Landa}\
  \bibnamefont {L.~Shirizly}, \bibfnamefont {G.~Misguich}},\
  }\bibfield  {title} {\enquote {\bibinfo {title} {Dissipative dynamics of
  graph-state stabilizers with superconducting qubits},}\ }\href@noop {} {\
  }\Eprint {http://arxiv.org/abs/2308.01860} {arXiv:2308.01860 [quant-ph]}
  \BibitemShut {NoStop}%

\bibitem [{\citenamefont {Llano}\ and\ \citenamefont
  {Annett}(2007)}]{Llano2007}%
  \BibitemOpen
  \bibfield  {author} {\bibinfo {author} {\bibfnamefont {M.~De}\ \bibnamefont
  {Llano}}\ and\ \bibinfo {author} {\bibfnamefont {J.~F.}\ \bibnamefont
  {Annett}},\ }\bibfield  {title} {\enquote {\bibinfo {title} {Generalized
  Cooper pairing in superconductors}}\ }\href {\doibase
  10.1142/s0217979207037661} {\bibfield  {journal} {\bibinfo  {journal}
  {International Journal of Modern Physics B}\ }\textbf {\bibinfo {volume}
  {21}},\ \bibinfo {pages} {3657--3686} (\bibinfo {year} {2007})}\BibitemShut
  {NoStop}%
\bibitem [{\citenamefont {Anderson}(1958)}]{Anderson1958}%
  \BibitemOpen
  \bibfield  {author} {\bibinfo {author} {\bibfnamefont {P.~W.}\ \bibnamefont
  {Anderson}},\ }\bibfield  {title} {\enquote {\bibinfo {title} {Random-phase
  approximation in the theory of superconductivity}}\ }\href {\doibase
  10.1103/physrev.112.1900} {\bibfield  {journal} {\bibinfo  {journal}
  {Physical Review}\ }\textbf {\bibinfo {volume} {112}},\ \bibinfo {pages}
  {1900--1916} (\bibinfo {year} {1958})}\BibitemShut {NoStop}%
\end{thebibliography}

%

\end{document}